\documentclass[12pt]{article}

\usepackage[a4paper,pdftex]{geometry}
\usepackage[T1]{fontenc}
\usepackage[english]{babel}
\usepackage[authoryear]{natbib}
\usepackage[small]{titlesec}
\usepackage[justification=centering]{caption}
\usepackage{hyperref}

\usepackage{amsmath,amsthm,amssymb,graphicx,enumerate,booktabs,bigstrut,rotating,multirow,float,etoolbox,lmodern,comment,listings,qtree,a4wide,moresize,bbm,subcaption,tikz,pdfescape,mathtools,flafter,enumitem,setspace,ragged2e,colortbl,lineno,lscape,pdflscape,stackengine}
\usetikzlibrary{snakes,decorations.pathreplacing,positioning, arrows.meta,shapes}

\DeclareMathOperator{\E}{\mathbb{E}}

\newcolumntype{L}[1]{>{\raggedright\let\newline\\\arraybackslash\hspace{0pt}}m{#1}}
\newcolumntype{C}[1]{>{\centering\let\newline\\\arraybackslash\hspace{0pt}}m{#1}}
\newcolumntype{R}[1]{>{\raggedleft\let\newline\\\arraybackslash\hspace{0pt}}m{#1}}

\hypersetup{colorlinks,linkcolor={red},citecolor={blue},urlcolor={blue}}

\makeatletter
\renewcommand\subsubsection{\@startsection{subsubsection}{3}{\z@}%
	{-3.25ex\@plus -1ex \@minus -.2ex}%
	{-1.5ex \@plus -.2ex}%
	{\normalfont\normalsize\bfseries}
}
\def\@biblabel#1{\hspace*{-\labelsep}}

\makeatother

\def\sym#1{\ifmmode^{#1}\else\(^{#1}\)\fi}

\pdfpageheight\paperheight
\pdfpagewidth\paperwidth

\theoremstyle{definition}
\newtheorem{prediction}{Prediction}

\newcommand*\circled[1]{\tikz[baseline=(char.base)]{
		\node[shape=circle,draw,inner sep=2pt] (char) {#1};}}

\begin{document}

	\title{Coarse Wage-Setting and Behavioral Firms} 
	
	\author{Germán Reyes\thanks{\noindent Department of Economics, Middlebury College, 014 Warner Hall, Middlebury, VT 05753 (email: greyes@middlebury.edu). I thank especially my advisor Ted O'Donoghue for invaluable guidance. For helpful comments, I am grateful to Nava Ashraf, Levon Barseghyan, Michele Belot, Douglas Bernheim, Aviv Caspi, Neel Datta, Stefano DellaVigna, Christa Deneault, Rebecca Deranian, Michael Grubb, Ori Heffetz, Guy Ishai, Patrick Kline, Philipp Kircher, David Laibson, Francesca Molinari, Ricardo Perez-Truglia, Marcel Preuss, Alex Rees-Jones, Evan Riehl, Seth Sanders, Jason Somerville, Dmitry Taubinsky, David Wasser, participants in the Russell Sage Foundation summer institute in behavioral economics, participants in the Cornell behavioral economics group, and participants in the UC Berkeley psychology and economics group. Errors and omissions are my own.}}
	
	\renewcommand{\today}{\ifcase \month \or January\or February\or March\or %
		April\or May \or June\or July\or August\or September\or October\or November\or %
		December\fi \ \number \year} 
	\date{\today}
	
	\maketitle

	\begin{abstract} 

		\noindent This paper shows that the bunching of wages at round numbers is partly driven by firm coarse wage-setting. Using data from over 200 million new hires in Brazil, I first establish that contracted salaries tend to cluster at round numbers. Then, I show that firms that tend to hire workers at round-numbered salaries have worse market outcomes. Next, I develop a wage-posting model in which optimization costs lead to the adoption of coarse rounded wages and provide evidence supporting two model predictions using two research designs. Finally, I examine some consequences of coarse wage-setting for relevant economic outcomes. 
	\end{abstract}

	\clearpage	
	\section{Introduction}

The wage-formation process is a central element of many economic models. While standard wage-formation models assume that workers and firms behave optimally, recent findings challenge this assumption. Both survey and administrative data reveal a tendency for wages to cluster---or ``bunch''---at round numbers \citep{riddles_handling_2016, dube_monopsony_2020}. This puzzling finding suggests non-standard behavior by some market participants. According to one view, the bunching is driven by strategic behavior on the firm side, whereby firms pay round-numbered wages to exploit a worker behavioral bias (e.g., left-digit bias). Alternatively, the bunching might reflect the behavior of firms engaging in non-standard wage-setting, possibly due to misoptimization.\footnote{Throughout the paper, I use the term ``non-standard'' to refer to any behavior that departs from the predictions of the neoclassical model. Specifically, ``non-standard wage-setting'' refers to firm wage-setting practices that depart from the first-order condition of canonical wage-formation models. See Appendix \ref{app:theory} for a description of the two main classes of wage-setting models.}

In this paper, I use rich worker-firm matched data to assess whether the wage bunching is partly due to firm non-standard wage-setting. First, I establish the existence of substantial bunching at round-numbered salaries in the data. Then, I provide a set of reduced-form results compatible with firms engaging in coarse wage-setting and inconsistent with firms paying round-numbered wages to exploit a worker bias. Motivated by the reduced-form findings, I develop a wage-posting model in which firms pay coarse rounded salaries due to optimization costs. The model delivers two predictions for which I find support using two research designs. Finally, I quantify some of the downstream consequences of coarse wage-setting for within-firm wage inequality, nominal wage stickiness, and policies that affect the wage distribution, such as changes in the minimum wage.

I use an administrative employee-employer matched dataset covering the universe of formal-sector firms in Brazil from 2003 to 2017. This dataset contains the salary at which firms hire workers (``contracted salary''). I use data on the contracted monthly salary of over 200 million new hires. In addition, I use a sample of over 300,000 firms that includes information on all of their employees. Using this data, I document the existence of substantial bunching at round-numbered salaries (i.e., those divisible by 10) in the distribution of contracted salaries. For example, 33.8\% of new hires' contracted salaries are round numbers (a uniform distribution would imply 10\%). My findings stand in opposition to the predictions of canonical wage-determination models, in which market-level wages should be smoothly distributed. 

Then, I present a series of reduced-form results to shed light on whether worker or firm non-standard behavior drives the bunching. I begin by identifying a set of firms (``bunching firms'') that tend to hire workers at round-numbered salaries. Next, I compare the market outcomes of bunching firms with those of non-bunching firms. If bunching firms pay round-numbered wages to exploit a worker bias, one would expect these firms to have better market performance than non-bunching firms. However, I find that conditional on a large set of controls, bunching firms tend to experience worse outcomes. For instance, they have worse worker-firm matches as measured by new hires' separation likelihood, a lower job growth rate, and they are more likely to exit the market. 

These reduced-form results suggest that bunching firms do not pay round salaries to exploit a worker bias. Alternatively, firms may pay round-numbered salaries as a simple but coarse approximation when they are uncertain about what the fully-optimal salary is. When hiring a new worker, firms face considerable uncertainty about a worker's marginal revenue product (or ``productivity''). Estimating a worker's contribution to the firm requires answering complex questions: What are all of the possible tasks that the new hire will perform? How does each of these tasks affect the firm's bottom line? How likely is the prospective employee to successfully accomplish each of these tasks? Instead of attempting to gather all of the information required to compute worker productivity, firms might rely on a rule-of-thumb or heuristic as an approximation---a form of pricing that I refer to as ``coarse wage-setting.''

As a suggestive reduced-form test for the use of coarse wage-setting, I assess whether bunching firms also rely on coarse figures when deciding on salary increases. This is a different environment in which firms also face uncertainty about the optimal action. The canonical model predicts that workers' wage increase depends on their realized productivity \citep{jovanovic1979job}. Since this variable is difficult to measure, some firms might use coarse approximations as salary increases, such as integer numbers if the salary increase is measured in percentage terms or round numbers if the increase is measured in monetary units. I find that bunching firms also rely on coarse approximations while deciding wage increases. Bunching firms are 26 percentage points more likely to offer a round-numbered salary increase in monetary units (from a baseline of 20.4\%) and 9 percentage points more likely to offer an integer salary increase in percent terms (from a baseline of 12.9\%).

The reduced-form results motivate the hypothesis that coarse wage-setting is partly what drives the bunching observed in the data. To further explore this hypothesis, I build a wage-posting model in which coarse wage-setting is a consequence of optimization costs. The goal of the model is to account for the bunching observed in the data and to generate ancillary predictions that should hold if firms engage in coarse wage-setting. The model relies on two key assumptions that I motivate based on numerical cognition research: first, that firms use a rounding heuristic to form an initial estimate of the fully-optimal salary (the salary firms would pay if there were no optimization costs), and second, that firms can generate a more precise estimate of the fully-optimal salary by paying some cost. The standard wage-posting model is a special case of the model with frictions, in which the optimization cost is zero.

The model delivers two testable predictions that characterize the conditions under which firms are more likely to hire workers at coarse round-numbered wages. First, a smaller expected gap between the coarse wage and the fully-optimal wage should increase the likelihood of firms paying the coarse wage. In the model, the firm's benefit of fully optimizing is proportional to this gap. Hence, as this benefit decreases, firms are less likely to pay the cost of computing the fully-optimal wage. Second, firms with lower optimization costs should be less likely to pay coarse wages and more likely to pay the fully-optimal wage. 

I test the model's predictions using two research designs. The first empirical strategy is a bunching design that uses standard techniques in the bunching literature \citep{kleven_bunching_2016}. This design consists of correlating the fraction of workers hired through a coarse rounding heuristic with firm and worker characteristics. I partition the data based on firm and worker characteristics and recover the fraction of workers hired through a coarse wage-setting using the ``excess mass'' in the density of workers earning round-numbered salaries. Second, I estimate linear probability models with firm fixed effects, where the dependent variable is an indicator for paying a round-numbered salary to a new hire. This design allows me to control for a large set of confounding factors, including unobserved heterogeneity at the firm level. The two research designs deliver similar results in support of the model's predictions.

While my findings are consistent with non-standard firm wage-setting, they do not imply that firms are making mistakes. For instance, firms may be adopting management practices that, despite leading to coarse wage-setting, may improve overall firm performance. Still, these findings are important for three main reasons. %
First, round-numbered wages make up a disproportionate share of all wages. Thus, understanding why firms pay round-numbered wages sheds light on overall firm wage-setting and can inform the modeling assumptions of wage-setting models. %
Second, research designs that infer parameter values from firm optimality conditions might yield biased estimates due to coarse wage-setting. Researchers within the structural tradition often infers unobservable variables using the firm's first-order conditions (FOC).\footnote{For example, in the context of a wage-posting model, a researcher equipped with wage data and an estimate of worker productivity could use a firm's FOC to identify the labor supply elasticity. This particular strategy has gained traction in recent years as researchers are increasingly interested in understanding imperfect competition in the labor market \citep[e.g.,][]{lamadon2022imperfect}.} However, if firms do not fully optimize with respect to wages, the FOC may not characterize firms' pricing decisions. %
Third, coarse wage-setting can have downstream consequences for important economic outcomes. I show that coarse wage-setting may lead to within-firm wage compression and may increase wage stickiness. In addition, in the presence of firms that engage in coarse wage-setting, policies that affect the earnings distribution can affect firm wage-optimization behavior.

This paper contributes to empirical studies of firm wage-setting.\footnote{See, among others, \cite{hall2012evidence,caldwell2019outside, hjort2020across, derenoncourt2021spillover, lachowska2021wage, cullen2022s, hazell2022national}.} At least since \cite{jones1896round}, labor economists have documented the bunching of salaries at round numbers. Some work has suggested that this bunching arises from measurement error \citep[e.g.,][]{schweitzer_rounding_1996}. I contribute by documenting bunching in an administrative dataset where earnings are not self-reported, which shows that round-number bunching is a real feature of labor markets.\footnote{Other papers have also documented the bunching of earnings at round numbers \citep[e.g.,][]{kleven2013using, devereux2014elasticity, mavrokonstantis2022bunching}. In these papers, the excess mass of salaries at round numbers is a nuance parameter, not the object of interest.} The most closely related paper is \cite{dube_monopsony_2020}. Using unemployment insurance records from the US, they document substantial bunching at \$10 per hour and show that this pattern is not explained by worker left-digit bias. This raises the possibility that firm non-standard behavior drives the bunching. I contribute in two main ways, first by establishing a novel set of stylized facts about firms that frequently hire workers at round-numbered wages, and second by quantifying some downstream consequences of such coarse wage-setting on important outcomes.

This paper also contributes to an emerging empirical literature on simplified firm pricing strategies. The view that firms set prices based on heuristics and simplified rules dates back to \cite{simon1962new}, who noted that ``price setting involves an enormous burden of information gathering and computation that precludes the use of any but simple rules of thumb as guiding principles.'' Recent empirical work provides evidence to support Simon's claim. For example, \cite{cho2010flat} show that car companies charge a uniform rental price across cars with heterogeneous odometer values, \cite{cavallo2014currency} find that global retailers engage in uniform pricing across heterogeneous countries, and \cite{dellavigna2019uniform} show that US retail chains engage in uniform pricing across heterogeneous outlets.\footnote{Other work shows that many firms follow coarse pricing policies \citep[e.g.,][]{matejka_rationally_2016, stevens2020coarse}.} While these papers focus on the goods market, I contribute by documenting a form of simplified pricing in the labor market.

Finally, this paper contributes to the literature studying market outcomes in the presence of behavioral firms. Compared to the ever-growing number of papers that document biases in individuals' behavior, work on firm heuristics and biases is scarce.\footnote{See \cite{heidhues_behavioral_2018} for a theoretical overview of this literature and \cite{kremer2019behavioral} for work on behavioral firms in developing countries. Among the empirical papers that study behavioral firms, previous work has shown that entrepreneurs are overconfident regarding future growth \citep{landier2008financial}, restaurant owners do not account for the transitory nature of weather shocks \citep{goldfarb2019transitory}, car dealerships exhibit loss aversion \citep{pierce2020negative}, and retailers underestimate the degree of consumers' left-digit bias \citep{strulov2019more}. A closely related body of work documents firms' failure to maximize profits \citep[e.g.,][]{hanna2014learning, bloom2013does, almunia2021strategic}.} This is partly due to data limitations. Most of the heuristics and biases body of work studies individuals' behavior in carefully-controlled lab environments. There is not a straightforward way of conducting the same type of experiments using firms as research subjects. I contribute by providing field evidence on firm non-standard behavior in a high-stakes setting. 
	
	\section{Institutional Context, Data, and Descriptive Statistics} \label{sec:context-data}

This section provides institutional context on Brazil's labor market, describes the administrative dataset, and provides descriptive statistics of the samples.

\subsection{Brazil's Labor Market} \label{subsec_context}

Brazil's labor market has both a formal and an informal sector (see Appendix \ref{app:inform}). I focus on the formal sector, which employs about 80\% of wage employees and has a strict labor code. The contracts of formal-sector workers are governed by the Brazilian Labor Code, which mandates provisions such as a relatively high minimum wage, an extra monthly salary annually, one month of paid leave each year, and high firing costs.

\subsection{Data: Employee-Employer Matched Information} \label{sub:data}

The main data source is the \textit{Relação Anual de Informações Sociais} (RAIS), an employee-employer matched dataset covering the universe of formal-sector jobs in Brazil from 2003 to 2017. This administrative dataset is assembled yearly by the Ministry of Labor with information provided by firms. Accurate reporting in the RAIS is required for workers to receive payments from some government programs, and firms face financial penalties for not reporting.\footnote{The main drawback of the RAIS is that it only contains information on formal-sector workers and firms. Thus, the analysis is not representative of the informal sector. Appendix \ref{app:inform} uses data from the Brazilian Household Hurvey, which includes data on informal-sector workers, to compare workers in the RAIS to workers in the entire labor force.}

The RAIS contains both firm- and worker-level information. Firms' data include the number of employees, industry, and location. Workers' data include educational attainment, occupation, and employment information, including the hiring date, recruitment type (e.g., new hire, transfer, etc.), and contracted salary.

The contracted salary of a worker is central to the empirical analysis. It is the salary contained in an individual's ``Work and Social Security Card'' (or CTPS) at the end of each year.\footnote{Appendix \ref{app:handbook} provides an example of a CTPS and the information it contains.} The CTPS documents a worker's employment history, including the initial salary at the firm and any subsequent modifications. For a new hire, the contracted salary is the initial salary at which the firm hired the worker. For other workers, the contracted salary might differ from the initial salary due to a raise or promotion, for instance.

\subsection{Samples and Descriptive Statistics} \label{sub:summ}

\textit{New-hires sample.} For much of the empirical analysis, I use a \textit{new-hires sample}, in which each observation represents a new hire (defined by a worker-firm-hiring date triplet). To construct this sample, I only include new workers hired each year by private-sector firms. For workers holding multiple positions at a firm in a given year, I only keep the highest-paying position. I exclude new hires without a valid identification number or a reported contracted salary below the federal monthly minimum wage. Finally, I only keep new hires with a monthly earnings contract. This excludes, for example, workers who bill by the hour or per day worked, which constitute a small fraction of workers in the data. After imposing these restrictions, the database contains information on the contracted salary of 206 million hires (henceforth, ``contracts'' or ``workers'' for short) from 2003 to 2017.\footnote{Appendix \ref{app:samp-rest} provides more detail on each of these steps and show the fraction of excluded observations after each sample restriction.}

\textit{Firm random sample.} To conduct any analysis that requires exploiting the panel structure of the dataset, I select a random sample of firms. To construct this sample, I create a registry of all private-sector firms ever observed in the RAIS who hired at least one worker at a monthly earnings contract during 2003--2017. Due to computational constraints, I randomly select 5\% of them. I track all of the employees of these firms over time (both new hires and other employees). This sample includes over 300,000 firms, 1.8 million firm-years, and 31.8 million worker-years.

\textit{Descriptive statistics.} Table \ref{tab_rais_summ} presents summary statistics of workers in the RAIS and the samples. The average worker in the new-hires sample is 30.3 years old. Most workers are male (63.5\%), white (57.4\%), and have completed high school (53.2\%). The average monthly salary is R\$807 (approximately \$373). Most workers are employed by the retail industry (35.9\%), followed by the services industry (26.5\%). Workers in the firm random sample have similar characteristics.

	\section{Bunching in the Distribution of Contracted Salaries}  \label{sec:anomalies}

The two main classes of wage-formation models in labor economics are wage-posting models and wage-bargaining models \citep{manning_imperfect_2011}. Under standard assumptions, both types of models predict a smooth distribution of wages at the market level (see Appendix \ref{app:labor-models}). 

\begin{figure}[H]
	\caption{Bunching at round numbers in the salary distribution}\label{fig_bunching}
	\centering
		\centering
	\begin{subfigure}[t]{.48\textwidth}
		\caption*{Panel A. Distribution of contracted earnings in R\$1 bins}
		\centering
		\includegraphics[width=\linewidth]{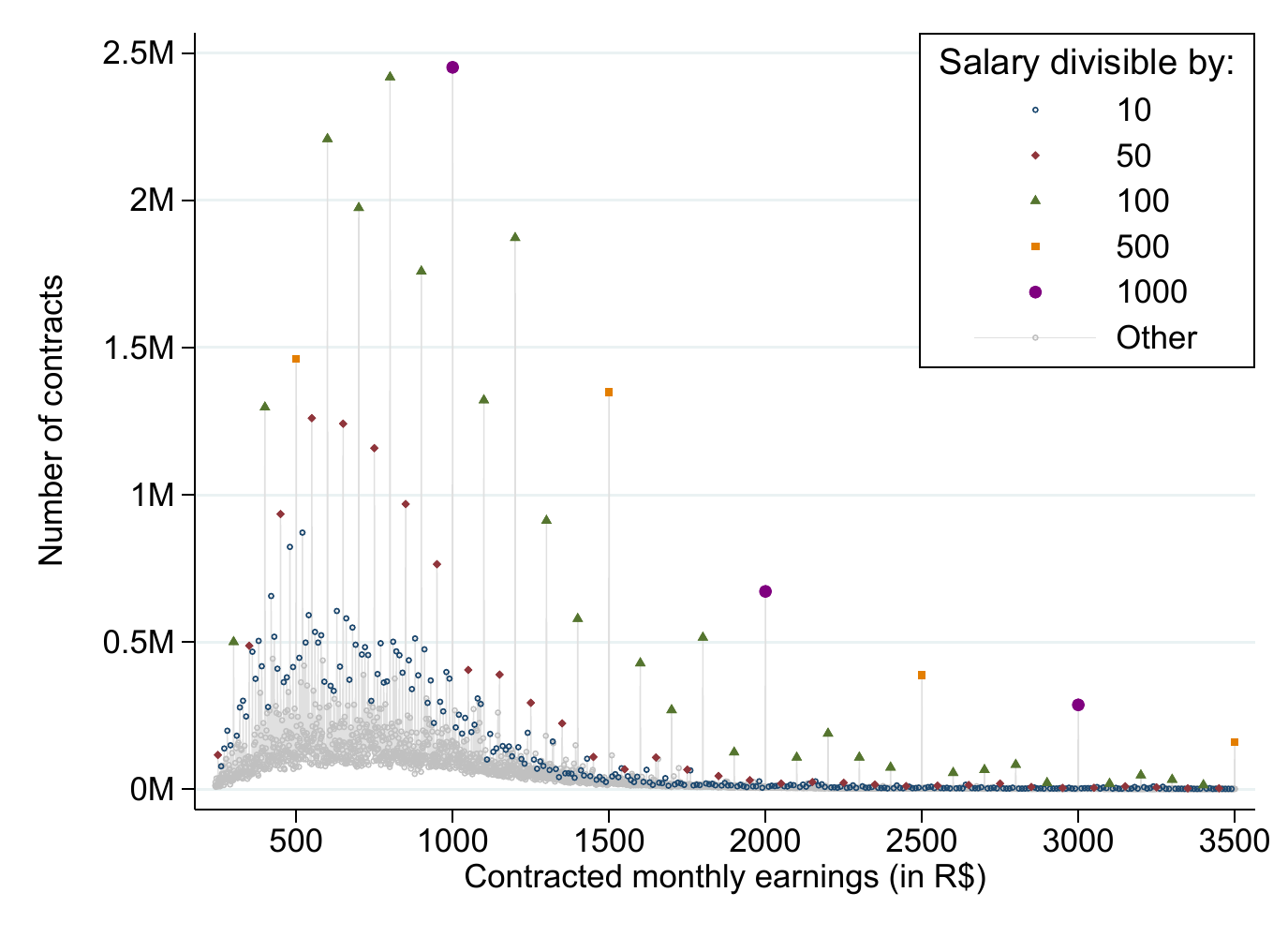}
	\end{subfigure}
	\hfill		
	\begin{subfigure}[t]{0.48\textwidth}
		\caption*{Panel B. Fraction of salaries divisible by round numbers: observed vs. uniform}
		\centering
		\includegraphics[width=\linewidth]{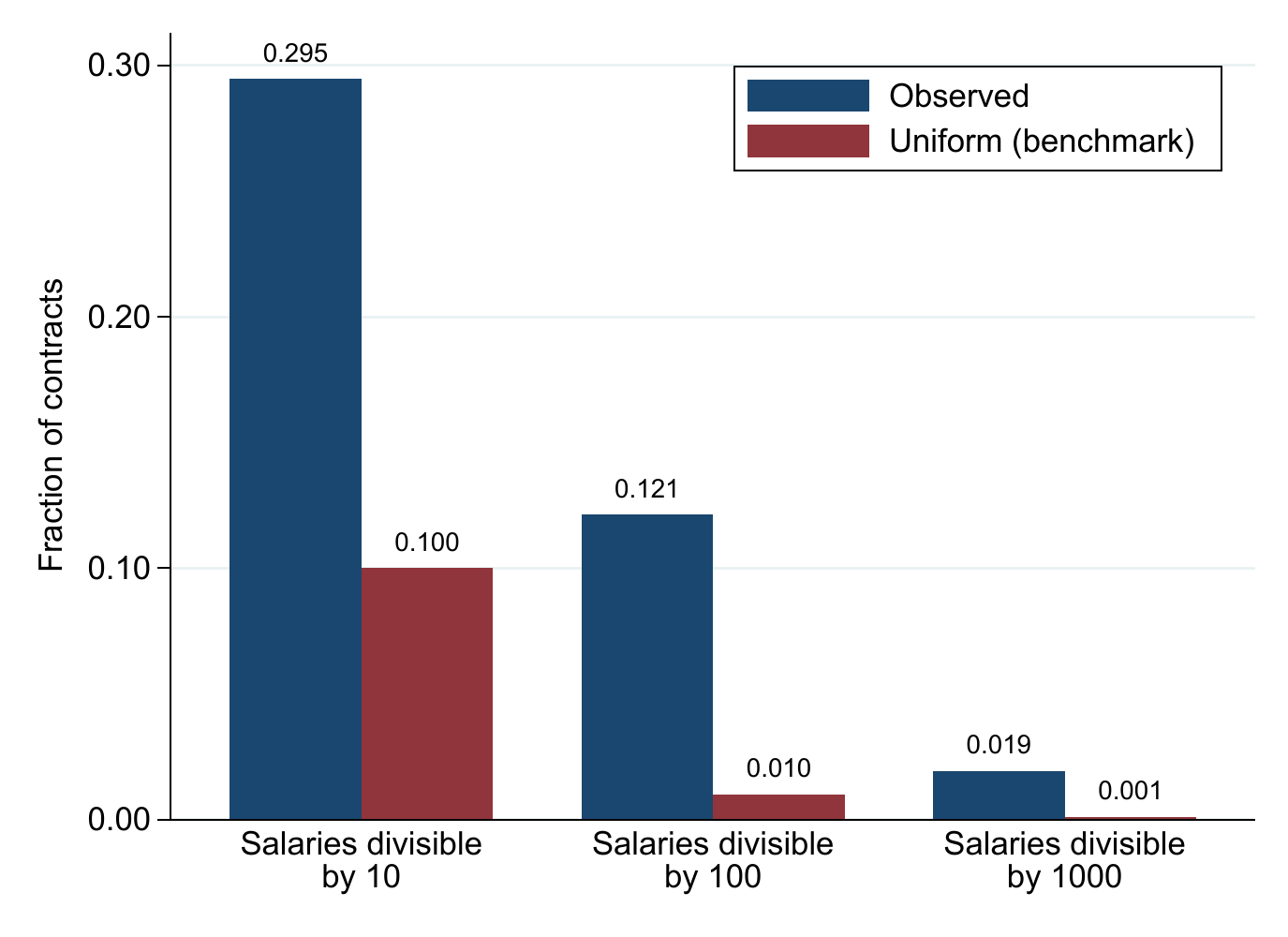}
	\end{subfigure}
	\hfill	
  	\footnotesize
  	\singlespacing \justify \textit{Notes:} Panel A shows the distribution of contracted salaries in the new-hires sample pooling all of the years during 2003--2017. To construct this figure, I group workers in R\$1 bins and count the number of workers in each bin. Workers whose contracted salary is a round number are denoted with colored markers. The figure only displays workers with earnings above the minimum wage and below R\$3,500 (which corresponds roughly to the 99th percentile of the distribution of earnings above the minimum wage).
  	  	
  	Panel B shows the fraction of contracted salaries divisible by 10, 100, and 1,000 in the new-hires sample (blue bars) and the fraction that would be observed if the distribution of the last digits of salaries were uniform (red bars). The figure excludes workers hired at the minimum wage. See Appendix \ref{app:data} for the sample restrictions.
\end{figure}

The data unequivocally rejects this prediction. Figure \ref{fig_bunching}, Panel A plots the distribution of contracted salaries in the new-hires sample. The earnings distribution exhibits stark bunching at round numbers (i.e., numbers divisible by 10). For example, workers are fifteen times more likely to earn exactly R\$3,000 per month than any other salary between R\$3,001 and R\$3,010. The modal monthly salary in the new-hires sample is R\$1,000, followed by R\$800, and R\$600 (jointly accounting for over six million contracts), with all being round numbers.

The bunching is also manifested in a non-uniform distribution of the last digit of salaries. Figure \ref{fig_bunching}, Panel B shows the fraction of salaries that are divisible by 10, 100, and 1,000. About one-third of the salaries (29.5\%) in the new-hires sample are divisible by 10 (see also Appendix Figure \ref{fig_last_digits}). This figure would be 10\% if the last digits of salaries were uniformly distributed. Over one-tenth of salaries (12.1\%) are divisible by 100 (a uniform distribution would imply 1\%), and 1.9\% are divisible by 1,000 (a uniform distribution would imply 0.1\%). 

These figures likely underestimate the true degree of bunching for several reasons. First, the contracted salary might be a round number at a different periodicity. For instance, over one-tenth (10.7\%) of the salaries that are not round numbers at the monthly level are round numbers at the yearly level. Similarly, some non-round-numbered salaries might be due to firms setting the wages of new hires equal to the wage of current employees. Some of these wages might have started initially as a round number but were updated over time into non-round-numbered wages.

\subsection{Bunching of Salaries at Round Numbers in Four Other Datasets}

To provide further evidence on the existence and magnitude of bunching, I study the earnings distribution in four additional datasets: the 2013 Brazilian Household Survey (\textit{Pesquisa Nacional por Amostra de Domicílios}, PNAD), the 2013 Brazilian Labor Force Survey (\textit{Pesquisa Mensal de Emprego}, PME), the 2010 Brazilian Population Census (\textit{Censo Demográfico}) and the 2013 Social Programs Registry of Individuals (\textit{Cadastro Único}).\footnote{The PNAD is a nationally-representative survey conducted annually by the National Statistics Office to measure several characteristics of the population, such as household composition, education, and income. The PME is a monthly survey conducted in six large metropolitan areas to provide frequent updates on the unemployment rate and other labor-market variables. The census is conducted approximately every ten years to count the population in the country, but it also includes earnings information. Finally, the Social Programs Registry contains information on all beneficiaries of government programs, including their earnings.} In all datasets, I focus on the monthly salary of full-time workers aged 18--65. I exclude workers employed by public-sector firms and individuals who work without remuneration. 

The advantage of these datasets is that they include information on informal-sector workers, while the main disadvantage is that earnings are self-reported. Hence, worker salaries might be measured with error due to---for example---recollection bias or social-desirability bias. Another drawback is that the labor income measure refers to the earnings during the month before the survey was conducted and not the contracted earnings when the employer hired the worker. 

\begin{figure}[H]
	\caption{Fraction of salaries divisible by round numbers in four Brazilian datasets}\label{fig_rounding_datasets}  \centering
	\centering
	\includegraphics[width=.75\linewidth]{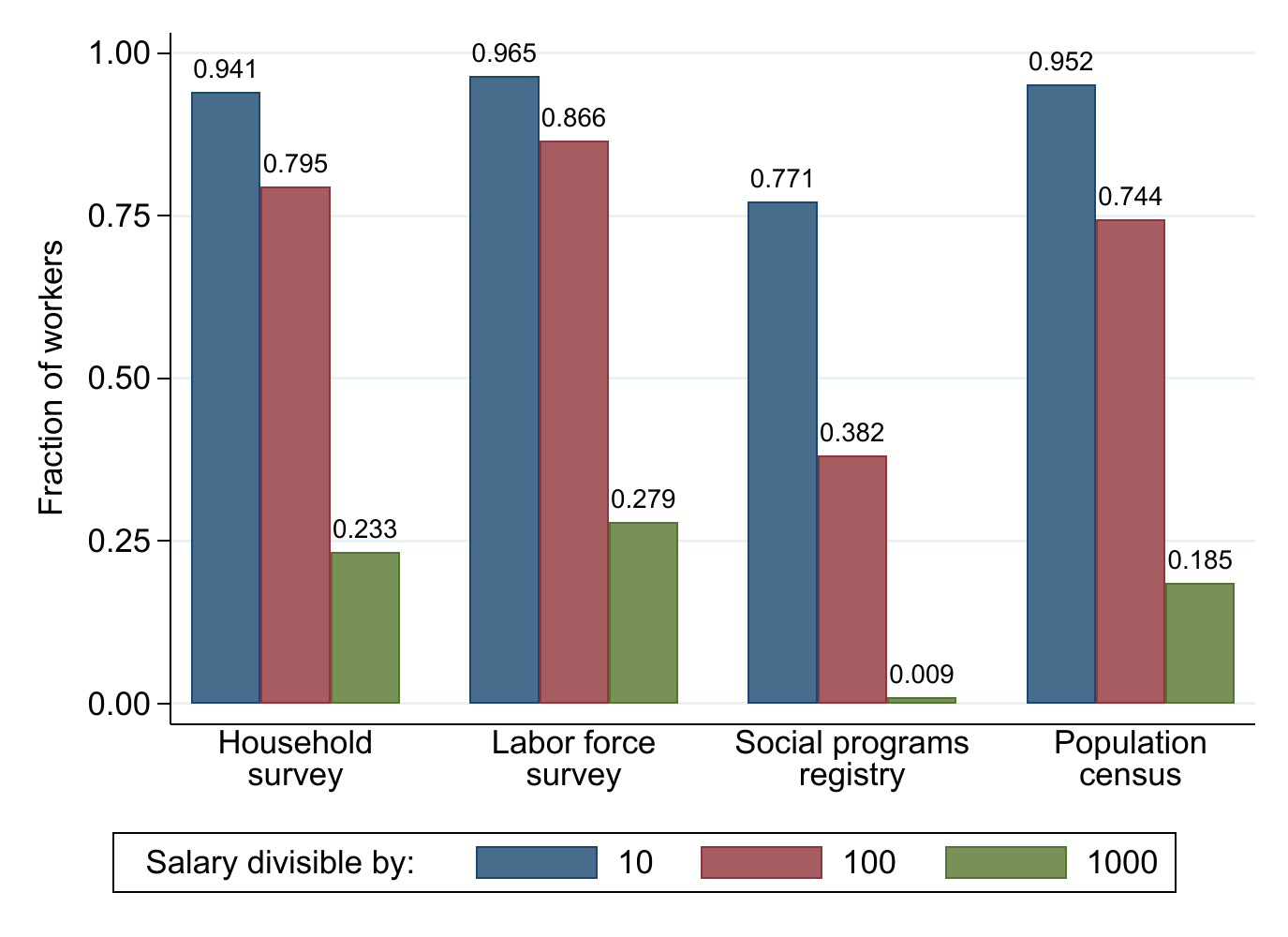}
	\footnotesize
	\singlespacing \justify	\textit{Notes:} This figure shows the fraction of monthly salaries divisible by 10, 100, and 1,000 observed in four datasets. The datasets are the 2013 Brazilian Household Survey (\textit{Pesquisa Nacional por Amostra de Domicílios}, abbreviated PNAD), the 2013 Brazilian Labor Force Survey (\textit{Pesquisa Mensal de Emprego}, abbreviated PME), the 2010 Brazilian Population Census (\textit{Censo Demográfico}), and the 2013 Social Programs Registry of Individuals (\textit{Cadastro Único}). The sample comprises full-time workers aged 18--65. I exclude public-sector workers and individuals that who without remuneration. 
	
\end{figure}

Figure \ref{fig_rounding_datasets} shows the fraction of monthly earnings divisible by 10, 100, and 1,000 in each dataset (see Appendix Figure \ref{fig_bunching_data} for the  earnings distribution). All datasets exhibit stark bunching at round numbers. For example, 96.1\% of monthly earnings in the census are divisible by 10. The corresponding figure in the Household Survey is 94.1\%, in the Labor Force Survey it is 96.5\%, and in the Social Programs Registry it is 79.2\%. This provides additional evidence against the hypothesis that salaries are smoothly distributed. The fact that we do not observe such an extreme bunching in the RAIS is consistent with previous research showing that the bunching in surveys is partly driven by recollection bias from the respondent side, although it could also reflect informal-sector firms paying round-numbered salaries at a higher rate.

Taken together, the results of this section show that bunching at round numbers is a ubiquitous feature of labor markets and not simply a consequence of measurement error.	
	\section{Firm Non-standard Behavior and Wage Bunching} \label{sec:firm-behavior}

This section investigates whether the bunching observed in the data is driven by non-standard behavior of workers or firms. The approach taken here involves studying the characteristics and outcomes of firms that tend to hire workers at round numbers and assessing whether these are consistent with the hypothesis that the firms that pay round numbers do so to exploit a worker bias. 

\subsection{Defining Bunching Firms} 

I start by measuring a firm's propensity to hire workers at round-numbered salaries. As a simple and intuitive measure, I compute the fraction of a firm's new hires over 2003--2017 whose initial salary is a round number.\footnote{I exclude from this computation workers hired at or below the minimum wage and those without a monthly earnings contract.} I focus on hiring at salaries divisible by 10 for consistency with previous research on the clustering of wages at round numbers \citep[e.g.,][]{riddles_handling_2016}, and show that the results below are robust to defining bunching firms using coarser salaries (e.g., those divisible by 100).

Round-number wage-setting is highly heterogeneous across firms, with many firms only hiring workers at round-numbered salaries (Appendix Figure \ref{fig_hist_bunch_firms}). In the data, one in six firms (16.9\%) exclusively hired workers at round salaries. I refer to these as \textit{bunching firms}. Appendix Table \ref{reg_firm_char} compares the characteristics of bunching and non-bunching firms.

The fraction of bunching firms is higher under less stringent definitions. For instance, 33.2\% [27.1\%] of firms hired more than half [two-thirds] of their new workers at a round salary. In Appendix Tables \ref{reg_firm_performance_size}--\ref{reg_salary_increase_def}, I show that the results below are robust to these alternative definitions of bunching firms. The results are also robust to excluding small firms (i.e., firms that employ fewer than five workers) and using the \textit{yearly} salary of new hires to define bunching firms (instead of the monthly salary). 

\subsection{The Market Outcomes of Bunching Firms} \label{sub:outcomes-bunchers}

A potential rationale for why many firms pay round-numbered salaries is to extract surplus from workers who have non-standard preferences. If this hypothesis holds true, one should see the consequences reflected in better firm outcomes. To evaluate this, I consider four outcomes: worker separation and resignation likelihoods during the hiring year or the following year, which I use as proxies of a poor worker-firm match; the growth rate of the firm's size, as measured by its number of employees; and a binary variable indicating if a firm leaves the market. While I do not observe firm profit, the firm growth and survival rates are functions of realized profits. 
 
In Table \ref{reg_firm_performance}, I estimate regressions of the form:
\begin{align} \label{reg:firm-outcomes}
	y_{ijt} = \alpha + \beta \text{BunchingFirm}_{j} + \psi X_{it} + \delta Z_{jt} + \varepsilon_{ijt},
\end{align}
where $i$ denotes workers, $j$ firms, and $t$ years; $y_{ijt}$ is one of the four outcomes; and $\text{BunchingFirm}_{j}$ equals one if firm $j$ hired all new employees at a round salary in the sample. 

The regression includes $X_{it}$, a vector of worker characteristics (age, gender, race, and occupation dummies), and $Z_{jt}$, a vector of fixed and time-varying firm characteristics that are typically associated with firm sophistication (see Appendix \ref{app:var-def} for variable definitions). These characteristics are the presence of a human resources (HR) department, the share of employees with a high-school and college degree, educational attainment of the manager, firm age, mean earnings of the firm employees, firm size, and firm hiring experience. I flexibly control for firm size and hiring experience by including fixed effects for the number of workers hired and the mean number of workers employed. $Z_{jt}$ also includes industry-by-year-by-microregion fixed effects.\footnote{A microregion is a geographical area that groups together economically integrated contiguous municipalities with similar productive structures. There are about 500 microregions in Brazil, each of which can be thought of as a local labor market (see \citealp{dix2017trade}). The boundaries of these microregions are defined by the National Statistics Office of Brazil.}

To analyze worker separation likelihood, I estimate the regressions at the worker-by-firm-by-year level. To analyze the firm growth and survival rates, I estimate the regressions at the firm-by-year level (and exclude the worker controls). I cluster the standard errors at the firm level.

Table \ref{reg_firm_performance} shows that bunching firms tend to have worse outcomes. Columns 1 and 2 show that new hires in bunching firms are, on average, 4.1 percentage points (an 11.6\% increase relative to the sample mean) and 1.2 percentage points (or 10.4\%) more likely to separate and resign, respectively, than new hires in non-bunching firms ($p<0.01$). Column 3 shows that bunching firms have a 3.4 percentage points lower growth rate, on average, than non-bunching firms ($p<0.01$). Column 4 shows that bunching firms are 1.1 percentage points (or 10.8\%) more likely to exit the market ($p<0.01$). 

These results are robust to excluding small firms, varying the set of controls, and using alternative definitions of bunching firms. Appendix Table \ref{reg_firm_performance_size} estimates the baseline specification for the subset of firms that employ more than five workers, on average across all years. The same results hold for these large firms. Appendix Table \ref{reg_firm_performance_lev} estimates all specifications at the worker level and additionally controls for the wage level by including wage fixed effects (in R\$100 bins). The point estimates are quantitatively similar to those of the baseline specification. Finally, Appendix Table \ref{reg_firm_performance_def} shows that the results are robust to defining bunching firms in several alternative ways.

In Appendix \ref{app:outcomes}, I study additional outcomes. Specifically, I analyze whether the higher separation rates of new workers hired by bunching firms persist over time, whether the higher separation rates are driven by high- or low-skilled workers, and assess whether the lower job growth rate of bunching firm is driven by high- or low-skilled workers. I find that the higher separation rates persist up to three years after new hires join bunching firms (Appendix Table \ref{reg_firm_performance2}). The higher separation and resignation rates of new workers hired by bunching firms are mainly driven by high-skilled workers (Appendix Table \ref{reg_firm_performance3}). Furthermore, bunching firms experience lower job growth rates for both high-skilled and low-skilled workers, as well as for high- and low-paid employees (Appendix Table \ref{reg_firm_performance4}).

These results admit several interpretations. First, the worse outcomes experienced by bunching firms may partly be a consequence of paying round-numbered salaries. For example, the higher separation likelihood might be due to a poor worker-firm match caused by paying a non-optimal wage. In addition, the results can also be explained by bunching firms being less sophisticated in other unobserved dimensions, which in turn might drive their worse outcomes. For example, in addition to having non-standard pay-setting practices, bunching firms may have less efficient production processes, which may be the cause of the lower survival rates. Finally, the round-number wage-setting may be the results of the constrained optimization problem faced by firms. For example, managers might (optimally) prioritize spending resources to improve production efficiency over pay-setting strategies. Regardless of the right explanation, the results at odds with the hypothesis that sophisticated firms pay round-numbered salaries to exploit a worker bias.

\subsection{Behavior of Bunching Firms in a Different Decision Environment}  \label{sub:heuristics-bunchers}

Another possible explanation for why many firms pay round-numbered salaries is that they are uncertain about what the fully-optimal salary is and use round-numbered salaries as a simple but coarse approximation.\footnote{Simplified pricing strategies have been documented in several environments \citep[e.g.][]{cho2010flat, dellavigna2019uniform, stevens2020coarse}.} For example, firms might be uncertain about worker productivity, which is a central determinant of the optimal salary in wage-determination models.

If this hypothesis is true, one would also expect these firms to rely on similar coarse approximations in other environments where they also face uncertainty. To evaluate this, I use salary increases as a different domain to explore the potential use of coarse pay-setting. In this domain, firms face uncertainty about employee realized productivity. The canonical Bayesian model of wage formation predicts that a worker's wage increase depends on her realized productivity \citep{jovanovic1979job, tervio2009superstars}. By contrast, coarse pricing suggests that firms determine raises based on coarse approximations, such as integer numbers if the salary increase is measured in percentage terms or round numbers if the increase is measured in monetary units.

Table \ref{reg_salary_increase} shows estimates of equation \eqref{reg:firm-outcomes} using two measures of coarse pay-setting as outcomes: first, a dummy that equals one if a new hire received a round-numbered salary increase in Brazilian Reals (e.g., R\$310 as opposed to R\$314), and second, a dummy that equals one if a new hire received an integer salary increase in percentage terms (e.g., 3\% as opposed to 3.14\%).

Firms that tend to hire workers at round salaries also tend to rely on coarse figures when deciding salary increases. Columns 1 and 3 show that bunching firms are 29.3 percentage points more likely to offer a round-numbered salary increase in Brazilian Reals (from a baseline of 35.4\%, $p<0.01$) and 16.2 percentage points more likely to offer an integer salary increase in percent terms (from a baseline of 34.4\%, $p<0.01$). Column 5 shows that bunching firms are about 28.6 percentage points more likely to engage in either of the two behaviors (from a baseline of 42.1\%, $p<0.01$). 

These results are robust to excluding workers whose salaries remained constant in nominal terms (columns 2, 4, and 6). They are also robust to excluding small firms (Appendix Table \ref{reg_salary_increase_size}), controlling for the wage level (Appendix Table \ref{reg_salary_increase_lev}), and defining bunching firms in several alternative ways (Appendix Table \ref{reg_salary_increase_def}).

	\section{A Wage-Setting Model with Optimization Frictions} \label{sec:model}

The evidence in Section \ref{sec:firm-behavior} motivates the hypothesis that firm coarse wage-setting is partly behind the rounding observed in the data. To further explore this hypothesis, in this section I build a wage-posting model in which coarse wage-setting is a consequence of optimization frictions. The goal of the model is to account for the bunching observed in the data and to generate additional testable predictions. 

This section first reviews evidence from numerical cognition research to support the model's assumptions. Next, I present a summary of the model and discuss the model's testable predictions. Then, I describe the two research designs that I use to test the predictions and present the results.

\subsection{Stylized Facts from Numerical Cognition Research} \label{sub_psych}

Round numbers are ubiquitous in open numerical judgments. For example, in contingent valuation studies, individuals often report round numbers \citep{whynes_think_2005}. Similarly, in judging the likelihood of future events, subjects often report round-numbered probabilities \citep{manski2010rounding}. According to numerical cognition research, this phenomenon occurs because the mental computation cost of round numbers is lower, and thus, round numbers are the first that ``come to mind.'' I use this finding to motivate one of the assumptions of the model, namely that firms use a round number as an initial estimate of the worker fully-optimal salary.

Numerical cognition research also sheds light on how individuals generate more precise estimates. According to prominence theory \citep{albers1983prominence, albers_prominence_2001}, individuals start from a round number and sequentially refine it by adding and subtracting smaller round numbers until they reach a satisfactory estimate.\footnote{Consistent with this theory, \cite{converse_role_2018} show that individuals are more likely to use ``prominent numbers'' (a subset of the round numbers) in numerical judgments when they are induced to quickly make a judgment and when they are under a high cognitive load. Relatedly, \cite{giustinelli2018tail} show that individuals with high cognitive ability are less likely to give round-numbered responses in expectations surveys, possibly because they have a lower cost of refining their estimates.} Following these findings, in the model I assume that firms can pay some cost to refine their initial estimate of the optimal salary.\footnote{The notion that it is costly to obtain more precise estimates of a target value also has parallels in mathematics and computer science. For example, improving the precision of a Taylor expansion approximation (i.e., computing more decimals) requires increasing the number of expansion terms, requiring more computational power and memory to store the additional terms.}

\subsection{Summary of the Model}

This section presents an abbreviated version of the model, focusing on its key assumptions and predictions. Appendix \ref{app:model} provides a complete description of the model.

In the model, monopsonistic firms decide what wage to offer to prospective workers. In the textbook formulation of the wage-posting model, firms know the marginal revenue product (MRP) of hiring an additional worker and offer a wage proportional to it. The difference between the textbook model and the one presented here is that I depart from the assumption that firms observe worker MRP.

The model rests on two key assumptions. First, I assume that firms form an estimate of the fully-optimal salary (the salary that firms would pay if they had perfect information) based on a coarse rounding heuristic. This assumption is supported by the research on numerical cognition reviewed above. For simplicity, I model hiring decisions around a single round number. Appendix \ref{app:varying-precision} considers an extension where firms can approximate the fully-optimal salary with different degrees of precision.

The second key assumption is that by paying an ``optimization cost,'' firms can generate a more precise estimate of the fully-optimal salary. This reduced-form cost likely reflects a range of underlying mechanisms, including information-gathering costs, attention costs, and the cost of integrating the data available.\footnote{The compensation reports sold by pay-consulting firms such as ADP or PayScale provide a market-based approach to quantifying all these costs. These reports provide advice on how much a firm should pay a prospective employee with given characteristics. Appendix Figure \ref{fig_payscale} shows an example of a compensation report. After gathering information on the prospective employee, such as job title, educational attainment, and years of experience, these firms provide a distribution of suggested compensations. It is noteworthy that the suggested compensations in Appendix Figure \ref{fig_payscale} are not round numbers.} I say that a worker is hired through coarse wage-setting (or a coarse rounding heuristic) if the firm does not pay the optimization cost and instead hires the worker at the round-numbered salary.

Under these two assumptions, the market-level distribution of wages comes from a mixture of two distributions: one distribution with the same support as the distribution of fully-optimal wages and one with support on the set of round numbers. The (endogenous) mixture weight is the fraction of workers hired through coarse wage-setting, a variable denoted by $\theta$. Hence, the cross-section distribution of wages in the model exhibits bunching at round numbers. The standard wage-posting model is a special case of the model with optimization frictions, in which the optimization cost is zero (which implies $\theta = 0$).

\subsection{Model Predictions} 

Firms pay the fully-optimal salary whenever the benefit of doing so exceeds the optimization cost; otherwise, they rely on the coarse rounding heuristic and pay a round-numbered salary. The comparative statics generate the following testable predictions:

\begin{prediction} \label{pred:wedge}
	As the value of the expected gap between the coarse wage and the fully-optimal wage decreases, firms are more likely to rely on the coarse rounding heuristic and pay a round-numbered salary. This is the case because the profit return to generating a more precise estimate of the fully-optimal wage is proportional to the distance between this wage and the coarse wage.
\end{prediction}

To test Prediction \ref{pred:wedge}, I exploit changes in the purchasing power of gaps over time and across regions in the country. As inflation erodes the purchasing power of money, the real monetary cost of mispricing a fixed gap decreases. Intuitively, ``getting the wage right'' is less profitable in real terms. 

\begin{prediction} \label{pred:cost}
	Firms with a higher optimization cost are more likely to rely on coarse wage-setting and thus pay round-numbered salaries. Intuitively, as finding the fully-optimal salary becomes costlier, firms are more likely to rely on a coarse approximation.
\end{prediction}

The firm optimization cost is unobservable. I use two proxies of firm optimization cost: firm size and hiring experience. Larger firms and firms with more hiring experience might have a lower optimization cost because they are more likely to have an HR department or structured management practices \citep{cornwell2019building}. Thus, to test Prediction \ref{pred:cost}, I evaluate whether a given firm is less likely to pay a round-numbered salary as it grows larger over time or accumulates more hiring experience.

\subsection{Testing the Model Predictions} \label{sec:testing-pred}

I use two research designs to test the model's predictions: a bunching design and a regression design. First, I describe each design and then present the results.

\subsubsection{Bunching Design.} \label{sub:bunching}

The first research design consists of estimating $\theta$, the fraction of workers hired through a coarse rounding heuristic, for each value taken by an observable variable, such as firm size or worker educational attainment, and testing whether the sign of the correlations aligns with the model's predictions. 

By definition, $\theta$ can be written as the ratio between $B$, the number of workers hired through coarse wage-setting, and $N$, the total number of new hires:
\begin{align} \label{eq_theta_ratio}
	\theta = \frac{B}{N}.
\end{align}

While $B$ is not observed in the data, it can be estimated by assuming that the excess mass of workers at round numbers in the earnings distribution represents workers hired through coarse wage-setting. To compute $\hat{B}$, it is necessary to estimate a counterfactual distribution in which there is no bunching, which I obtain using standard techniques of the bunching literature \citep{kleven_bunching_2016}. Appendix \ref{app:ctfl-dist} describes the methodology in detail.

The estimated excess number of workers at round number $r$, $\hat{B}_r$, equals the difference between the number of workers earning $r$ in the actual and the counterfactual distribution, $\hat{B}_r = C_r - \hat{C_r}$. To estimate $B$, I integrate the excess mass across all round numbers:  
\begin{align} \label{eq_int_exc_mass}
	\hat{B} = \sum_{r \in R} \hat{B}_r,
\end{align}
where $R = \Big\{w \ \Big| \  w = 10k \text{ for some } k \in \mathbb{Z}\Big\}$ is the set of round-numbered salaries. I estimate $\theta$ by replacing $B$ in equation \eqref{eq_theta_ratio} with its empirical counterpart, $\hat{B}$:
\begin{align} \label{eq_theta_hat}
	\hat{\theta} = \frac{\hat{B}}{N} = \frac{1}{N} \sum_{r \in R} \hat{B_r}.
\end{align}

To test the predictions of the model, I estimate $\theta$ conditioning on the values taken by a given covariate. For example, I calculate the excess number of workers in the distribution of college-educated workers and then compute the ratio between this estimate and the total number of college-educated workers. This ratio represents the fraction of college-educated workers who were hired through coarse wage-setting. I repeat this process for workers with only a high-school diploma, etc. More generally, this procedure yields estimates of $B$ and $\theta$ for each value taken by a covariate of interest. I use this procedure to examine whether $\hat{\theta}$ is correlated with characteristics of the firm (e.g., size or hiring experience).

To evaluate whether a decrease in the value of the gap increases coarse wage-setting (Prediction \ref{pred:wedge}), I calculate the correlation between $\hat{\theta}$ (estimated for each metropolitan region-month-year triplet) and the log of the consumer price index (CPI) of the corresponding region-month-year in which the worker was hired.\footnote{Metropolitan region is the most disaggregated geographical level at which CPI data is available. The Brazilian National Statistics Office collects inflation data at the monthly level for eleven metropolitan regions. Each metropolitan region is a collection of several municipalities.} To assess whether a lower optimization cost reduces coarse wage-setting (Prediction \ref{pred:cost}), I test for a negative correlation between $\hat{\theta}$, firm size, and firm hiring experience. In the bunching design, these correlations are mainly identified off of cross-section variation in firm size and hiring experience. Under the assumption that larger firms and firms with more hiring experience have lower optimization costs, we should observe a negative correlation between these variables and $\hat{\theta}$.

\subsubsection{Regression Design.}  \label{sub:lpm}

The second research design is a regression design that allows me to control for a large set of variables, including unobserved heterogeneity at the firm level. I estimate linear probability models of the form:
\begin{align} \label{reg:lpm}
	\mathbbm{1}\{w_{ijsmt} \in R\} &= \pi \log\text{CPI}_{smt} + \beta_1 \text{FirmSize}_{jt} + \beta_2 \text{HiringExp}_{jt} + \delta X_{it} +  \notag  \\ &+ \gamma_j + \gamma_t + \gamma_s + \varepsilon_{ijsmt},
\end{align}	
where the dependent variable, $\mathbbm{1}\{w_{ijsmt} \in R\}$, equals one if the contracted salary of new hire $i$ employed by firm $j$ in metropolitan region $s$ during month $m$ in year $t$ is a round number, and zero otherwise, $\text{FirmSize}_{jt}$ is the (log) number of employees, and $\text{HiringExp}_{jt}$ is the (log) number of employees hired since the firm first appeared in the sample. Equation \eqref{reg:lpm} also includes $X_{it}$, a vector of worker-level characteristics (gender, education, working experience, and occupation), and region, year, and firm fixed effects. I cluster standard errors at the firm level and normalize all covariates by their standard deviation so that their corresponding coefficients can be interpreted as partial correlations. This normalization makes the results of the regression design comparable to those of the bunching design.

I use the coefficients of equation \eqref{reg:lpm} to test the model's predictions. To assess whether a smaller gap---in real terms---reduces coarse wage-setting (Prediction \ref{pred:wedge}), I test whether $\hat{\pi} > 0$. Given the region and year fixed effects, identification mainly comes from within-region changes in the price level over time. To assess whether a higher optimization costs increases coarse wage-setting (Prediction \ref{pred:cost}), I test whether $\hat{\beta}_1 < 0$ and $\hat{\beta}_2 < 0 $. Since equation \eqref{reg:lpm} includes firm fixed effects, these coefficients are identified off of variation in the size and number of workers hired by a given firm over time. 

\subsubsection{Results.}

Table \ref{tab_predictions} shows the results. Columns 1--2 present the results of the bunching design and columns 3--4 the results of the regression design.

The first row shows the relation between hiring workers at a coarse salary and the log CPI (Prediction \ref{pred:wedge}). The bunching design shows a positive and statistically significant relationship between the fraction of workers hired through coarse wage-setting and the log CPI ($p<0.01$). Consistent with this, the regression design shows that---ceteris paribus---an increase in the inflation rate increases the likelihood of a given firm paying a round-numbered salary to new hires ($p<0.01$).

The second and third rows display the relation between hiring workers at a coarse salary and the two proxies of the optimization cost (Prediction \ref{pred:cost}). Larger firms and firms that have hired more workers have a lower likelihood of relying on coarse salaries ($p<0.01$). Consistent with this, the regression design shows that as firms grow larger in size and accumulate more hiring experience, they become less likely to hire workers at round-numbered salaries ($p<0.01$).

A possible concern is that some of the correlations might be partly driven by the fact that firms that tend to hire workers at round-numbered salaries are more likely to exit the market. This type of selective attrition could explain the negative association between firm size and paying coarse wages. To deal with this, in columns 2 and 4 of Table \ref{tab_predictions}, I re-estimate all specifications using a fixed sample of firms that I observe in all fifteen years of the data. The coefficients tend to be similar to the baseline results, albeit in some cases the magnitudes are smaller and the estimates more imprecise.

As an additional robustness test, in Appendix Table \ref{tab_predictions_rob} I test the model's predictions using alternative measures of the dependent variable. Instead of using all round-numbered salaries, I measure the dependent variable using salaries divisible by 100 (Panel A) or 1,000 (Panel B). The results are remarkably consistent across specifications. For example, the correlation between the fraction of workers hired through coarse wage-setting and firm hiring experience is $-$0.88 in the baseline specification, compared to $-$0.87 when using salaries divisible by 100, and $-$0.64 when using salaries divisible by 1,000.

In summary, I find evidence in support of the two predictions of the model using two different research designs. This is consistent with the hypothesis that the bunching of wages at round numbers observed in the data is partly due to coarse wage-setting. Appendix \ref{app:alt-exp} evaluates whether several alternative explanations are compatible with the bunching observed in the data and the stylized facts documented in this section. The alternative explanations that I discuss are worker left-digit bias, focal points in wage bargaining, collective bargaining agreements, fairness concerns, round wages as a signal of job quality, and changes in marginal tax rates. While some of these explanations have explanatory power in accounting for some features of the data, I conclude that none of them can provide a cohesive account of the entire pattern of results.
	
	\section{Implications for Other Economic Outcomes} \label{sec:implic}

In this section, I explore some of the downstream consequences of firm coarse wage-setting for important economic outcomes.

\subsection{Within-Firm Wage Inequality}

Understanding the drivers of wage inequality is an important research agenda in public and labor economics. Previous research has found that firm wage-setting policies influence wage inequality \citep{card2018firms}. One might expect firm coarse wage-setting to affect wage dispersion among new hires.\footnote{Ex-ante, the direction is ambiguous. To see this, consider a firm that pays workers their fully-optimal salary rounded to the nearest 1,000th deciding the wages of two new hires. If the workers' fully-optimal salaries are R\$700 and R\$1,400, but the firm pays both of them R\$1,000, then the coarse pricing generates wage compression. Instead, if the first worker's fully-optimal salary is R\$1,700, the paid salaries would be R\$2,000 and R\$1,000, respectively. In this case, the coarse wage-setting increases wage dispersion.} To assess this, I estimate equation \eqref{reg:firm-outcomes} using as outcomes the Gini coefficient and ratios between the contracted salary at the 90th and 10th percentile, 90th and 50th percentile, and 50th and 10th percentile.\footnote{The Gini measures overall inequality in the contracted salary distribution, while the ratios measure inequality at different parts of the distribution (e.g., top- or low-end inequality, see \citealp{lemieux2008changing}).} Since equation \eqref{reg:firm-outcomes} includes fixed effects for the number of workers hired, the research design compares the within-firm wage inequality of two firms that hired the same number of workers using a different decision rule to determine their initial pay.

Figure \ref{fig_wage_comp} shows that bunching firms tend to compress wage differentials among new hires (see Appendix Table \ref{reg_wage_compression} for the corresponding regression coefficients). The average Gini coefficient among non-bunching firms is 0.112 (Panel A).\footnote{By country standards, this is a very low level of inequality. The most egalitarian countries in the world---typically, the Nordic countries---have a Gini coefficient on the order of 0.25. Two reasons explain the difference in magnitudes. First, country-level inequality is typically measured using household \textit{consumption} per capita as the welfare measure, whereas I compute the Gini using earnings. Second, I calculate the Gini among new hires of a given firm, which is likely a more homogeneous population than the overall population of a country.} Bunching firms have a 0.01 lower Gini coefficient (or 8.9\% of the baseline value). The decline in overall wage inequality is driven by mostly top- and mid-end inequality (Panel B). The ratio between the 90th and 10th percentile is, on average, 3.9\% lower (from a baseline ratio of 1.76) in bunching firms relative to the rest of the firms. Similarly, bunching firms have, on average, a 2.8\% lower 90th to 50th percentile ratio and a 1.6\% lower 50th to 10th percentile ratio than non-bunching firms (from baseline ratios of 1.39 and 1.24, respectively). These effects are robust to excluding small firms (Appendix Figure \ref{fig_wage_comp_bigf}). 

\subsection{Nominal Wage Rigidity}

Nominal wage stickiness influences the effects of monetary policy \citep{barattieri2014some}. Previous work has documented that behavioral considerations such as inertia \citep{eichenbaum2011reference}, managerial inattention \citep{ellison2018costs}, and fairness norms \citep{kaur2019nominal} influence nominal rigidities. Coarse wage-setting might contribute to wage rigidity if it makes firms less likely to change the initial salary of their new hires. To assess this, I estimate equation \eqref{reg:firm-outcomes} using as the dependent variable a dummy that equals one if the nominal salary of a new hire remained constant in nominal terms during the year following the hiring, and zero otherwise. 

\begin{figure}[H]
	\caption{Wage compression and wage stickiness in the salaries of new hires} \label{fig_wage_comp}
	\centering
	\begin{subfigure}[t]{.48\textwidth}
		\caption*{Panel A. Outcome: Gini coefficient}\label{fig_wage_comp_gini}
		\centering
		\includegraphics[width=\linewidth]{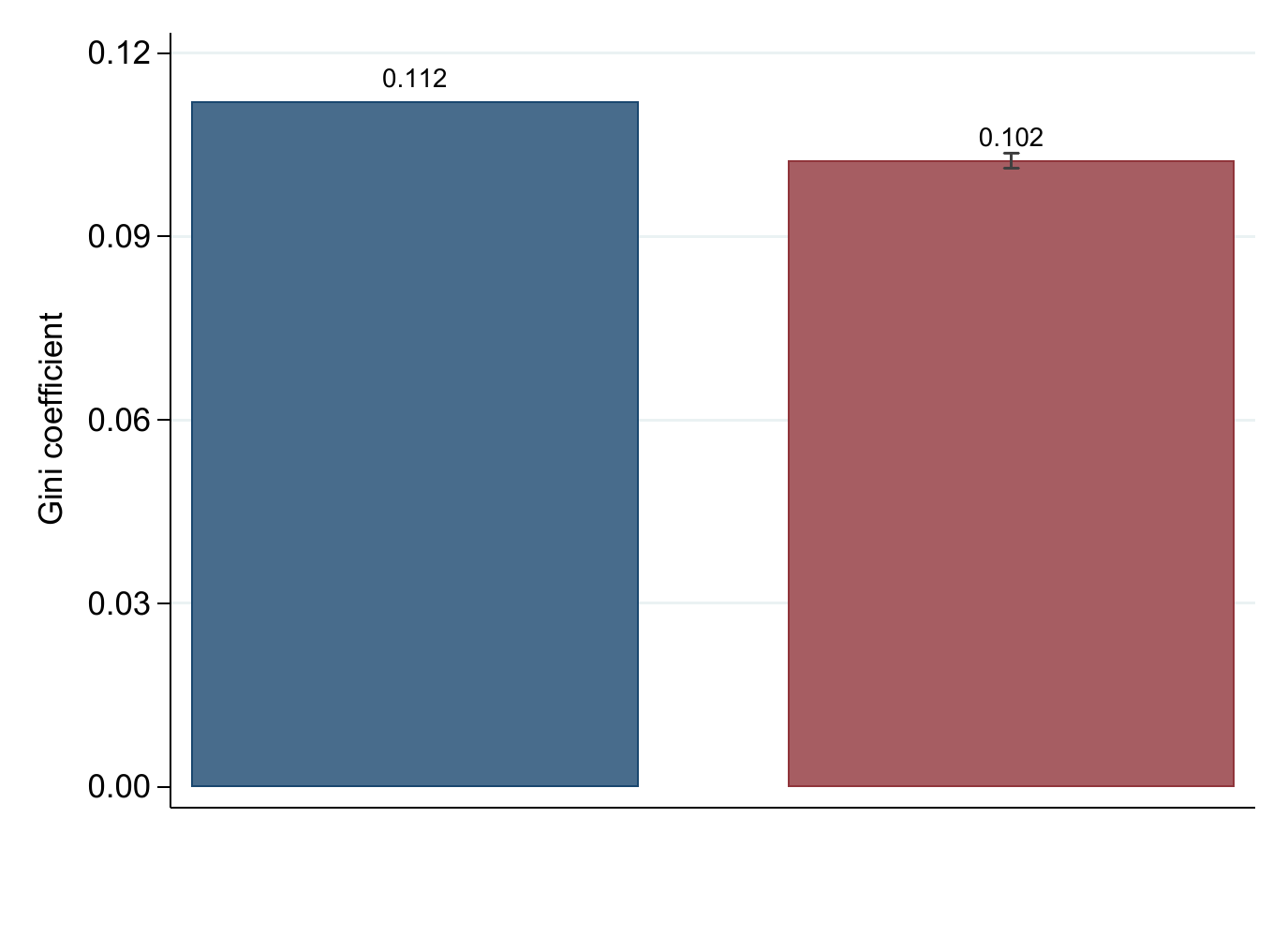}
	\end{subfigure}
	\hfill		
	\begin{subfigure}[t]{0.48\textwidth}
		\caption*{Panel B. Outcome: Percentiles ratios}\label{fig_wage_comp_ratios}
		\centering
		\includegraphics[width=\linewidth]{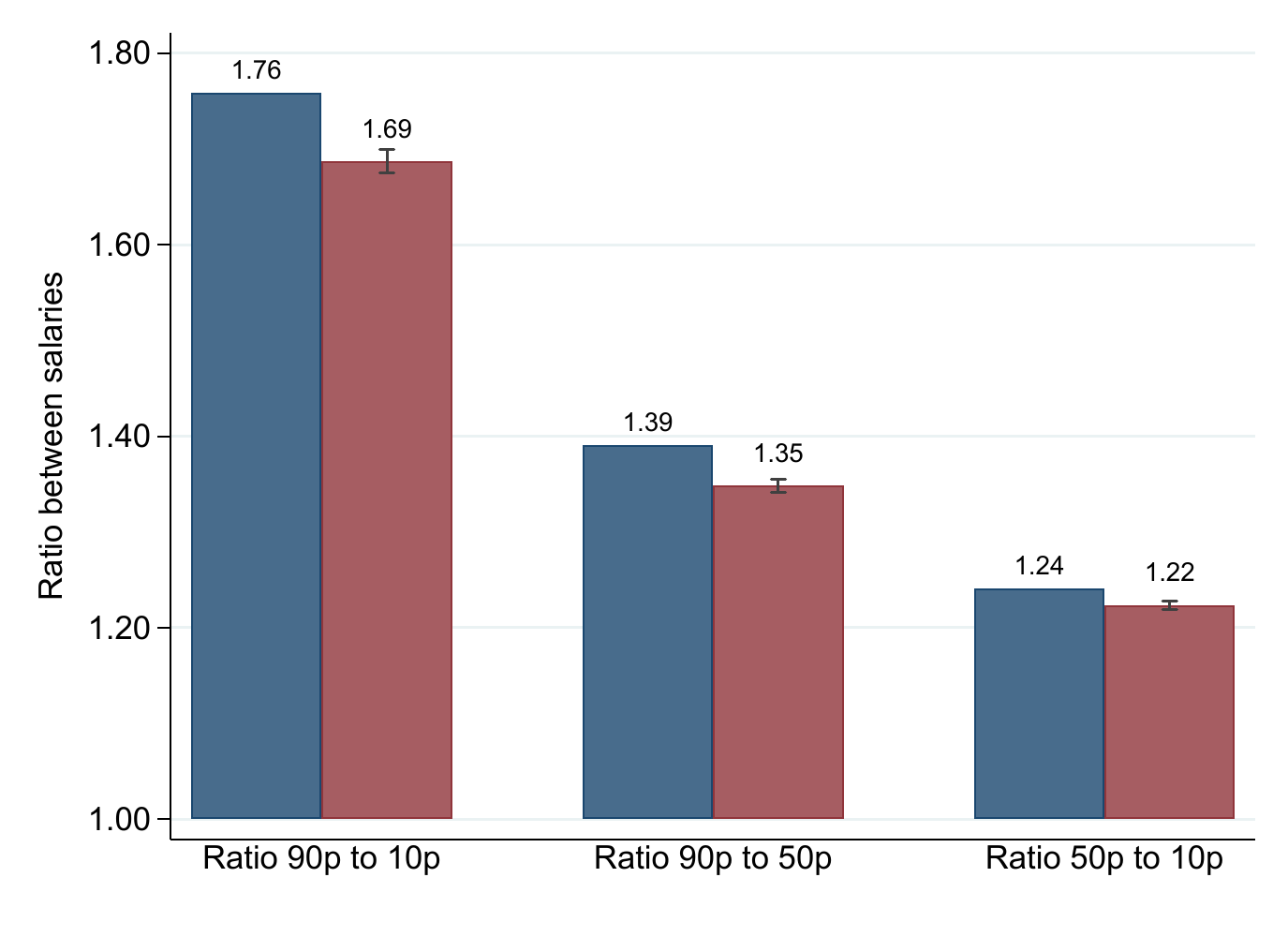}
	\end{subfigure}
	
	\centering	
	\begin{subfigure}[t]{0.48\textwidth}
		\centering	
		\caption*{Panel C. Outcome: Initial wage remained constant in nominal terms}
		\centering
		\includegraphics[width=\linewidth]{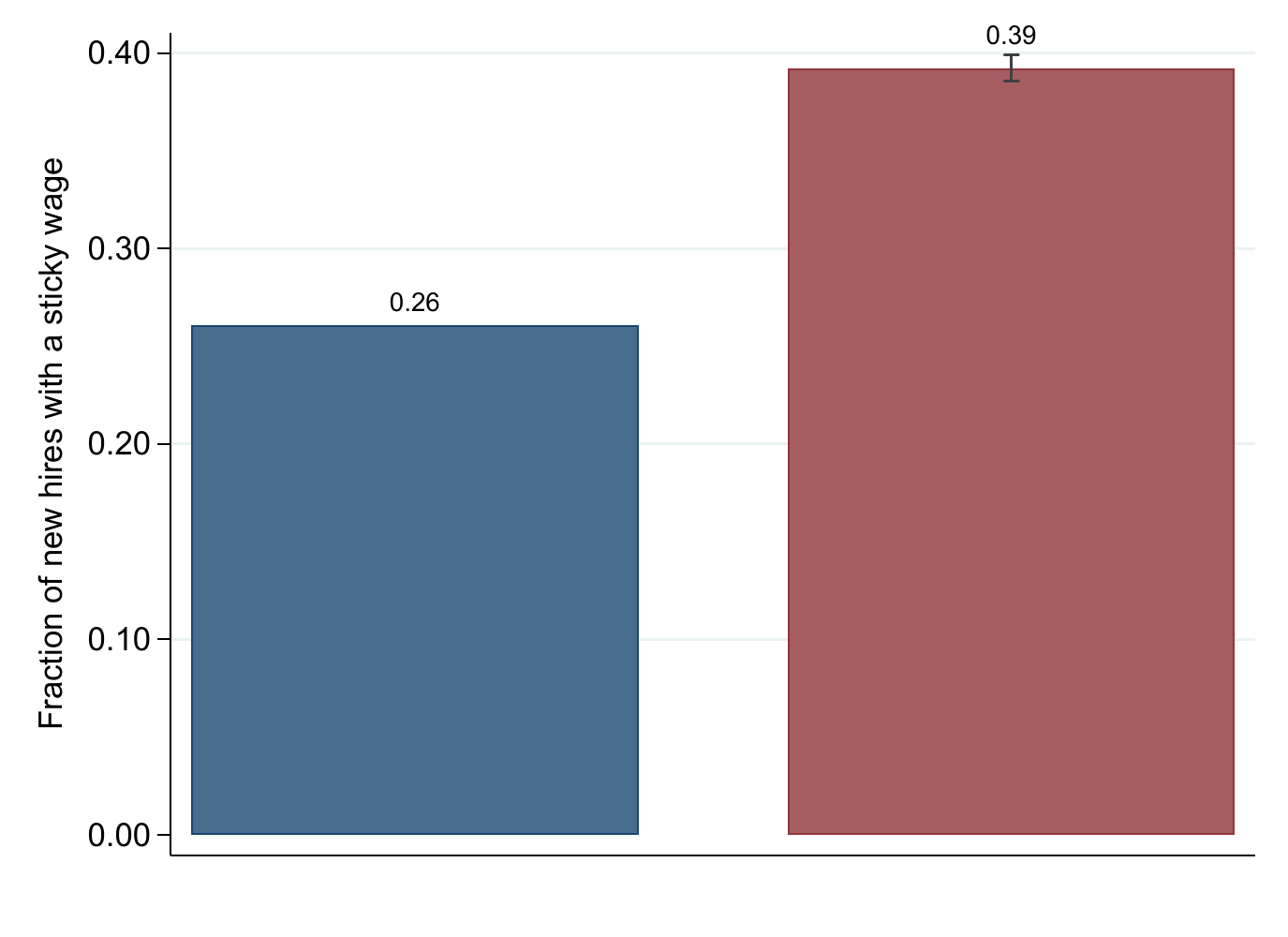}
	\end{subfigure}
	
	\vspace{-.3cm}
				
	\begin{subfigure}[t]{\textwidth}		
		\centering
		\includegraphics[width=0.75\linewidth]{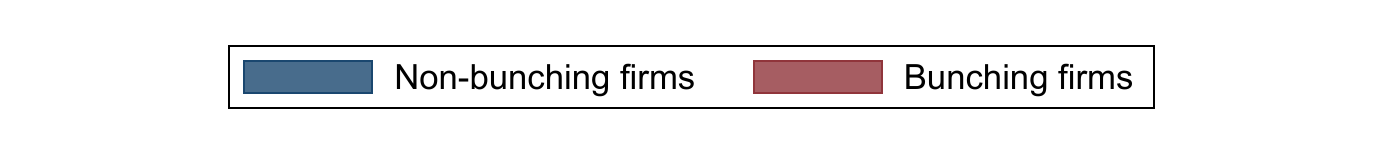}
	\end{subfigure} \vspace{-.5cm}
	{\footnotesize
		\singlespacing \justify
		
		\textit{Notes:} Blue bars plot the average of the variable listed in the panel title for non-bunching firms. Red bars plot the sum of this average and the estimated bunching firm effect (i.e., the estimated $\hat{\beta}$ from equation \eqref{reg_firm_performance}). To calculate the effect of bunching firms on each outcome, I estimate equation \eqref{reg_firm_performance} at the firm level using as the dependent variable one of the four measures of inequality or the measure of wage stickiness. In addition to the bunching firm dummy, the regressions control for: firm age, share of employees with completed high school, share of employees with completed college, educational attainment of the firm manager, a dummy for having an HR department, the mean earnings of the firm employees, and fixed effects for firm size, number of hires, and industry-by-microregion fixed effects. The wage inequality regressions are estimated at the firm level for firms that hired at least two workers in the sample. The wage rigidity regressions are estimated at the worker-by-firm-by-year level and additionally control for worker gender, race, and occupation. The vertical lines denote the 95\% confidence interval on the bunching firm dummy using heteroskedasticity-robust standard errors clustered at the firm level.
								
	}
\end{figure}

The initial salaries of bunching firms' workers tend to be stickier (Figure \ref{fig_wage_comp}, Panel C). From a baseline of 26.0\%, workers employed by bunching firms have a 13-percentage-point increase in the probability of experiencing no salary change. Thus, relative to new hires of non-bunching firms, those employed by bunching firms are about 50\% more likely to exhibit nominal wage stickiness. This effect is robust to excluding small firms (Appendix Figure \ref{fig_wage_comp_bigf}).

\subsection{Minimum Wage Spillovers}

\cite{dube_monopsony_2020} hypothesize that in the presence of firms that pay round-numbered wages, a change in the minimum wage could generate a novel spillover effect if the new minimum wage crosses a round number. Intuitively, a change in the minimum wage might cause firms that initially pay a round-numbered wage to fully optimize. However, their data does not allow them to test this hypothesis. In my sample, I observe hiring decisions under fifteen different federal minimum salaries, seven of which are round numbers. I also observe the year $t+1$ salary of workers hired in year $t$, which allows me to assess the importance of this potential spillover effect. 

I describe the methodology and results in detail in Appendix \ref{app:min-wage}. In short, using a differences-in-differences approach comparing salaries directly affected by the change in the minimum salary and those not directly affected by it, I find that an increase in the minimum salary reduces the share of round-numbered salaries by 5.4 percentage points (or 11.3\%). This finding suggests that changes in the minimum wage can have sizable spillover effects on firm wage-optimization behavior.

	\section{Discussion} \label{sec:conclusions}

Setting the right wage is challenging. To estimate the fully-optimal wage prescribed by economic models, a firm needs substantial information, including an estimate of the worker's contribution to the firm. Most workers have multiple goals and no measured output, which makes productivity difficult to estimate. This paper posits that the stark bunching at round numbers in the earnings distribution partly reflects the challenges associated with optimal labor pricing. In the data, millions of workers are hired at round-numbered salaries, reflecting a behavior that cannot be accommodated by existing wage-setting models. The evidence presented in this paper indicates that this behavior is partly due to firms engaging in coarse wage-setting. 

An important unresolved question is whether the coarse wage-setting is suboptimal. Setting optimal pay-setting practices likely requires substantial resources. If these costs are large, offering a coarse wage might lead to better outcomes. Nonetheless, the findings have intrinsic value for understanding how firms set wages. Coarse wage-setting may also ahve consequences for wage inequality, nominal wage rigidity, and may interact with policies that affect the wage distribution.

Future work could also explore the extent to which rounding reflects the quality of management practices. Management quality is often not available in traditional datasets (the World Management Survey is a notable exception, see \citealp{bloom2007measuring}). If coarse pricing partly reflects how human resources are managed at the firm, researchers could use the type of salaries offered to new hires as a proxy for overall HR management quality.

	\clearpage
	\begin{singlespace}
		\bibliographystyle{chicago}
		\bibliography{0-coarse-wage-setting}

\begin{thebibliography}{}

\bibitem[\protect\citeauthoryear{Albers}{Albers}{2001}]{albers_prominence_2001}
Albers, W. (2001).
\newblock Prominence theory as a tool to model boundedly rational decisions.
\newblock In {\em Bounded rationality: {The} adaptive toolbox}, pp.\  297--317.
  Cambridge, MA, US: The MIT Press.

\bibitem[\protect\citeauthoryear{Albers and Albers}{Albers and
  Albers}{1983}]{albers1983prominence}
Albers, W. and G.~Albers (1983).
\newblock On the prominence structure of the decimal system.
\newblock In {\em Advances in Psychology}, Volume~16, pp.\  271--287. Elsevier.

\bibitem[\protect\citeauthoryear{Almunia, Hjort, Knebelmann, and Tian}{Almunia
  et~al.}{2022}]{almunia2021strategic}
Almunia, M., J.~Hjort, J.~Knebelmann, and L.~Tian (2022).
\newblock Strategic or confused firms? evidence from “missing” transactions
  in uganda.
\newblock {\em Review of Economics and Statistics\/}, 1--35.

\bibitem[\protect\citeauthoryear{Barattieri, Basu, and Gottschalk}{Barattieri
  et~al.}{2014}]{barattieri2014some}
Barattieri, A., S.~Basu, and P.~Gottschalk (2014).
\newblock Some evidence on the importance of sticky wages.
\newblock {\em American Economic Journal: Macroeconomics\/}~{\em 6\/}(1),
  70--101.

\bibitem[\protect\citeauthoryear{Bloom, Eifert, Mahajan, McKenzie, and
  Roberts}{Bloom et~al.}{2013}]{bloom2013does}
Bloom, N., B.~Eifert, A.~Mahajan, D.~McKenzie, and J.~Roberts (2013).
\newblock Does management matter? evidence from india.
\newblock {\em The Quarterly Journal of Economics\/}~{\em 128\/}(1), 1--51.

\bibitem[\protect\citeauthoryear{Bloom and Van~Reenen}{Bloom and
  Van~Reenen}{2007}]{bloom2007measuring}
Bloom, N. and J.~Van~Reenen (2007).
\newblock Measuring and explaining management practices across firms and
  countries.
\newblock {\em The Quarterly Journal of Economics\/}~{\em 122\/}(4),
  1351--1408.

\bibitem[\protect\citeauthoryear{Caldwell and Harmon}{Caldwell and
  Harmon}{2019}]{caldwell2019outside}
Caldwell, S. and N.~Harmon (2019).
\newblock Outside options, bargaining, and wages: Evidence from coworker
  networks.
\newblock {\em Unpublished manuscript, Univ. Copenhagen\/}, 203--207.

\bibitem[\protect\citeauthoryear{Camacho and Conover}{Camacho and
  Conover}{2011}]{camacho2011manipulation}
Camacho, A. and E.~Conover (2011).
\newblock Manipulation of social program eligibility.
\newblock {\em American Economic Journal: Economic Policy\/}~{\em 3\/}(2),
  41--65.

\bibitem[\protect\citeauthoryear{Card, Cardoso, Heining, and Kline}{Card
  et~al.}{2018}]{card2018firms}
Card, D., A.~R. Cardoso, J.~Heining, and P.~Kline (2018).
\newblock Firms and labor market inequality: Evidence and some theory.
\newblock {\em Journal of Labor Economics\/}~{\em 36\/}(S1), S13--S70.

\bibitem[\protect\citeauthoryear{Cavallo, Neiman, and Rigobon}{Cavallo
  et~al.}{2014}]{cavallo2014currency}
Cavallo, A., B.~Neiman, and R.~Rigobon (2014).
\newblock Currency unions, product introductions, and the real exchange rate.
\newblock {\em The Quarterly Journal of Economics\/}~{\em 129\/}(2), 529--595.

\bibitem[\protect\citeauthoryear{Chetty, Friedman, Olsen, and
  Pistaferri}{Chetty et~al.}{2011}]{chetty_adjustment_2011}
Chetty, R., J.~N. Friedman, T.~Olsen, and L.~Pistaferri (2011).
\newblock Adjustment {Costs}, {Firm} {Responses}, and {Micro} vs. {Macro}
  {Labor} {Supply} {Elasticities}: {Evidence} from {Danish} {Tax} {Records}.
\newblock {\em The Quarterly Journal of Economics\/}~{\em 126\/}(2), 749--804.

\bibitem[\protect\citeauthoryear{Chetty, Looney, and Kroft}{Chetty
  et~al.}{2009}]{chetty_salience_2009}
Chetty, R., A.~Looney, and K.~Kroft (2009).
\newblock Salience and {Taxation}: {Theory} and {Evidence}.
\newblock {\em American Economic Review\/}~{\em 99\/}(4), 1145--1177.

\bibitem[\protect\citeauthoryear{Cho and Rust}{Cho and
  Rust}{2010}]{cho2010flat}
Cho, S. and J.~Rust (2010).
\newblock The flat rental puzzle.
\newblock {\em The Review of Economic Studies\/}~{\em 77\/}(2), 560--594.

\bibitem[\protect\citeauthoryear{Converse and Dennis}{Converse and
  Dennis}{2018}]{converse_role_2018}
Converse, B.~A. and P.~J. Dennis (2018).
\newblock The role of “{Prominent} {Numbers}” in open numerical judgment:
  {Strained} decision makers choose from a limited set of accessible numbers.
\newblock {\em Organizational Behavior and Human Decision Processes\/}~{\em
  147}, 94--107.

\bibitem[\protect\citeauthoryear{Cornwell, Schmutte, and Scur}{Cornwell
  et~al.}{2021}]{cornwell2019building}
Cornwell, C., I.~M. Schmutte, and D.~Scur (2021).
\newblock Building a productive workforce: The role of structured management
  practices.
\newblock {\em Management Science\/}~{\em 67\/}(12), 7308--7321.

\bibitem[\protect\citeauthoryear{Cullen, Li, and Perez-Truglia}{Cullen
  et~al.}{2022}]{cullen2022s}
Cullen, Z.~B., S.~Li, and R.~Perez-Truglia (2022).
\newblock What's my employee worth? the effects of salary benchmarking.
\newblock Technical report, National Bureau of Economic Research.

\bibitem[\protect\citeauthoryear{DellaVigna and Gentzkow}{DellaVigna and
  Gentzkow}{2019}]{dellavigna2019uniform}
DellaVigna, S. and M.~Gentzkow (2019).
\newblock Uniform pricing in us retail chains.
\newblock {\em The Quarterly Journal of Economics\/}~{\em 134\/}(4),
  2011--2084.

\bibitem[\protect\citeauthoryear{Derenoncourt, Noelke, Weil, and
  Taska}{Derenoncourt et~al.}{2021}]{derenoncourt2021spillover}
Derenoncourt, E., C.~Noelke, D.~Weil, and B.~Taska (2021).
\newblock Spillover effects from voluntary employer minimum wages.
\newblock Technical report, National Bureau of Economic Research.

\bibitem[\protect\citeauthoryear{Devereux, Liu, and Loretz}{Devereux
  et~al.}{2014}]{devereux2014elasticity}
Devereux, M.~P., L.~Liu, and S.~Loretz (2014).
\newblock The elasticity of corporate taxable income: New evidence from uk tax
  records.
\newblock {\em American Economic Journal: Economic Policy\/}~{\em 6\/}(2),
  19--53.

\bibitem[\protect\citeauthoryear{Dix-Carneiro and Kovak}{Dix-Carneiro and
  Kovak}{2017}]{dix2017trade}
Dix-Carneiro, R. and B.~K. Kovak (2017).
\newblock Trade liberalization and regional dynamics.
\newblock {\em American Economic Review\/}~{\em 107\/}(10), 2908--46.

\bibitem[\protect\citeauthoryear{Dube, Manning, and Naidu}{Dube
  et~al.}{2020}]{dube_monopsony_2020}
Dube, A., A.~Manning, and S.~Naidu (2020).
\newblock Monopsony and {Employer} {Mis}-optimization {Explain} {Why} {Wages}
  {Bunch} at {Round} {Numbers}.
\newblock NBER Working Paper \#24991.

\bibitem[\protect\citeauthoryear{Eichenbaum, Jaimovich, and Rebelo}{Eichenbaum
  et~al.}{2011}]{eichenbaum2011reference}
Eichenbaum, M., N.~Jaimovich, and S.~Rebelo (2011).
\newblock Reference prices, costs, and nominal rigidities.
\newblock {\em American Economic Review\/}~{\em 101\/}(1), 234--62.

\bibitem[\protect\citeauthoryear{Ellison, Snyder, and Zhang}{Ellison
  et~al.}{2018}]{ellison2018costs}
Ellison, S.~F., C.~Snyder, and H.~Zhang (2018).
\newblock Costs of managerial attention and activity as a source of sticky
  prices: Structural estimates from an online market.

\bibitem[\protect\citeauthoryear{Giustinelli, Manski, and Molinari}{Giustinelli
  et~al.}{2020}]{giustinelli2018tail}
Giustinelli, P., C.~F. Manski, and F.~Molinari (2020).
\newblock Tail and center rounding of probabilistic expectations in the health
  and retirement study.
\newblock {\em Journal of Econometrics\/}.

\bibitem[\protect\citeauthoryear{Goldfarb and Xiao}{Goldfarb and
  Xiao}{2019}]{goldfarb2019transitory}
Goldfarb, A. and M.~Xiao (2019).
\newblock Transitory shocks, limited attention, and a firm’s decision to
  exit.
\newblock {\em Mimeo\/}.

\bibitem[\protect\citeauthoryear{Hall and Krueger}{Hall and
  Krueger}{2012}]{hall2012evidence}
Hall, R.~E. and A.~B. Krueger (2012).
\newblock Evidence on the incidence of wage posting, wage bargaining, and
  on-the-job search.
\newblock {\em American Economic Journal: Macroeconomics\/}~{\em 4\/}(4),
  56--67.

\bibitem[\protect\citeauthoryear{Hanna, Mullainathan, and Schwartzstein}{Hanna
  et~al.}{2014}]{hanna2014learning}
Hanna, R., S.~Mullainathan, and J.~Schwartzstein (2014).
\newblock Learning through noticing: Theory and evidence from a field
  experiment.
\newblock {\em The Quarterly Journal of Economics\/}~{\em 129\/}(3),
  1311--1353.

\bibitem[\protect\citeauthoryear{Hazell, Patterson, Sarsons, and Taska}{Hazell
  et~al.}{2022}]{hazell2022national}
Hazell, J., C.~Patterson, H.~Sarsons, and B.~Taska (2022).
\newblock National wage setting.
\newblock Technical report, National Bureau of Economic Research.

\bibitem[\protect\citeauthoryear{Heidhues and K\H{o}szegi}{Heidhues and
  K\H{o}szegi}{2018}]{heidhues_behavioral_2018}
Heidhues, P. and B.~K\H{o}szegi (2018).
\newblock Behavioral {Industrial} {Organization}.
\newblock In B.~D. Bernheim, S.~DellaVigna, and D.~Laibson (Eds.), {\em
  Handbook of {Behavioral} {Economics}: {Applications} and {Foundations} 1},
  Volume~1 of {\em Handbook of {Behavioral} {Economics} - {Foundations} and
  {Applications} 1}, pp.\  517--612. North-Holland.

\bibitem[\protect\citeauthoryear{Hjort, Li, and Sarsons}{Hjort
  et~al.}{2020}]{hjort2020across}
Hjort, J., X.~Li, and H.~Sarsons (2020).
\newblock Across-country wage compression in multinationals.
\newblock Technical report, National Bureau of Economic Research.

\bibitem[\protect\citeauthoryear{Jones}{Jones}{1896}]{jones1896round}
Jones, E.~D. (1896).
\newblock Round numbers in wages and prices.
\newblock {\em Publications of the American Statistical Association\/}~{\em
  5\/}(35-36), 111--141.

\bibitem[\protect\citeauthoryear{Jovanovic}{Jovanovic}{1979}]{jovanovic1979job}
Jovanovic, B. (1979).
\newblock Job matching and the theory of turnover.
\newblock {\em Journal of Political Economy\/}~{\em 87\/}(5, Part 1), 972--990.

\bibitem[\protect\citeauthoryear{Kaur}{Kaur}{2019}]{kaur2019nominal}
Kaur, S. (2019).
\newblock Nominal wage rigidity in village labor markets.
\newblock {\em American Economic Review\/}~{\em 109\/}(10), 3585--3616.

\bibitem[\protect\citeauthoryear{Kleven}{Kleven}{2016}]{kleven_bunching_2016}
Kleven, H.~J. (2016).
\newblock Bunching.
\newblock {\em Annual Review of Economics\/}~{\em 8\/}(1), 435--464.

\bibitem[\protect\citeauthoryear{Kleven and Waseem}{Kleven and
  Waseem}{2013}]{kleven2013using}
Kleven, H.~J. and M.~Waseem (2013).
\newblock Using notches to uncover optimization frictions and structural
  elasticities: Theory and evidence from pakistan.
\newblock {\em The Quarterly Journal of Economics\/}~{\em 128\/}(2), 669--723.

\bibitem[\protect\citeauthoryear{Korvorst and Damian}{Korvorst and
  Damian}{2008}]{korvorst_differential_2008}
Korvorst, M. and M.~F. Damian (2008).
\newblock The differential influence of decades and units on multidigit number
  comparison.
\newblock {\em Quarterly Journal of Experimental Psychology\/}~{\em 61\/}(8),
  1250--1264.

\bibitem[\protect\citeauthoryear{Kremer, Rao, and Schilbach}{Kremer
  et~al.}{2019}]{kremer2019behavioral}
Kremer, M., G.~Rao, and F.~Schilbach (2019).
\newblock Behavioral development economics.
\newblock In {\em Handbook of Behavioral Economics: Applications and
  Foundations 1}, Volume~2, pp.\  345--458. Elsevier.

\bibitem[\protect\citeauthoryear{Lacetera, Pope, and Sydnor}{Lacetera
  et~al.}{2012}]{lacetera_heuristic_2012}
Lacetera, N., D.~G. Pope, and J.~R. Sydnor (2012).
\newblock Heuristic {Thinking} and {Limited} {Attention} in the {Car} {Market}.
\newblock {\em American Economic Review\/}~{\em 102\/}(5), 2206--2236.

\bibitem[\protect\citeauthoryear{Lachowska, Mas, Saggio, and
  Woodbury}{Lachowska et~al.}{2022}]{lachowska2021wage}
Lachowska, M., A.~Mas, R.~Saggio, and S.~Woodbury (2022).
\newblock Wage posting or wage bargaining? a test using dual jobholders.
\newblock {\em Journal of Labor Economics (forthcoming)\/}~(w28409).

\bibitem[\protect\citeauthoryear{Lagos}{Lagos}{2023}]{lagos2023labor}
Lagos, L. (2023).
\newblock Labor market institutions and the composition of firm compensation:
  Evidence from brazilian collective bargaining.
\newblock {\em Working Paper\/}.

\bibitem[\protect\citeauthoryear{Lamadon, Mogstad, and Setzler}{Lamadon
  et~al.}{2022}]{lamadon2022imperfect}
Lamadon, T., M.~Mogstad, and B.~Setzler (2022).
\newblock Imperfect competition, compensating differentials, and rent sharing
  in the us labor market.
\newblock {\em American Economic Review\/}~{\em 112\/}(1), 169--212.

\bibitem[\protect\citeauthoryear{Landier and Thesmar}{Landier and
  Thesmar}{2008}]{landier2008financial}
Landier, A. and D.~Thesmar (2008).
\newblock Financial contracting with optimistic entrepreneurs.
\newblock {\em The Review of Financial Studies\/}~{\em 22\/}(1), 117--150.

\bibitem[\protect\citeauthoryear{Lemieux}{Lemieux}{2008}]{lemieux2008changing}
Lemieux, T. (2008).
\newblock The changing nature of wage inequality.
\newblock {\em Journal of population Economics\/}~{\em 21\/}(1), 21--48.

\bibitem[\protect\citeauthoryear{Manning}{Manning}{2011}]{manning_imperfect_2011}
Manning, A. (2011).
\newblock Imperfect {Competition} in the {Labor} {Market}.
\newblock In {\em Handbook of {Labor} {Economics}}, Volume~4, pp.\  973--1041.
  Elsevier.

\bibitem[\protect\citeauthoryear{Manski and Molinari}{Manski and
  Molinari}{2010}]{manski2010rounding}
Manski, C.~F. and F.~Molinari (2010).
\newblock Rounding probabilistic expectations in surveys.
\newblock {\em Journal of Business \& Economic Statistics\/}~{\em 28\/}(2),
  219--231.

\bibitem[\protect\citeauthoryear{Matejka}{Matejka}{2016}]{matejka_rationally_2016}
Matejka, F. (2016, July).
\newblock Rationally {Inattentive} {Seller}: {Sales} and {Discrete} {Pricing}.
\newblock {\em The Review of Economic Studies\/}~{\em 83\/}(3), 1125--1155.

\bibitem[\protect\citeauthoryear{Mavrokonstantis and Seibold}{Mavrokonstantis
  and Seibold}{2022}]{mavrokonstantis2022bunching}
Mavrokonstantis, P. and A.~Seibold (2022).
\newblock Bunching and adjustment costs: Evidence from cypriot tax reforms.
\newblock {\em Journal of Public Economics\/}~{\em 214}, 104727.

\bibitem[\protect\citeauthoryear{McCall}{McCall}{1970}]{mccall1970economics}
McCall, J.~J. (1970).
\newblock Economics of information and job search.
\newblock {\em The Quarterly Journal of Economics\/}~{\em 84}, 113--126.

\bibitem[\protect\citeauthoryear{Pierce, Rees-Jones, and Blank}{Pierce
  et~al.}{2020}]{pierce2020negative}
Pierce, L., A.~Rees-Jones, and C.~Blank (2020).
\newblock The negative consequences of loss-framed performance incentives.
\newblock {\em NBER Working Paper\/}~(w26619).

\bibitem[\protect\citeauthoryear{Riddles, Lohr, Brick, Langetieg, Payne, and
  Plumley}{Riddles et~al.}{2016}]{riddles_handling_2016}
Riddles, M.~K., S.~L. Lohr, J.~M. Brick, P.~T. Langetieg, J.~M. Payne, and
  A.~H. Plumley (2016).
\newblock Handling {Respondent} {Rounding} of {Wages} {Using} the {IRS} and
  {CPS} {Matched} {Dataset}.

\bibitem[\protect\citeauthoryear{Saez}{Saez}{2010}]{saez_taxpayers_2010}
Saez, E. (2010).
\newblock Do {Taxpayers} {Bunch} at {Kink} {Points}?
\newblock {\em American Economic Journal: Economic Policy\/}~{\em 2\/}(3),
  180--212.

\bibitem[\protect\citeauthoryear{Schweitzer and Severance-Lossin}{Schweitzer
  and Severance-Lossin}{1996}]{schweitzer_rounding_1996}
Schweitzer, M.~E. and E.~Severance-Lossin (1996).
\newblock Rounding in {Earnings} {Data}.
\newblock {\em Working Papers\/}~(WP 96-12).
\newblock Publisher: Federal Reserve Bank of Cleveland.

\bibitem[\protect\citeauthoryear{Simon}{Simon}{1962}]{simon1962new}
Simon, H.~A. (1962).
\newblock New developments in the theory of the firm.
\newblock {\em The American Economic Review\/}~{\em 52\/}(2), 1--15.

\bibitem[\protect\citeauthoryear{Stevens}{Stevens}{2020}]{stevens2020coarse}
Stevens, L. (2020).
\newblock Coarse pricing policies.
\newblock {\em The Review of Economic Studies\/}~{\em 87\/}(1), 420--453.

\bibitem[\protect\citeauthoryear{Stiving}{Stiving}{2000}]{stiving_price-endings_2000}
Stiving, M. (2000).
\newblock Price-{Endings} {When} {Prices} {Signal} {Quality}.
\newblock {\em Management Science\/}~{\em 46\/}(12), 1617--1629.

\bibitem[\protect\citeauthoryear{Strulov-Shlain}{Strulov-Shlain}{2023}]{strulov2019more}
Strulov-Shlain, A. (2023).
\newblock More than a penny’s worth: Left-digit bias and firm pricing.
\newblock {\em Review of Economic Studies\/}~{\em 90\/}(5), 2612--2645.

\bibitem[\protect\citeauthoryear{Tervi{\"o}}{Tervi{\"o}}{2009}]{tervio2009superstars}
Tervi{\"o}, M. (2009).
\newblock Superstars and mediocrities: Market failure in the discovery of
  talent.
\newblock {\em The Review of Economic Studies\/}~{\em 76\/}(2), 829--850.

\bibitem[\protect\citeauthoryear{Whynes, Philips, and Frew}{Whynes
  et~al.}{2005}]{whynes_think_2005}
Whynes, D.~K., Z.~Philips, and E.~Frew (2005).
\newblock Think of a number… any number?
\newblock {\em Health Economics\/}~{\em 14\/}(11), 1191--1195.

\end{thebibliography}
	\end{singlespace}
	
	\clearpage 
	\clearpage
\section*{Figures and Tables}

\clearpage

\clearpage

	\clearpage
\begin{table}[H]
	\caption{\centering Summary statistics on workers in the RAIS, new-hires sample, and firm random sample} \label{tab_rais_summ}
	{\footnotesize
		\begin{centering} 
			\protect
			\begin{tabular}{lcccc}
				\addlinespace \addlinespace \midrule			
				&   & New-hires  & Firm random \\
				& RAIS & sample & sample \\ \cmidrule{2-4}
				& (1) & (2) & (3) \\
				\midrule 	
				
				\multicolumn{4}{l}{\hspace{-1em} \textbf{Panel A. Worker characteristics}} \\ 
				Average age &34.966 &30.304 &32.845 \\
Male &0.580 &0.635 &0.619 \\
White &0.606 &0.574 &0.600 \\
Elementary school or less &0.376 &0.410 &0.408 \\
Completed high school &0.471 &0.532 &0.510 \\
Completed university &0.153 &0.058 &0.082 \\
 \midrule
				
				\multicolumn{4}{l}{\hspace{-1em} \textbf{Panel B. Earnings}} \\ 
				Mean monthly salary (Reals) &1288.915 &807.502 &1066.613 \\
Median monthly salary (Reals) &742.429 &621.422 &686.848 \\
Share of mean earnings divisible by 10 &0.037 &0.072 &0.041 \\
Share of mean earnings divisible by 100 &0.017 &0.034 &0.018 \\
Share of mean earnings divisible by 1,000 &0.003 &0.006 &0.003 \\
Share of contracted earnings divisible by 10 &0.180 &0.295 &0.230 \\
Share of contracted earnings divisible by 100 &0.066 &0.121 &0.084 \\
Share of contracted earnings divisible by 1,000 &0.010 &0.019 &0.013 \\
 \midrule						
				
				\multicolumn{4}{l}{\hspace{-1em} \textbf{Panel C. Industry}} \\ 
				Primary sector &0.046 &0.034 &0.026 \\
Construction and utilities &0.140 &0.167 &0.195 \\
Manufacturing &0.121 &0.174 &0.153 \\
Retail &0.257 &0.359 &0.368 \\
Services &0.437 &0.265 &0.259 \\
 \midrule											
				
				\multicolumn{4}{l}{\hspace{-1em} \textbf{Panel D. Region}} \\ 					
				North &0.054 &0.050 &0.048 \\
Northeast &0.176 &0.149 &0.156 \\
Southeast &0.506 &0.521 &0.530 \\
South &0.173 &0.187 &0.189 \\
Midwest &0.092 &0.093 &0.077 \\
 \midrule

				Number of observations &842,196,095 &206,685,308 &28,087,000 \\
 \midrule \addlinespace \addlinespace

			\end{tabular}
			\par\end{centering}
		
		\begin{singlespace} \vspace{-.5cm}
			\noindent \justify \textit{Notes:} This table shows summary statistics on workers in the \textit{Relação Anual de Informações Sociais} (RAIS), the new-hires sample, and the firm random sample. See Section \ref{sec:context-data} for sample definitions. Earnings are expressed in Brazilian Reals.
			
		\end{singlespace}	
		
	}
\end{table}

\clearpage
\begin{table}[H]{\footnotesize
		\begin{center}
			\caption{Outcomes of firms that tend to hire workers at round numbers} \label{reg_firm_performance}
			\newcommand\w{2}
			\begin{tabular}{l@{}lR{\w cm}@{}L{0.5cm}R{\w cm}@{}L{0.5cm}R{\w cm}@{}L{0.5cm}R{\w cm}@{}L{0.5cm}}
				\midrule
				&& \multicolumn{8}{c}{Dependent variable:} \\ \cmidrule{3-10}
				&& New hire  && New hire    && Firm job    && Firm left  \\
				&& separated && resigned    && growth rate && market \\
				&& (1) && (2) && (3) && (4)  \\
				\midrule
				Bunching firm&&0.041&\sym{***}&0.012&\sym{***}&$-$0.034&\sym{***}&0.011&\sym{***}\\
&&(0.003&)&(0.001&)&(0.002&)&(0.001&)\\
Mean Dep. Var.&&0.353&&0.115&&0.016&&0.101&\\
N&&20,281,312&&20,281,312&&1,381,246&&1,752,411&\\
  \midrule
			\end{tabular}
		\end{center}
		\begin{singlespace}  \vspace{-.5cm}
			\noindent \justify \textit{Notes:} This table displays estimates of $\beta$ from equation \eqref{reg:firm-outcomes}. Each column shows the result of a regression using the dependent variable listed in the column header. In column 1, the outcome equals one if a new hire separated from the firm during the year she was hired (year $t$) or the following year (year $t+1$), and zero otherwise. Column 2 is defined analogously but using worker resignation likelihood instead of separation likelihood. In column 3, the dependent variable is the percent change in the number of workers employed between consecutive years. In column 4, the outcome is a dummy that equals one if the firm had no workers at the end of the year and zero otherwise.
			
			I use the firm random sample to estimate all regressions. In columns 1 and 2, the regressions are estimated at the worker-by-firm-by-year level and only using data from the year in which a worker was hired and the following year. In columns 3 and 4, the regressions are estimated at the firm-by-year level.
			
			The regressions control for firm age, share of employees with completed high school, share of employees with completed college, educational attainment of the firm manager, a dummy for having an HR department, the mean earnings of the firm employees, firm size fixed effects, number of hires fixed effects, and industry-by-microregion-by-year fixed effects. The specifications in columns 1 and 2 additionally control for worker gender, race, and occupation. 
			
			Heteroskedasticity-robust standard errors clustered at the firm level in parentheses. $^{***}$, $^{**}$ and $^*$ denote significance at the 1\%, 5\% and 10\% levels.
		\end{singlespace} 	
	}
\end{table}

\clearpage
\begin{table}[H]{\footnotesize
		\begin{center}
			\caption{The use of round-numbered salaries across decision environments} \label{reg_salary_increase}
			\newcommand\w{1.45}
			\begin{tabular}{l@{}lR{\w cm}@{}L{0.45cm}R{\w cm}@{}L{0.45cm}R{\w cm}@{}L{0.45cm}R{\w cm}@{}L{0.45cm}R{\w cm}@{}L{0.45cm}R{\w cm}@{}L{0.45cm}}
				\midrule
				&& \multicolumn{12}{c}{Dependent variable:} \\ \cmidrule{3-13}
				&& \multicolumn{3}{c}{Salary increase in R\$}  && \multicolumn{3}{c}{Salary increase in \%} && \multicolumn{3}{c}{Either a round} \\
				&& \multicolumn{3}{c}{is a round number}  && \multicolumn{3}{c}{is an integer} && \multicolumn{3}{c}{number or an integer} & \\
				\cmidrule{3-5} \cmidrule{7-9} \cmidrule{11-13}
				&& (1) && (2) && (3) && (4) && (5) && (6) \\
				\midrule
				Bunching firm&&0.293&\sym{***}&0.322&\sym{***}&0.162&\sym{***}&0.082&\sym{***}&0.286&\sym{***}&0.331&\sym{***}\\
&&(0.003&)&(0.004&)&(0.003&)&(0.003&)&(0.003&)&(0.004&)\\
Mean Dep. Var.&&0.354&&0.123&&0.344&&0.110&&0.421&&0.214&\\
N&&4,953,782&&3,646,479&&4,953,782&&3,646,479&&4,953,782&&3,646,479&\\
 
				Excl. zero growth && No && Yes && No && Yes && No && Yes \\ \midrule 
			\end{tabular}
		\end{center}
		\begin{singlespace} \vspace{-.5cm}
		
		\noindent \justify \textit{Notes:} This table displays estimates of $\beta$ from equation \eqref{reg:firm-outcomes}. In columns 1 and 2, the outcome is a dummy that equals one if the change in worker's $i$'s contracted salary between $t$ and $t+1$`, measured in Brazilian Reals, is a round number and zero otherwise. In columns 3 and 4, the outcome is a dummy that equals one if the percent change between $t$ and $t+1$ of worker $i$'s contracted salary is an integer and zero otherwise. In columns 5 and 6, the outcome equals one if either the absolute wage change is a round number or the percent change is an integer and zero otherwise. 
		
		I use the firm random sample to estimate all regressions. The regressions are estimated at the worker-by-firm-by-year level and only using data from the year in which a worker was hired and the following year. Even columns exclude new hires whose salary did not change in nominal terms.
		
		The regressions control for worker gender, worker race, worker occupation, firm age, share of employees with completed high school, share of employees with completed college, educational attainment of the firm manager, a dummy for having an HR department, the mean earnings of the firm employees, firm size fixed effects, number of hires fixed effects, and industry-by-microregion-by-year fixed effects. 
		
		Heteroskedasticity-robust standard errors clustered at the firm level in parentheses. $^{***}$, $^{**}$ and $^*$ denote significance at the 1\%, 5\% and 10\% levels.

		\end{singlespace}
		
	}
\end{table}

\clearpage
\begin{table}[H]{\footnotesize
		\begin{center}
			\caption{\centering Testing the predictions of the model} \label{tab_predictions}
			
			\newcommand\w{1.8}
			\begin{tabular}{l@{}lR{\w cm}@{}L{0.45cm}R{\w cm}@{}L{0.45cm}R{\w cm}@{}L{0.45cm}R{\w cm}@{}L{0.45cm}}
				\midrule
				& \multicolumn{8}{c}{Dependent variable} \\ \cmidrule{3-10} 
				& \multicolumn{4}{c}{Fraction of workers hired} && \multicolumn{4}{c}{Dummy for hiring a worker} \\
				& \multicolumn{4}{c}{through coarse wage-setting $(\hat{\theta} )$} && \multicolumn{4}{c}{at a round number ($\mathbbm{1}\{w_{i} \in R\}$)} \\
				\cmidrule{3-5} \cmidrule{7-10}
				&& (1) && (2) && (3) && (4)\\
				\midrule
				Consumer price index (logs)&&0.108&\sym{***}&$-$0.004&&0.036&\sym{***}&0.029&\sym{***}\\
&&(0.020&)&(0.021&)&(0.001&)&(0.001&)\\

				Firm size (logs)&&$-$0.972&\sym{***}&$-$0.879&\sym{***}&$-$0.012&\sym{***}&$-$0.032&\sym{***}\\
&&(0.034&)&(0.086&)&(0.001&)&(0.001&)\\

				Firm hiring experience (logs)&&$-$0.882&\sym{***}&$-$0.053&&$-$0.025&\sym{***}&$-$0.002&\sym{**}\\
&&(0.050&)&(0.140&)&(0.001&)&(0.001&)\\

				\midrule
				Fixed firm sample? && No && Yes && No && Yes \\
				\midrule
			\end{tabular}
		\end{center}
		\begin{singlespace} \vspace{-.5cm}
			\noindent \justify \textit{Notes:} This table shows linear correlations between the covariate listed in the row header and the outcome listed in the column header. 
			
			In columns 1--2, the dependent variable is the estimated fraction of workers hired through coarse wage-setting, $\hat{\theta}$. Section \ref{sub:bunching} describes how $\hat{\theta}$ is estimated. Each observation denotes the excess mass at each value taken by a covariate. Thus, the sample size varies by covariate. For log CPI, $N = 1,980$. For firm size and hiring experience, $N = 100$. 
			
			In columns 3--4, the dependent variable is a dummy that equals one for workers hired at a round-numbered salary ($\mathbbm{1}\{w_{i} \in R\}$). In addition to the variables listed in the table, the regressions control for worker gender, worker race, worker occupation, worker education, worker potential experience, firm fixed effects, year fixed effects, and metropolitan region fixed effects. I normalize variables by their standard deviation so that the coefficients of the regressions can be interpreted as the linear correlation coefficients. The sample size is $N = 3,988,606$ in column 3 and $N = 1,493,286$ in column 4.
			
			Heteroskedasticity-robust standard errors clustered at the firm level in parentheses. $^{***}$, $^{**}$ and $^*$ denote significance at the 1\%, 5\% and 10\% levels.		
			
		\end{singlespace} 	
	}
\end{table}

	\clearpage 
	\appendix 
	\clearpage 
\begin{center}
	\noindent {\LARGE \textbf{Appendix --- For Online Publication}}
\end{center}
\section{Appendix Figures and Tables} \label{app:add-figs}

\setcounter{table}{0}
\setcounter{figure}{0}
\setcounter{equation}{0}	
\renewcommand{\thetable}{A\arabic{table}}
\renewcommand{\thefigure}{A\arabic{figure}}
\renewcommand{\theequation}{A\arabic{equation}}

\begin{figure}[htp]
	\caption{Distribution of the last digits of new hires' contracted salaries} \label{fig_last_digits}
	\centering
	\begin{subfigure}[t]{.48\textwidth}
		\caption*{Panel A. Last two digits} \label{fig_last_2digits}
		\centering
		\includegraphics[width=\textwidth]{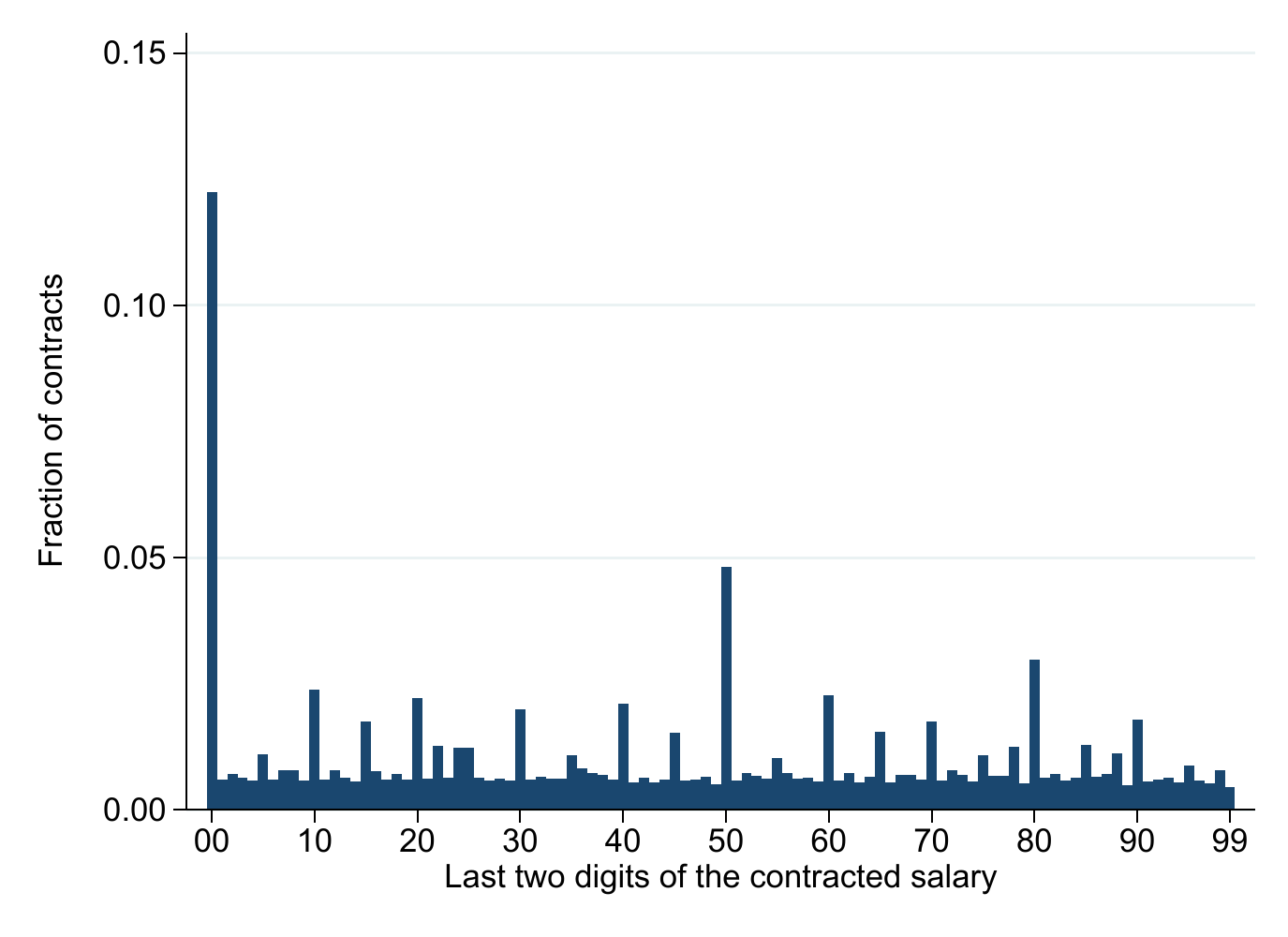}
	\end{subfigure}
	\hfill        
	\begin{subfigure}[t]{0.48\textwidth}
		\caption*{Panel B. Last three digits} \label{fig_last_3digits}
		\centering
		\includegraphics[width=\textwidth]{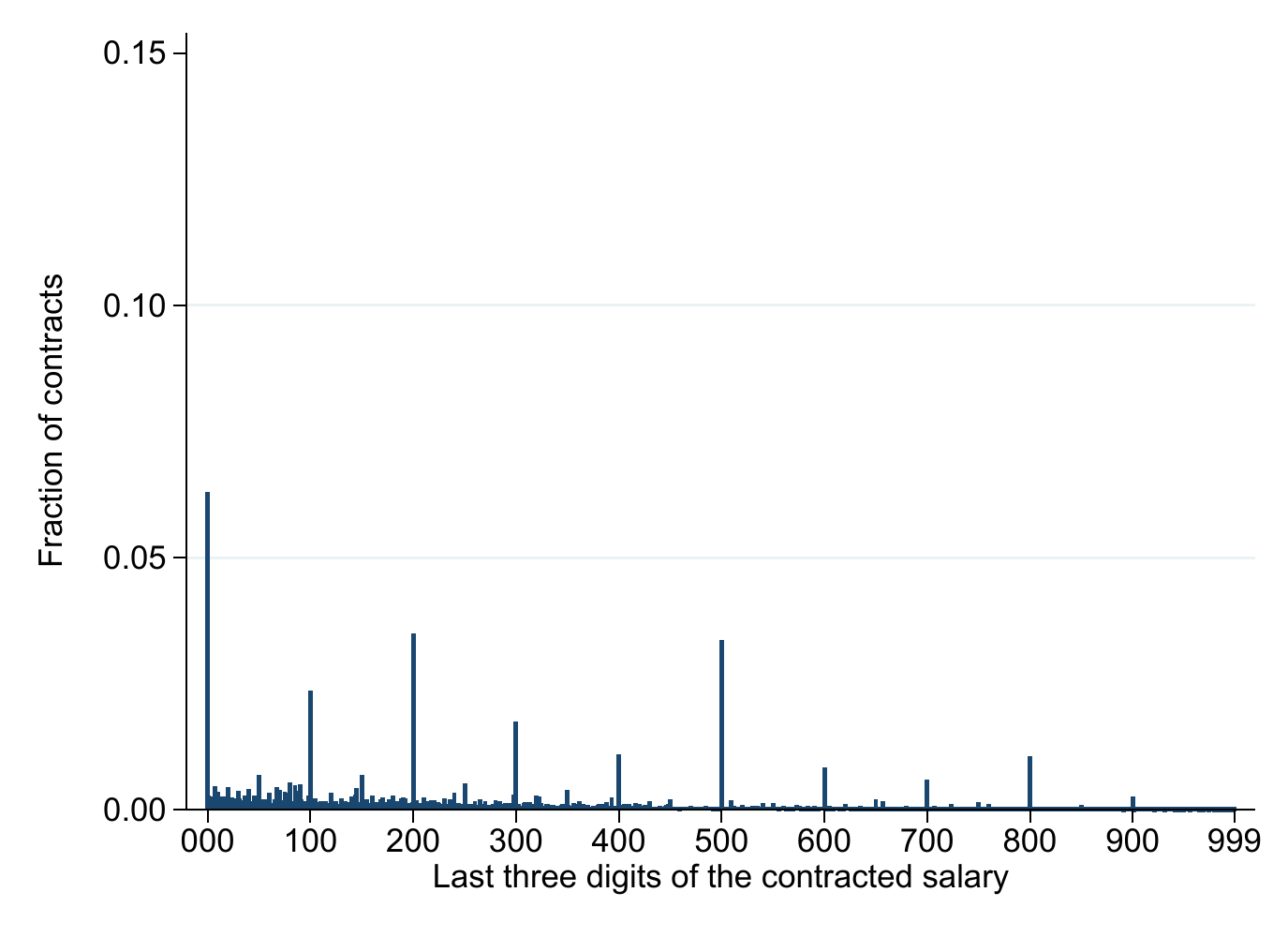}
	\end{subfigure}		
	\footnotesize \singlespacing \justify \textit{Notes:} Panel A shows the distribution of the last two digits of contracted earnings (in R\$1 bins) in the new-hires sample. Panel B shows the distribution of the last three digits (conditional on the salary having more than three digits).
\end{figure}

\clearpage
\begin{figure}[H]
	\caption{Distribution of monthly earnings in four Brazilian datasets}\label{fig_bunching_data}
	\centering
	\begin{subfigure}[t]{.48\textwidth}
		\caption*{Panel A. Household Survey (PNAD)}\label{fig_bunching_pnad}
		\centering
		\includegraphics[width=\linewidth]{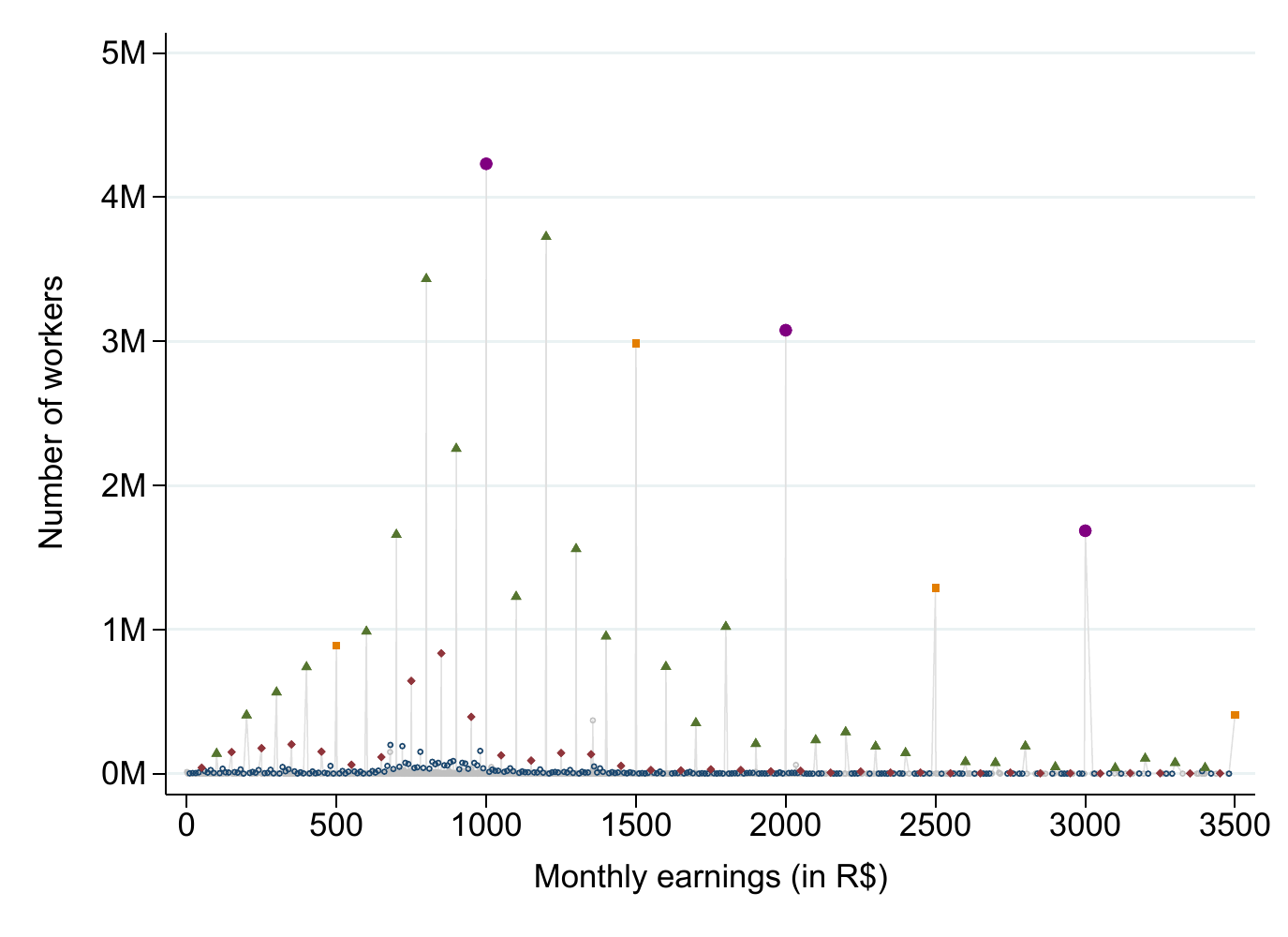}
	\end{subfigure}
	\hfill		
	\begin{subfigure}[t]{0.48\textwidth}
		\caption*{Panel B. Labor Force Survey (PME)}\label{fig_bunching_pme}
		\centering
		\includegraphics[width=\linewidth]{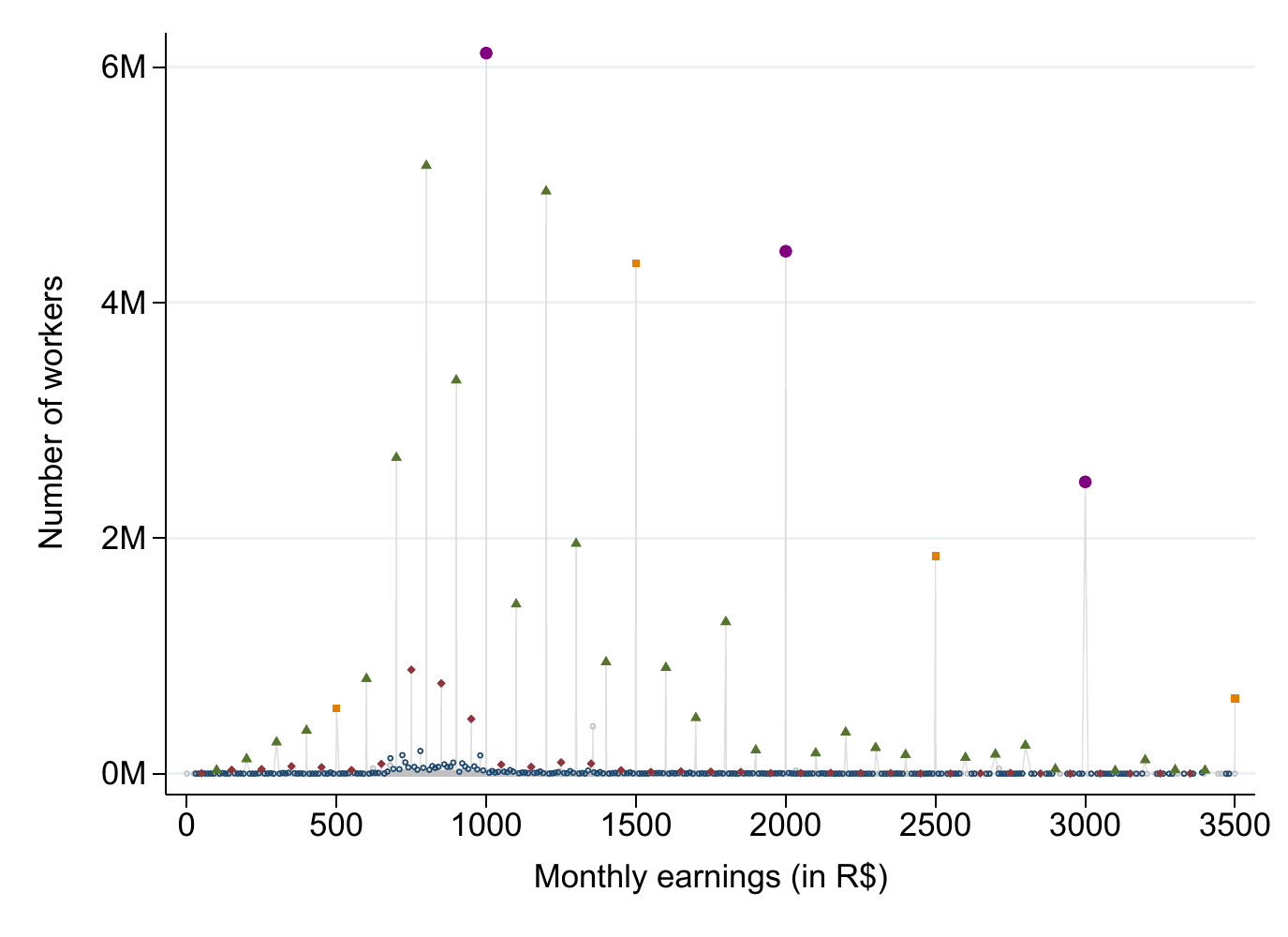}
	\end{subfigure}
	\hfill		
	\begin{subfigure}[t]{.48\textwidth}
		\caption*{Panel C. Population Census}\label{fig_bunching_cen}
		\centering
		\includegraphics[width=\linewidth]{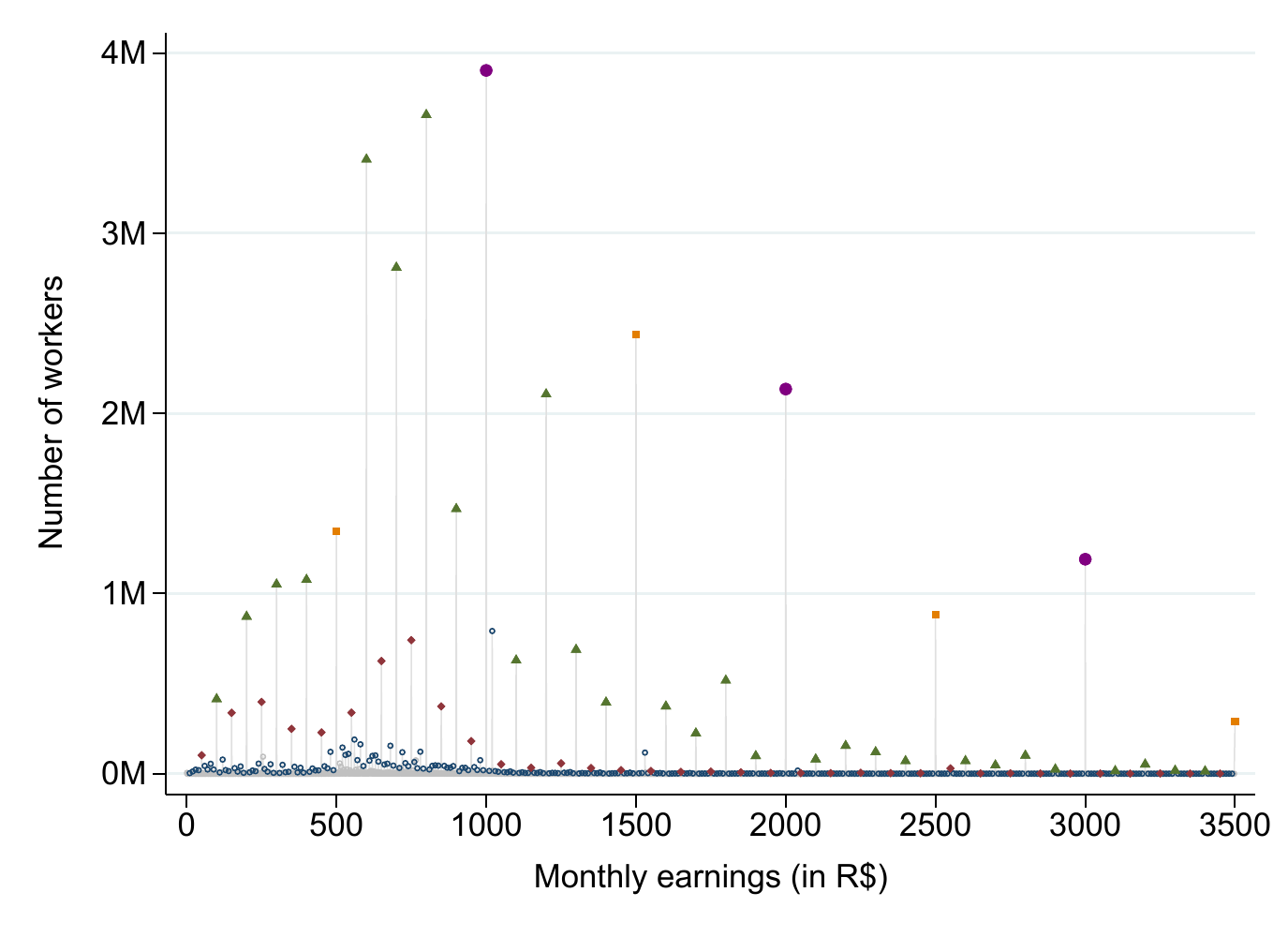}
	\end{subfigure}
	\hfill		
	\begin{subfigure}[t]{0.48\textwidth}
		\caption*{Panel D. Social Programs Registry}\label{fig_bunching_cad}
		\centering
		\includegraphics[width=\linewidth]{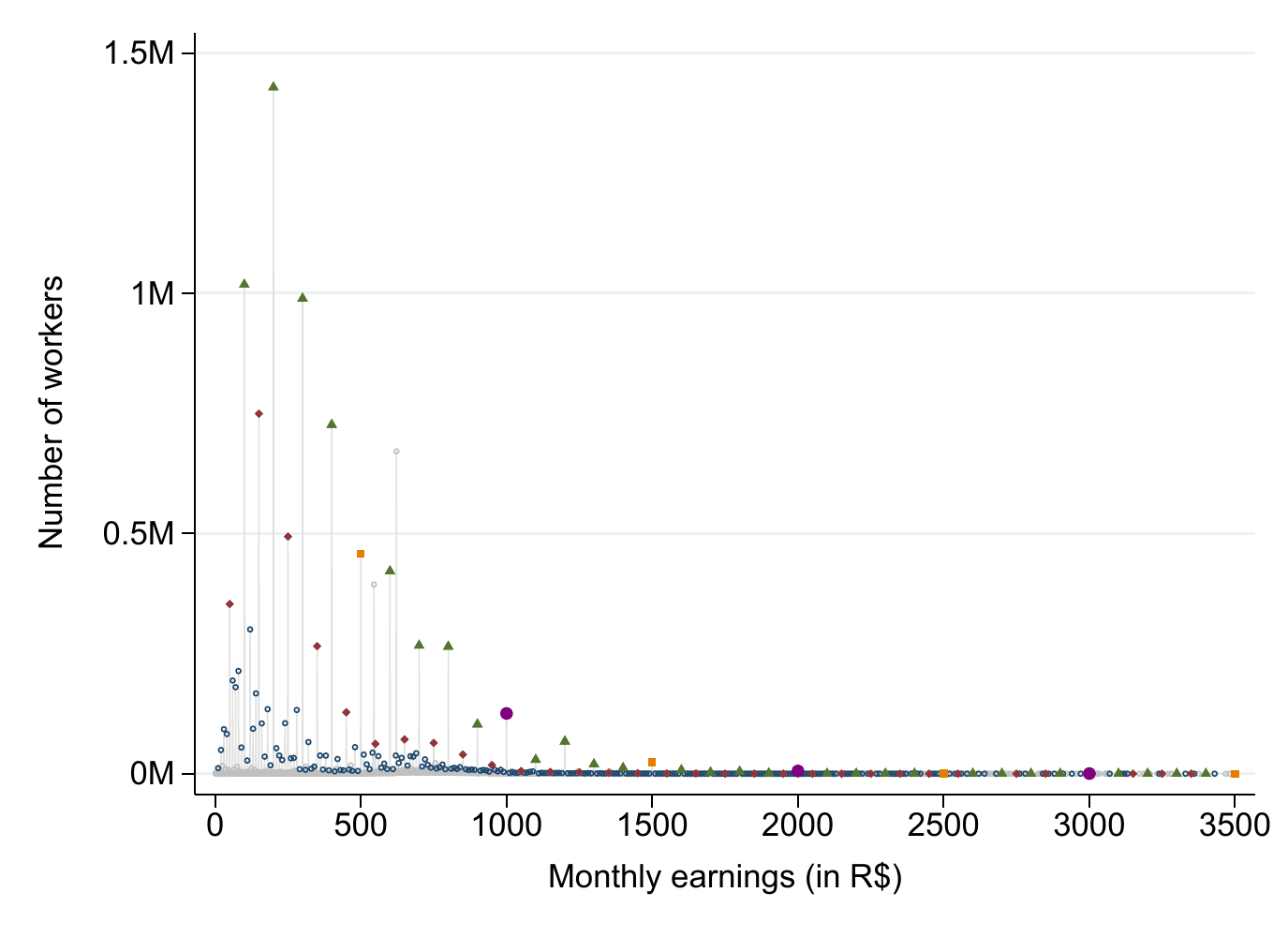}
	\end{subfigure}	
	\includegraphics[width=.9\linewidth]{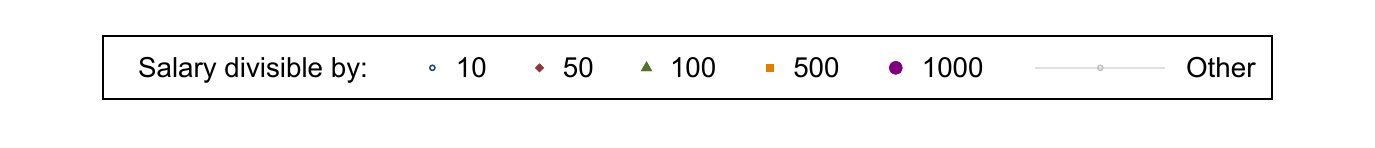}
	\hfill				
	{\footnotesize
		\singlespacing \justify
		
		\textit{Notes:} This figure shows the distribution of monthly earnings in the dataset listed in the panel title. The datasets are the 2013 Brazilian Household Survey (\textit{Pesquisa Nacional por Amostra de Domicílios}, abbreviated PNAD), the 2013 Brazilian Labor Force Survey (\textit{Pesquisa Mensal de Emprego}, abbreviated PME), the 2010 Brazilian Population Census (\textit{Censo Demográfico}), and the 2013 Social Programs Registry of Individuals (\textit{Cadastro Único}). I focus on the monthly earnings of full-time employed workers aged 18--65. I exclude workers employed by public-sector firms and individuals who work without remuneration. 
		
	}
\end{figure}

\clearpage
\begin{figure}[H]
	\caption{Histogram of the share of workers hired at a round salary in each firm} \label{fig_hist_bunch_firms}
	\centering
	\begin{subfigure}[t]{.48\textwidth} 
		\caption*{Panel A. All firms \\ ~ }
		\centering
		\includegraphics[width=\textwidth]{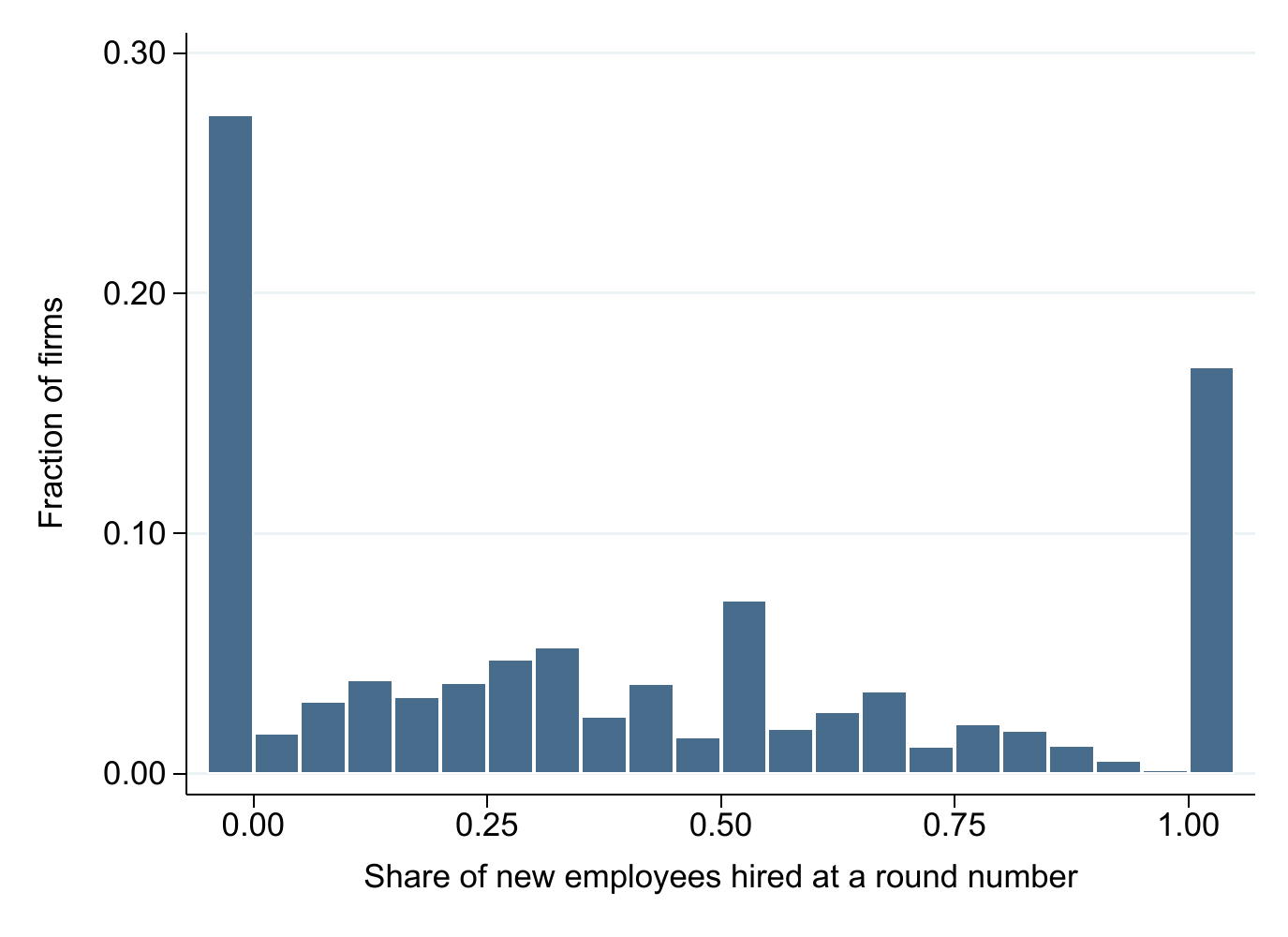} \label{fig_hist_bunch_firms_10}
	\end{subfigure}
	\hfill		
	\begin{subfigure}[t]{0.48\textwidth}
		\caption*{Panel B. Firms that hired five or more workers in the sample}
		\centering
		\includegraphics[width=\textwidth]{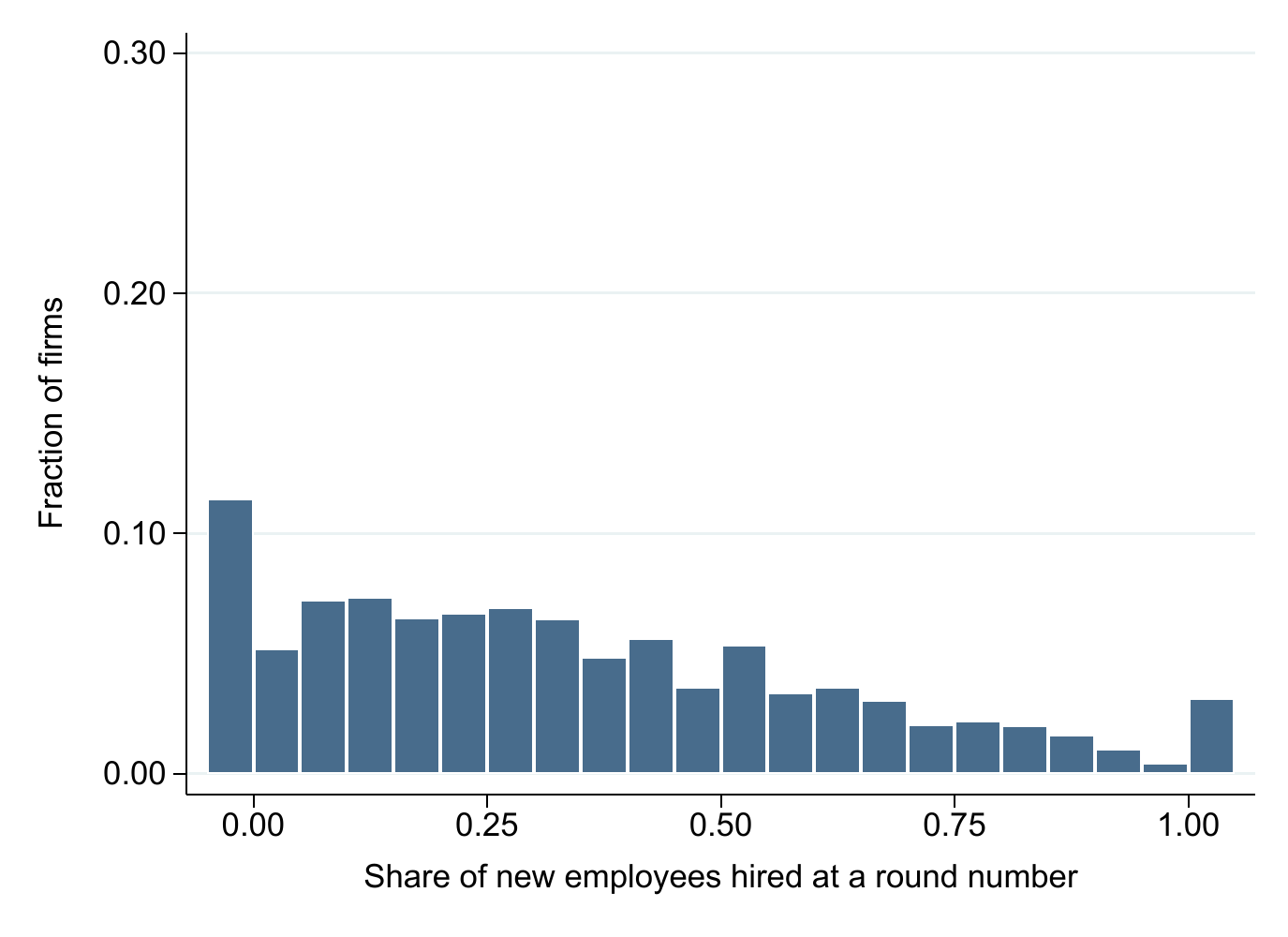} \label{fig_hist_bunch_firms_10_5hires}
	\end{subfigure}	
	\footnotesize \singlespacing \justify \textit{Notes:} This figure show histograms of the share of workers in each firm hired at a round-numbered salary in the firm random sample. Panel A shows the histogram for all firms. Panel B shows the histogram for the subset of firms that hired at least five workers during 2003--2017.
\end{figure}

\clearpage
\begin{figure}[H]
	\caption{Compensation report for an economist in Ithaca, NY, USA}\label{fig_payscale}
	\centering
	\vspace{.5cm}
	
	\begin{subfigure}[t]{1\textwidth}
		\caption*{Panel A. Factors that affect the compensation report}
		\centering
		\includegraphics[width=\textwidth]{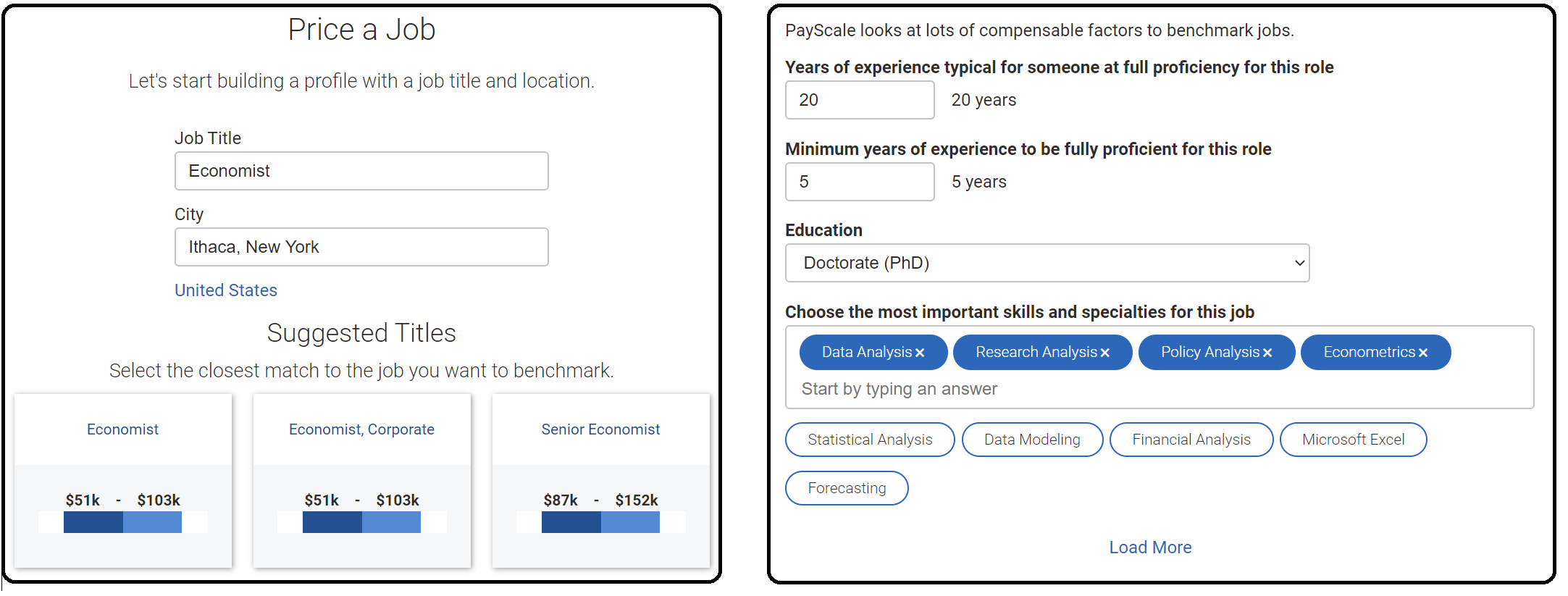}
	\end{subfigure} \vspace{.7cm}
	
	\begin{subfigure}[t]{0.7\textwidth}
		\caption*{Panel B. Suggested compensation}
		\centering
		\includegraphics[width=\textwidth]{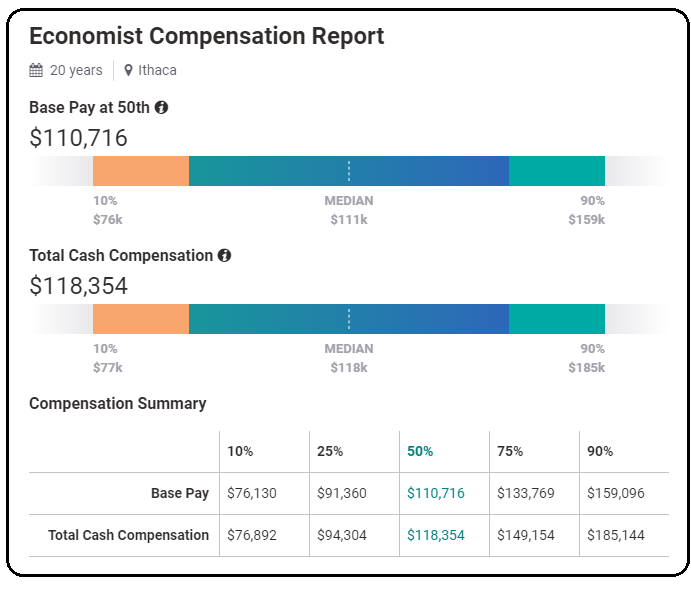} 
	\end{subfigure}
	\footnotesize \singlespacing \justify \textit{Notes:} This figure shows a compensation report provided by the firm PayScale, based on a query by the author. These compensation reports are advertised as the ``right pay'' for a prospective candidate.
\end{figure}

\clearpage
\begin{figure}[H]
	\caption{Wage compression and wage stickiness in the salaries of large firms' new hires} \label{fig_wage_comp_bigf}
	\centering
	\begin{subfigure}[t]{.48\textwidth}
		\caption*{Panel A. Outcome: Gini coefficient}
		\centering
				\includegraphics[width=\linewidth]{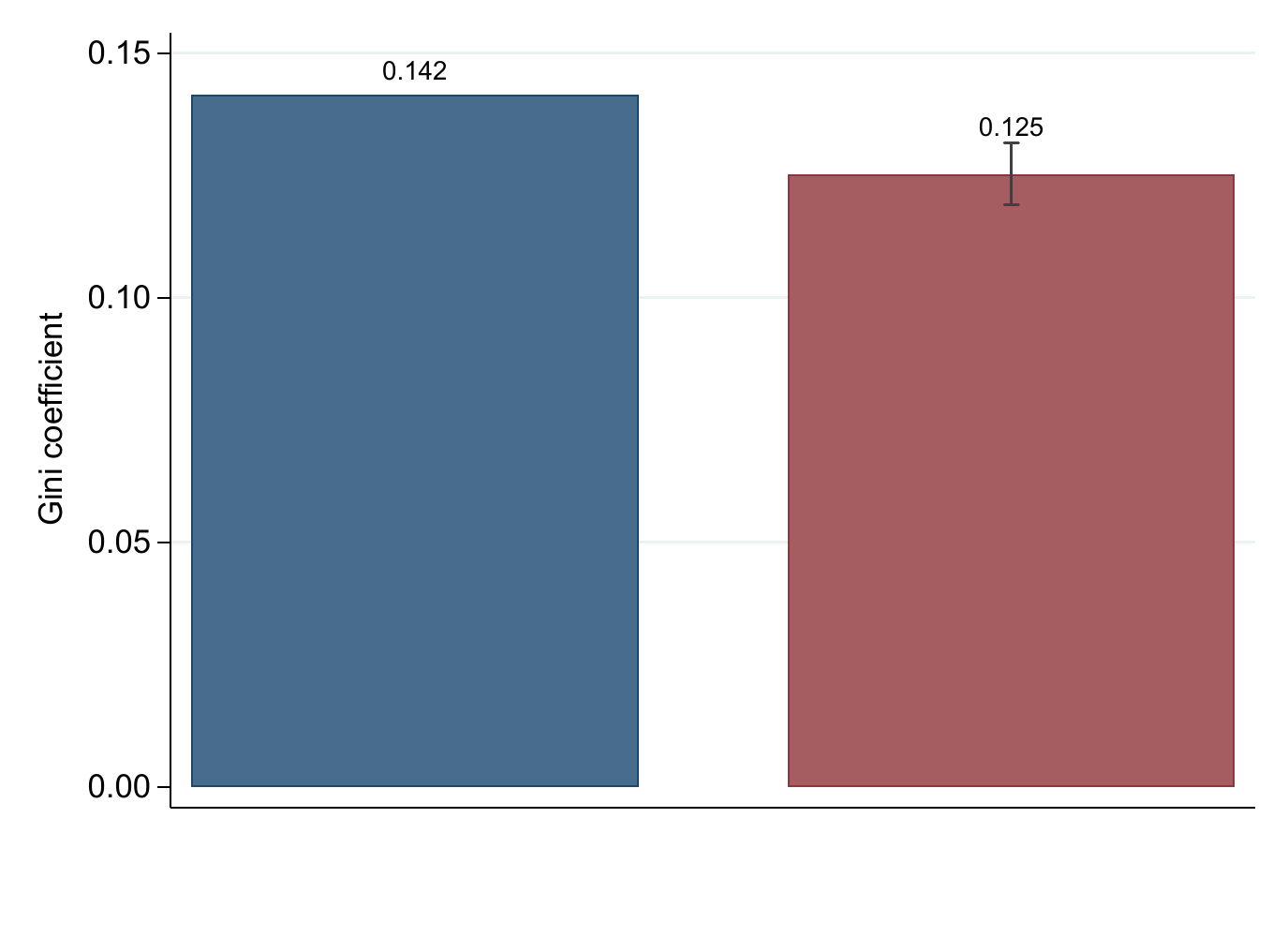}
	\end{subfigure}
	\hfill		
	\begin{subfigure}[t]{0.48\textwidth}
		\caption*{Panel B. Outcome: Percentiles ratios}
		\centering
				\includegraphics[width=\linewidth]{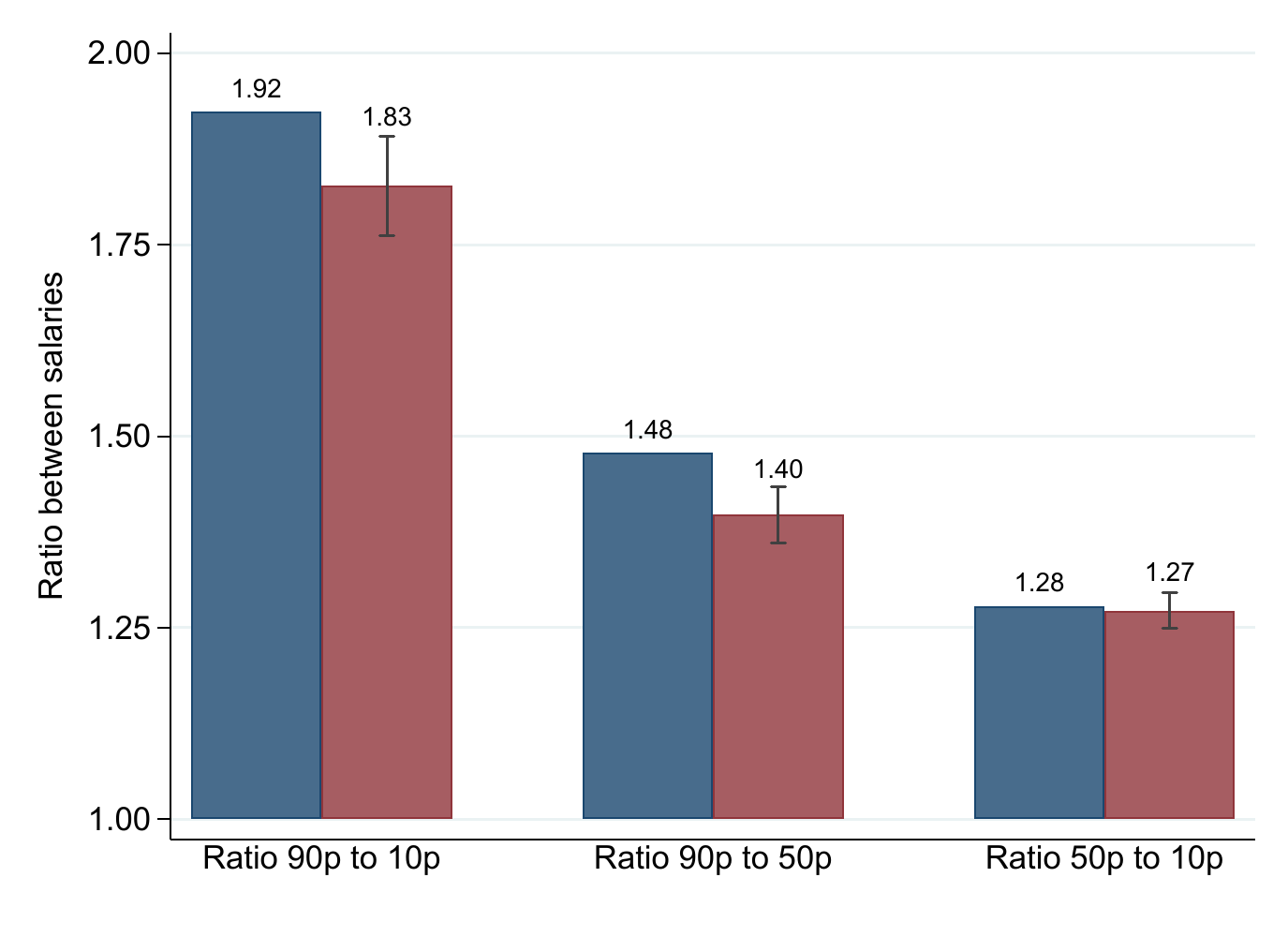}
	\end{subfigure}
	\centering	
	\begin{subfigure}[t]{0.48\textwidth}
		\centering	
		\caption*{Panel C. Outcome: Initial wage remained constant in nominal terms ratios}
		\centering
				\includegraphics[width=\linewidth]{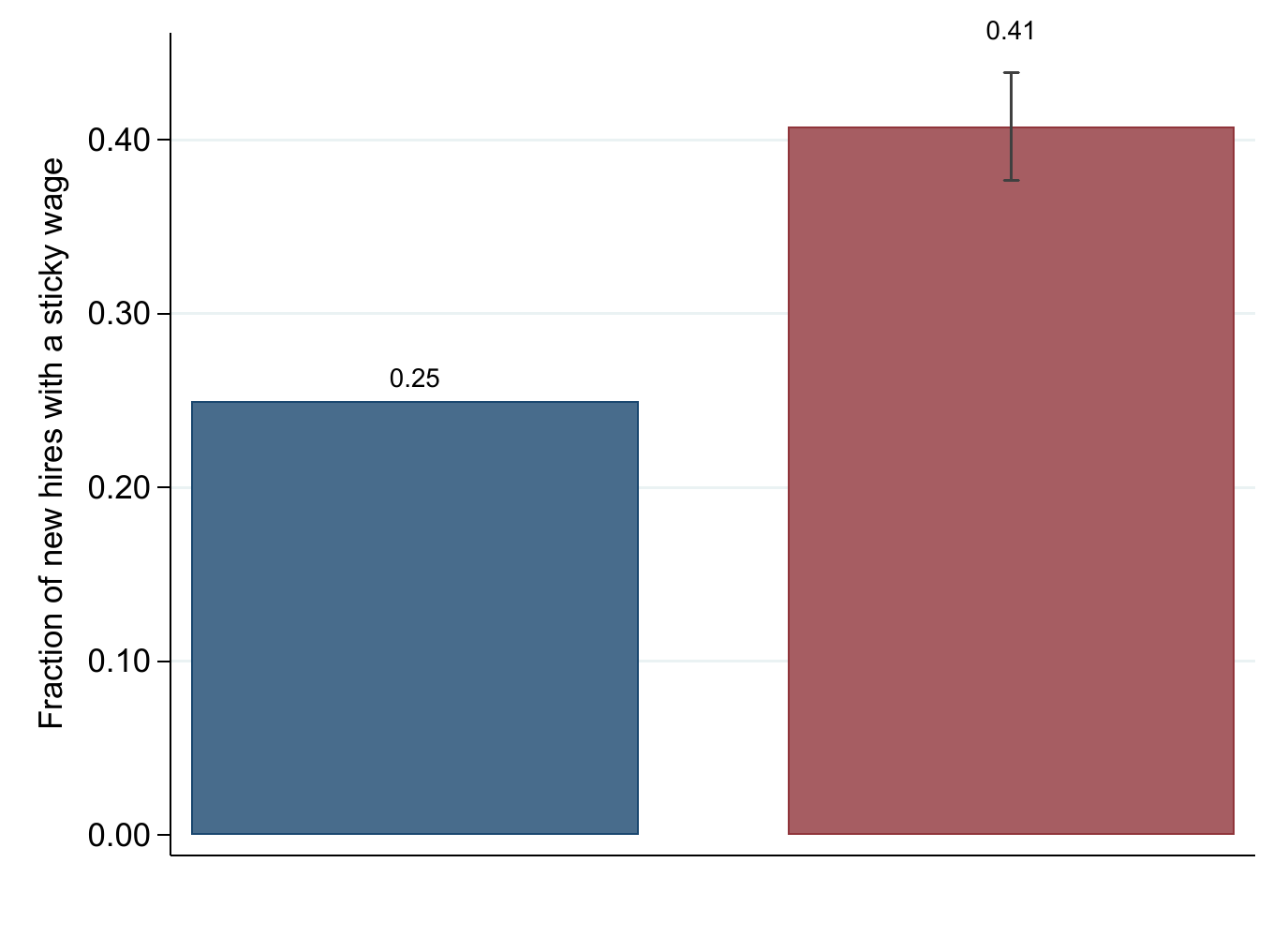}
	\end{subfigure}
	
	\vspace{-.3cm}
	
	\begin{subfigure}[t]{\textwidth}		
		\centering
				\includegraphics[width=0.75\linewidth]{results/fig_wage_comp_leg}
	\end{subfigure}
	\hfill \vspace{-.5cm}	
	{\footnotesize
		\singlespacing \justify
		
		\textit{Notes:} This figure is analogous to Figure \ref{fig_wage_comp}, but the estimates are conditional on firms who, on average across all years in the sample, employ more than five workers. See the notes to Figure \ref{fig_wage_comp} for details on how the figure is constructed, the set of control variables, the definition of the dependent variables, and sample restrictions.
		
	}
\end{figure}

\clearpage
\begin{table}[H]{\footnotesize
		\begin{center}
			\caption{The characteristics of bunching firms} \label{reg_firm_char}
			\newcommand\w{1.55}
			\begin{tabular}{l@{}lR{\w cm}@{}L{0cm}R{\w cm}@{}L{0cm}R{\w cm}@{}L{0.5cm}R{\w cm}@{}L{0cm}R{\w cm}@{}L{0cm}R{\w cm}@{}L{0.5cm}}
				\midrule
				&& \multicolumn{6}{c}{All firms} & \multicolumn{6}{c}{Large firms} \\ \cmidrule{3-7} \cmidrule{9-14}
				
				&& Bunching  && Non-      && Difference  && Bunching  && Non-         && Difference \\
				&& firms     && bunching  &&             && firms     && bunching      &&              \\
				&& (1) && (2)            && (3) && (4) && (5) && (6)  \\
				\midrule
				
				Hiring experience (logs)&&1.248&&2.482&&$-$1.234&\sym{***}&2.461&&3.926&&$-$1.465&\sym{***}\\
&&&&&&(0.004&)&&&&&(0.024&)\\
 
				Firm size (logs)&&0.220&&1.153&&$-$0.933&\sym{***}&2.228&&2.616&&$-$0.387&\sym{***}\\
&&&&&&(0.004&)&&&&&(0.013&)\\
 
				Firm age (years)&&2.283&&4.981&&$-$2.698&\sym{***}&4.328&&7.896&&$-$3.568&\sym{***}\\
&&&&&&(0.020&)&&&&&(0.145&)\\
 
				Has an HR department&&0.023&&0.072&&$-$0.049&\sym{***}&0.089&&0.186&&$-$0.097&\sym{***}\\
&&&&&&(0.001&)&&&&&(0.007&)\\
 
				Education manager&&6.568&&6.626&&$-$0.057&\sym{***}&6.595&&6.821&&$-$0.225&\sym{***}\\
&&&&&&(0.007&)&&&&&(0.037&)\\
 
				Average salary (logs)&&6.451&&6.523&&$-$0.072&\sym{***}&6.569&&6.688&&$-$0.119&\sym{***}\\
&&&&&&(0.002&)&&&&&(0.013&)\\
 
				\midrule
			\end{tabular}
		\end{center}
		\begin{singlespace}  \vspace{-.5cm}
			\noindent \justify \textit{Notes:} This table shows average firm characteristics of bunching firms and non-bunching firms. Bunching firms are defined as firms that hired all new hires at a round-numbered salary in the sample. Large firms are those who employ, on average across all years, more than five workers in the sample. Heteroskedasticity-robust standard errors clustered at the firm level in parentheses. $^{***}$, $^{**}$ and $^*$ denote significance at the 1\%, 5\% and 10\% levels.
			
		\end{singlespace} 	
	}
\end{table}

\clearpage 
\begin{table}[H]{\footnotesize
		\begin{center}
			\caption{Robustness of market outcomes regressions to excluding small firms} \label{reg_firm_performance_size}
			\newcommand\w{2.5}
			\begin{tabular}{l@{}lR{\w cm}@{}L{0.45cm}R{\w cm}@{}L{0.45cm}R{\w cm}@{}L{0.45cm}R{\w cm}@{}L{0.45cm}R{\w cm}@{}L{0.45cm}}
				\midrule
				&& New hire  && New hire  && Firm job && Firm left  \\
				&& separated && resigned  && growth rate && market \\
				&& (1) && (2) && (3) && (4) \\
				\midrule
				Bunching firm&&0.057&\sym{***}&0.020&\sym{***}&$-$0.029&\sym{***}&0.012&\sym{***}\\
&&(0.012&)&(0.006&)&(0.011&)&(0.003&)\\
Mean Dep. Var.&&0.344&&0.116&&0.083&&0.030&\\
N&&17,074,154&&17,074,154&&485,483&&559,667&\\
 \midrule
				
			\end{tabular}
		\end{center}
		\begin{singlespace}  \vspace{-.5cm}
			\noindent \justify \textit{Notes:} This table displays estimates of $\beta$ from equation \eqref{reg:firm-outcomes}, estimated on the subset of firms that employ, on average across years, more than five workers. See notes to Table \ref{reg_firm_performance} for the list of controls and variable definitions. Heteroskedasticity-robust standard errors clustered at the firm level in parentheses. $^{***}$, $^{**}$ and $^*$ denote significance at the 1\%, 5\% and 10\% levels.
			
		\end{singlespace} 	
	}
\end{table}

\clearpage 
\begin{table}[H]{\footnotesize
		\begin{center}
			\caption{Robustness of market outcomes regressions to controlling for wage level} \label{reg_firm_performance_lev}
			\newcommand\w{2.5}
			\begin{tabular}{l@{}lR{\w cm}@{}L{0.45cm}R{\w cm}@{}L{0.45cm}R{\w cm}@{}L{0.45cm}R{\w cm}@{}L{0.45cm}R{\w cm}@{}L{0.45cm}}
				\midrule
				&& New hire  && New hire  && Firm job && Firm left  \\
				&& separated && resigned  && growth rate && market \\
				&& (1) && (2) && (3) && (4) \\
				\midrule
								Bunching firm&&0.047&\sym{***}&0.015&\sym{***}&$-$0.021&\sym{***}&0.014&\sym{***}\\
&&(0.003&)&(0.001&)&(0.006&)&(0.001&)\\
Mean Dep. Var.&&0.353&&0.115&&0.117&&0.024&\\
N&&20,281,068&&20,281,068&&28,409,956&&31,641,563&\\
  \midrule				
			\end{tabular}
		\end{center}
		\begin{singlespace}  \vspace{-.5cm}
			\noindent \justify \textit{Notes:} This table displays estimates of $\beta$ from equation \eqref{reg:firm-outcomes}. All specifications are estimated at the worker level and include the baseline worker-level controls described in the main text. Additionally, the specifications in this table control for the wage level by including wage fixed effects (in R\$100 bins). See notes to Table \ref{reg_firm_performance} for the list of controls and variable definitions. Heteroskedasticity-robust standard errors clustered at the firm level in parentheses. $^{***}$, $^{**}$ and $^*$ denote significance at the 1\%, 5\% and 10\% levels.

		\end{singlespace} 	
	}
\end{table}

\clearpage
\begin{table}[H]{\footnotesize
		\begin{center}
			\caption{Robustness of market outcomes regressions to alternative definitions of bunching firms} \label{reg_firm_performance_def}
			\newcommand\w{2.5}
			\begin{tabular}{l@{}lR{\w cm}@{}L{0.45cm}R{\w cm}@{}L{0.45cm}R{\w cm}@{}L{0.45cm}R{\w cm}@{}L{0.45cm}R{\w cm}@{}L{0.45cm}}
				\midrule
				&& New hire  && New hire  && Firm job && Firm left  \\
				&& separated && resigned  && growth rate && market \\
				&& (1) && (2) && (3) && (4) \\
				\midrule
				\multicolumn{10}{l}{\hspace{-1em} \textbf{Panel A. Bunching firm equals one if firm hired all workers at a round number (baseline)}}  \\
				Bunching firm&&0.041&\sym{***}&0.012&\sym{***}&$-$0.034&\sym{***}&0.011&\sym{***}\\
&&(0.003&)&(0.001&)&(0.002&)&(0.001&)\\
Mean Dep. Var.&&0.353&&0.115&&0.016&&0.101&\\
N&&20,281,312&&20,281,312&&1,381,246&&1,752,411&\\
  \midrule
				
				\multicolumn{10}{l}{\hspace{-1em} \textbf{Panel B.  Bunching firm equals one if firm hired all workers at salaries divisible by 100}}  \\
				Bunching firm&&0.043&\sym{***}&0.015&\sym{***}&$-$0.016&\sym{***}&0.000&\\
&&(0.004&)&(0.002&)&(0.004&)&(0.002&)\\
Mean Dep. Var.&&0.353&&0.115&&0.016&&0.101&\\
N&&20,281,312&&20,281,312&&1,381,246&&1,752,411&\\
 \midrule
				
				\multicolumn{10}{l}{\hspace{-1em} \textbf{Panel C. Bunching firm equals one if firm hired over 1/2 workers at a round number}}  \\				
				Bunching firm&&0.012&\sym{***}&0.003&\sym{***}&$-$0.008&\sym{***}&0.004&\sym{***}\\
&&(0.001&)&(0.001&)&(0.001&)&(0.001&)\\
Mean Dep. Var.&&0.353&&0.115&&0.016&&0.101&\\
N&&20,281,312&&20,281,312&&1,381,246&&1,752,411&\\
  \midrule
				
				\multicolumn{10}{l}{\hspace{-1em} \textbf{Panel D. Bunching firm equals one if firm hired over 2/3 workers at a round number}}  \\
				Bunching firm&&0.017&\sym{***}&0.005&\sym{***}&$-$0.015&\sym{***}&0.007&\sym{***}\\
&&(0.001&)&(0.001&)&(0.001&)&(0.001&)\\
Mean Dep. Var.&&0.353&&0.115&&0.016&&0.101&\\
N&&20,281,312&&20,281,312&&1,381,246&&1,752,411&\\
  \midrule
				
				\multicolumn{10}{l}{\hspace{-1em} \textbf{Panel E. Bunching firm dummy defined using \textit{yearly} salaries}}  \\
				Bunching firm&&0.041&\sym{***}&0.011&\sym{***}&$-$0.039&\sym{***}&0.014&\sym{***}\\
&&(0.002&)&(0.001&)&(0.002&)&(0.001&)\\
Mean Dep. Var.&&0.353&&0.115&&0.016&&0.101&\\
N&&20,281,312&&20,281,312&&1,381,246&&1,752,411&\\
  \midrule

			\end{tabular}
		\end{center}
		\begin{singlespace}  \vspace{-.5cm}
			\noindent \justify \textit{Notes:} This table displays estimates of $\beta$ from equation \eqref{reg:firm-outcomes}, using several alternative definitions of bunching firms. See notes to Table \ref{reg_firm_performance} for the list of controls and variable definitions. Heteroskedasticity-robust standard errors clustered at the firm level in parentheses. $^{***}$, $^{**}$ and $^*$ denote significance at the 1\%, 5\% and 10\% levels.
			
		\end{singlespace} 	
	}
\end{table}

\clearpage 
\begin{table}[H]{\footnotesize
		\begin{center}
			\caption{Robustness of coarse pay-setting across decision environments to excluding small firms} \label{reg_salary_increase_size}
			\newcommand\w{1.45}
			\begin{tabular}{l@{}lR{\w cm}@{}L{0.45cm}R{\w cm}@{}L{0.45cm}R{\w cm}@{}L{0.45cm}R{\w cm}@{}L{0.45cm}R{\w cm}@{}L{0.45cm}R{\w cm}@{}L{0.45cm}}
				\midrule
				&& \multicolumn{12}{c}{Dependent variable:} \\ \cmidrule{3-13}
				&& \multicolumn{3}{c}{Salary increase in R\$}  && \multicolumn{3}{c}{Salary increase in \%} && \multicolumn{3}{c}{Either a round} \\
				&& \multicolumn{3}{c}{is a round number}  && \multicolumn{3}{c}{is an integer} && \multicolumn{3}{c}{number or an integer} & \\
				\cmidrule{3-5} \cmidrule{7-9} \cmidrule{11-13}
				&& (1) && (2) && (3) && (4) && (5) && (6) \\
				\midrule
				Bunching firm&&0.391&\sym{***}&0.438&\sym{***}&0.179&\sym{***}&0.070&\sym{***}&0.362&\sym{***}&0.416&\sym{***}\\
&&(0.015&)&(0.022&)&(0.016&)&(0.012&)&(0.014&)&(0.021&)\\
Mean Dep. Var.&&0.336&&0.115&&0.333&&0.111&&0.406&&0.208&\\
N&&4,155,671&&3,112,904&&4,155,671&&3,112,904&&4,155,671&&3,112,904&\\
 
				Excl. zero growth && No && Yes && No && Yes && No && Yes \\ \midrule 
			\end{tabular}
		\end{center}
		\begin{singlespace}  \vspace{-.5cm}
			\noindent \justify \textit{Notes:} This table displays estimates of $\beta$ from equation \eqref{reg:firm-outcomes}, estimated on the subset of firms that employ, on average across years, more than five workers. Even columns exclude new hires whose salary did not change in nominal terms. See notes to Table \ref{reg_salary_increase} for the list of controls and variable definitions. Heteroskedasticity-robust standard errors clustered at the firm level in parentheses. $^{***}$, $^{**}$ and $^*$ denote significance at the 1\%, 5\% and 10\% levels.

		\end{singlespace} 	
	}
\end{table}

\clearpage 
\begin{table}[H]{\footnotesize
		\begin{center}
			\caption{Robustness of coarse pay-setting across decision environments to controlling for wage level} \label{reg_salary_increase_lev}
			\newcommand\w{1.45}
			\begin{tabular}{l@{}lR{\w cm}@{}L{0.45cm}R{\w cm}@{}L{0.45cm}R{\w cm}@{}L{0.45cm}R{\w cm}@{}L{0.45cm}R{\w cm}@{}L{0.45cm}R{\w cm}@{}L{0.45cm}}
				\midrule
				&& \multicolumn{3}{c}{Salary increase in R\$}  && \multicolumn{3}{c}{Salary increase in \%} && \multicolumn{3}{c}{Either a round} \\
				&& \multicolumn{3}{c}{is a round number}  && \multicolumn{3}{c}{is an integer} && \multicolumn{3}{c}{number or an integer} & \\
				\cmidrule{3-5} \cmidrule{7-9} \cmidrule{11-13}
				&& (1) && (2) && (3) && (4) && (5) && (6) \\
				\midrule
				Bunching firm&&0.297&\sym{***}&0.320&\sym{***}&0.166&\sym{***}&0.079&\sym{***}&0.289&\sym{***}&0.329&\sym{***}\\
&&(0.003&)&(0.004&)&(0.003&)&(0.003&)&(0.003&)&(0.004&)\\
Mean Dep. Var.&&0.354&&0.123&&0.344&&0.110&&0.421&&0.214&\\
N&&4,953,625&&3,646,329&&4,953,625&&3,646,329&&4,953,625&&3,646,329&\\
 			
				Excl. zero growth && No && Yes && No && Yes && No && Yes \\ \midrule 			
			\end{tabular}
		\end{center}
		\begin{singlespace}  \vspace{-.5cm}
			\noindent \justify \textit{Notes:} This table displays estimates of $\beta$ from equation \eqref{reg:firm-outcomes}. In addition to the baseline controls, the specifications in this table control for the wage level by including wage fixed effects (in R\$100 bins). Even columns exclude new hires whose salary did not change in nominal terms. See notes to Table \ref{reg_salary_increase} for the list of controls and variable definitions. Heteroskedasticity-robust standard errors clustered at the firm level in parentheses. $^{***}$, $^{**}$ and $^*$ denote significance at the 1\%, 5\% and 10\% levels.
			
		\end{singlespace} 	
	}
\end{table}

\clearpage
\begin{table}[H]{\footnotesize
		\begin{center}
			\caption{Robustness of coarse pay-setting across decision environments to alternative definitions of bunching firms} \label{reg_salary_increase_def}
			\newcommand\w{1.5}
			\begin{tabular}{l@{}lR{\w cm}@{}L{0.45cm}R{\w cm}@{}L{0.45cm}R{\w cm}@{}L{0.45cm}R{\w cm}@{}L{0.45cm}R{\w cm}@{}L{0.45cm}R{\w cm}@{}L{0.45cm}}
				\midrule
				&& \multicolumn{3}{c}{Salary increase in R\$}  && \multicolumn{3}{c}{Salary increase in \%} && \multicolumn{3}{c}{Either a round} \\
				&& \multicolumn{3}{c}{is a round number}  && \multicolumn{3}{c}{is an integer} && \multicolumn{3}{c}{number or an integer} & \\
				\cmidrule{3-5} \cmidrule{7-9} \cmidrule{11-13}
				&& (1) && (2) && (3) && (4) && (5) && (6) \\
				\midrule
				\multicolumn{14}{l}{\hspace{-1em} \textbf{Panel A. Bunching firm equals one if firm hired all workers at a round number (baseline)}}  \\
				Bunching firm&&0.293&\sym{***}&0.322&\sym{***}&0.162&\sym{***}&0.082&\sym{***}&0.286&\sym{***}&0.331&\sym{***}\\
&&(0.003&)&(0.004&)&(0.003&)&(0.003&)&(0.003&)&(0.004&)\\
Mean Dep. Var.&&0.354&&0.123&&0.344&&0.110&&0.421&&0.214&\\
N&&4,953,782&&3,646,479&&4,953,782&&3,646,479&&4,953,782&&3,646,479&\\
  \midrule
				
				\multicolumn{14}{l}{\hspace{-1em} \textbf{Panel B.  Bunching firm equals one if firm hired all workers at salaries divisible by 100}}  \\
				Bunching firm&&0.277&\sym{***}&0.246&\sym{***}&0.249&\sym{***}&0.174&\sym{***}&0.274&\sym{***}&0.274&\sym{***}\\
&&(0.005&)&(0.008&)&(0.005&)&(0.006&)&(0.005&)&(0.008&)\\
Mean Dep. Var.&&0.354&&0.123&&0.344&&0.110&&0.421&&0.214&\\
N&&4,953,782&&3,646,479&&4,953,782&&3,646,479&&4,953,782&&3,646,479&\\
 \midrule
				
				\multicolumn{14}{l}{\hspace{-1em} \textbf{Panel C. Bunching firm equals one if firm hired over 1/2 workers at a round number}}  \\
				Bunching firm&&0.178&\sym{***}&0.209&\sym{***}&0.066&\sym{***}&0.045&\sym{***}&0.177&\sym{***}&0.212&\sym{***}\\
&&(0.002&)&(0.002&)&(0.002&)&(0.001&)&(0.002&)&(0.002&)\\
Mean Dep. Var.&&0.354&&0.123&&0.344&&0.110&&0.421&&0.214&\\
N&&4,953,782&&3,646,479&&4,953,782&&3,646,479&&4,953,782&&3,646,479&\\
  \midrule
				
				\multicolumn{14}{l}{\hspace{-1em} \textbf{Panel D. Bunching firm equals one if firm hired over 2/3 workers at a round number}}  \\
				Bunching firm&&0.223&\sym{***}&0.262&\sym{***}&0.085&\sym{***}&0.052&\sym{***}&0.217&\sym{***}&0.260&\sym{***}\\
&&(0.002&)&(0.002&)&(0.002&)&(0.001&)&(0.002&)&(0.002&)\\
Mean Dep. Var.&&0.354&&0.123&&0.344&&0.110&&0.421&&0.214&\\
N&&4,953,782&&3,646,479&&4,953,782&&3,646,479&&4,953,782&&3,646,479&\\
  \midrule
				
				\multicolumn{14}{l}{\hspace{-1em} \textbf{Panel E. Bunching firm dummy defined using \textit{yearly} salaries}}  \\
				Bunching firm&&0.239&\sym{***}&0.258&\sym{***}&0.118&\sym{***}&0.050&\sym{***}&0.231&\sym{***}&0.259&\sym{***}\\
&&(0.003&)&(0.003&)&(0.003&)&(0.002&)&(0.003&)&(0.004&)\\
Mean Dep. Var.&&0.354&&0.123&&0.344&&0.110&&0.421&&0.214&\\
N&&4,953,782&&3,646,479&&4,953,782&&3,646,479&&4,953,782&&3,646,479&\\
  \midrule
				
				Excl. zero growth && No && Yes && No && Yes && No && Yes \\ \midrule
			\end{tabular}
		\end{center}
		\begin{singlespace} \vspace{-.5cm}
			\noindent \justify \textit{Notes:} This table displays estimates of $\beta$ from equation \eqref{reg:firm-outcomes}, using alternative definitions of bunching firms. Even columns exclude new hires whose salary did not change in nominal terms. See notes to Table \ref{reg_salary_increase} for the list of controls and variable definitions. Heteroskedasticity-robust standard errors clustered at the firm level in parentheses. $^{***}$, $^{**}$ and $^*$ denote significance at the 1\%, 5\% and 10\% levels.
			
		\end{singlespace}
		
	}
\end{table}

\clearpage
\begin{landscape}
	\begin{table}[H]{\footnotesize
		\begin{center}
			\caption{\centering Testing the predictions of the model using alternative measures of coarse salaries} \label{tab_predictions_rob}
			
			\newcommand\w{2}
			\begin{tabular}{l@{}lR{\w cm}@{}L{0.45cm}R{\w cm}@{}L{0.45cm}R{\w cm}@{}L{0.45cm}R{\w cm}@{}L{0.45cm}R{\w cm}@{}L{0.45cm}R{\w cm}@{}L{0.45cm}}
				\midrule
				& \multicolumn{12}{c}{Dependent variable} \\ \cmidrule{3-13} 
				& \multicolumn{6}{c}{Fraction of workers hired} && \multicolumn{6}{c}{Dummy for hiring a worker} \\
				& \multicolumn{6}{c}{through coarse wage-setting $(\hat{\theta} )$} && \multicolumn{6}{c}{at a round number ($\mathbbm{1}\{w_{i} \in R\}$)} \\
				\cmidrule{3-7} \cmidrule{9-14}
								
				&& (1) && (2) && (3) && (4) && (5) && (6) \\ \addlinespace
				
				Consumer price index (logs)&&0.108&\sym{***}&0.329&\sym{***}&0.640&\sym{***}&0.036&\sym{***}&0.043&\sym{***}&0.017&\sym{***}\\
&&(0.020&)&(0.022&)&(0.016&)&(0.001&)&(0.001&)&(0.001&)\\

				Firm size (logs)&&$-$0.972&\sym{***}&$-$0.973&\sym{***}&$-$0.943&\sym{***}&$-$0.012&\sym{***}&$-$0.004&\sym{***}&$-$0.005&\sym{***}\\
&&(0.034&)&(0.021&)&(0.034&)&(0.001&)&(0.001&)&(0.001&)\\

				Firm hiring experience (logs)&&$-$0.882&\sym{***}&$-$0.876&\sym{***}&$-$0.642&\sym{***}&$-$0.025&\sym{***}&$-$0.023&\sym{***}&$-$0.007&\sym{***}\\
&&(0.050&)&(0.042&)&(0.071&)&(0.001&)&(0.001&)&(0.001&)\\
				
				\midrule
				
				Measure of coarse salary: && Div. by 10 (baseline) && Div. by 100 \textcolor{white}{~~~~~~~~} && Div. by 1000 \textcolor{white}{~~~~~~~~} && Div. by 10 (baseline) && Div. by 100 \textcolor{white}{~~~~~~~~} && Div. by 1,000 \textcolor{white}{~~~~~~~~}\\ \addlinespace

			\end{tabular}
		\end{center}
		\begin{singlespace} \vspace{-.5cm}
			\noindent \justify \textit{Notes:} This table shows linear correlations between the covariate listed in the row header and the outcome listed in the column header using different measures of coarse salaries. Columns 1 and 4 show the results of the baseline specification, using salaries divisible by 10. Columns 2 and 5 use salaries divisible by 100. Columns 3 and 6 use salaries divisible by 1000.
			
			See notes to Table \ref{tab_predictions} for variable definitions and additional details. Heteroskedasticity-robust standard errors clustered at the firm level in parentheses. $^{***}$, $^{**}$ and $^*$ denote significance at the 1\%, 5\% and 10\% levels.
			
		\end{singlespace} 	
	}
\end{table}
\end{landscape}

\clearpage
\begin{table}[H]{\footnotesize
		\begin{center}
			\caption{Wage compression among new hires of bunching firms} \label{reg_wage_compression}
			\newcommand\w{2}
			\begin{tabular}{l@{}lR{\w cm}@{}L{0.45cm}R{\w cm}@{}L{0.45cm}R{\w cm}@{}L{0.45cm}R{\w cm}@{}L{0.45cm}}
				\midrule
				&& && \multicolumn{6}{c}{Ratio of initial salary percentiles:}  \\
				\cmidrule{5-10}    
				&& Gini && 90th to 10th && 90th to 50th && 50th to 10th  \\
				&& (1) && (2) && (3) && (4) \\
				\midrule
				\multicolumn{6}{l}{\hspace{-1em} \textbf{Panel A. All firms}}  \\
				Bunching firm&&$-$0.010&\sym{***}&$-$0.071&\sym{***}&$-$0.043&\sym{***}&$-$0.018&\sym{***}\\
&&(0.001&)&(0.006&)&(0.004&)&(0.002&)\\
Mean Dep. Var.&&0.109&&1.739&&1.380&&1.236&\\
N&&1,622,410&&1,622,410&&1,622,410&&1,622,410&\\
 \midrule
				\multicolumn{6}{l}{\hspace{-1em} \textbf{Panel B. Firms with more than five workers}} \\
				Bunching firm&&$-$0.016&\sym{***}&$-$0.097&\sym{***}&$-$0.081&\sym{***}&$-$0.006&\\
&&(0.003&)&(0.033&)&(0.019&)&(0.012&)\\
Mean Dep. Var.&&0.141&&1.923&&1.478&&1.278&\\
N&&557,112&&557,112&&557,112&&557,112&\\
  \midrule
			\end{tabular}
		\end{center}
		\begin{singlespace} \vspace{-.5cm}
			\noindent \justify \textit{Notes:} This table displays estimates of $\beta$ from equation \eqref{reg:firm-outcomes}. Each column shows the results using a different dependent variable. In column 1, the dependent variable is the Gini coefficient. In column 2, the ratio between the 90th and 10th percentiles of the contracted salary distribution among all the new hires in each firm. In column 3, the ratio between the 90th and the 50th percentiles. In column 4, the ratio between the 50th and 10th percentiles. 
			
			I use the firm random sample to estimate all regressions. The regressions are estimated at the firm-by-year level on firms that hired at least two workers in my sample.
			
			The regressions control for firm age, share of employees with completed high school, share of employees with completed college, educational attainment of the firm manager, a dummy for having an HR department, the mean earnings of the firm employees, firm size fixed effects, number of hires fixed effects, and industry-by-microregion-by-year fixed effects. 
			
			Heteroskedasticity-robust standard errors clustered at the firm level in parentheses. $^{***}$, $^{**}$ and $^*$ denote significance at the 1\%, 5\% and 10\% levels.
			
		\end{singlespace}
	}
\end{table}

	\clearpage 
\section{Empirical Appendix} \label{app:add-results} 

\setcounter{table}{0}
\setcounter{figure}{0}
\setcounter{equation}{0}	
\renewcommand{\thetable}{B\arabic{table}}
\renewcommand{\thefigure}{B\arabic{figure}}
\renewcommand{\theequation}{B\arabic{equation}}

\subsection{Informality in Brazilian Labor Markets} \label{app:inform}

International organizations define informality in two main ways. Under the \textit{legal} definition, a worker is considered to be employed by the informal sector if she does not have the right to a pension when retired. Under the \textit{productive} definition, a worker is considered informal if (i) she is a salaried worker in a small firm (i.e., a firm that employs fewer than five workers), (ii) a non-professional self-employed, or (iii) a zero-income worker. The share of informal-sector workers in Brazil during 2013 was 35.9\% under the legal definition and 43.7\% according to the productive definition. Table \ref{tab_rais_pnad} shows summary statistics on workers in the national household survey (PNAD), which includes information on workers employed in the informal sector.

\begin{table}[H]
	\centering
	\caption{Summary statistics of workers in the RAIS and the PNAD during 2013} \label{tab_rais_pnad}
	\resizebox{1\textwidth}{!}{
		\begin{tabular}{lccccccccc}
			\midrule
			& RAIS &   & \multicolumn{7}{c}{Household Survey (PNAD)} \\
			\cmidrule{2-2}\cmidrule{4-10}      & \multicolumn{1}{c}{\multirow{2}[4]{*}{All workers}} &   & \multicolumn{1}{c}{\multirow{2}[4]{*}{All workers}} &   & \multicolumn{2}{c}{Legal definition} &   & \multicolumn{2}{c}{Productive definition} \\
			\cmidrule{6-7}\cmidrule{9-10}      &   &   &   &   & Formal & Informal &   & Formal & Informal \\
			\cmidrule{2-10}      & (1) &   & (2) &   & (3) & (4) &   & (5) & (6) \\
			\midrule
			
			\textbf{\hspace{-1em} Panel A. Workers' characteristics} &   &   &   &   &   &   &   &   &  \\
			Average age &34.916 & &38.169 & &37.548 &39.277 & &36.673 &40.096 \\
Male &0.573 & &0.568 & &0.563 &0.576 & &0.580 &0.552 \\
White &0.581 & &0.469 & &0.523 &0.372 & &0.525 &0.396 \\
Elementary school or less &0.336 & &0.481 & &0.362 &0.693 & &0.313 &0.698 \\
Complete high school &0.508 & &0.390 & &0.463 &0.261 & &0.472 &0.285 \\
Complete university & 0.156 & &0.129 & &0.176 &0.046 & &0.215 &0.017 \\
  \midrule

			\textbf{\hspace{-1em} Panel B. Earnings} &   &   &   &   &   &   &   &   &  \\
			Mean monthly salary (Reals) &1367.062 & &1350.902 & &1416.779 &1125.241 & &1534.599 &963.886 \\
Median monthly salary (Reals) &803.923 & &833.073 & &833.073 &694.228 & &902.496 &694.228 \\
  \midrule

			\textbf{\hspace{-1em} Panel C. Industry} &   &   &   &   &   &   &   &   &  \\
			Primary sector &0.041 & &0.127 & &0.050 &0.267 & &0.014 &0.274 \\
Manufacturing &0.136 & &0.129 & &0.155 &0.082 & &0.176 &0.068 \\
Construction and utilities &0.135 & &0.156 & &0.143 &0.181 & &0.149 &0.166 \\
Retail &0.264 & &0.223 & &0.227 &0.216 & &0.223 &0.223 \\
Services &0.424 & &0.364 & &0.425 &0.255 & &0.437 &0.270 \\
  \midrule

			\textbf{\hspace{-1em} Panel D. Region} &   &   &   &   &   &   &   &   &  \\
			North &0.055 & &0.078 & &0.057 &0.116 & &0.062 &0.099 \\
Northeast &0.176 & &0.252 & &0.183 &0.374 & &0.197 &0.323 \\
Southeast &0.503 & &0.432 & &0.491 &0.328 & &0.487 &0.362 \\
South &0.172 & &0.159 & &0.186 &0.110 & &0.174 &0.140 \\
Midwest &0.093 & &0.079 & &0.082 &0.072 & &0.081 &0.076 \\
  \midrule

			Observations (weighted) &68,589,569 & &87,446,610 & &56,021,547 &31,425,063 & &49,218,392 &38,228,218 \\
Observations (unweighted) &68,589,569 & &156,432 & &98,307 &58,125 & &86,631 &69,801 \\
  \midrule

		\end{tabular}
	}
	\begin{minipage}{\textwidth}  \vspace{0cm}
		\scriptsize\rule{0cm}{0cm} \noindent \textit{Notes:} This table shows summary statistics of workers in the \textit{Relação Anual de Informações Sociais} (RAIS) and the \textit{Pesquisa Nacional por Amostra de Domicílios} (PNAD), both during 2013. I restrict the PNAD sample to employed individuals aged 18--65. This excludes individuals out of the labor force and unemployed.
	\end{minipage}
	
\end{table}

Workers in the RAIS (column 1) are slightly younger, more educated, more likely to live in the Southeast (the wealthiest region), have higher earnings, and are significantly less likely to work in the primary sector than workers in the PNAD (column 2). Workers in the RAIS resemble workers in the formal sector of the PNAD (columns 3 and 5). As noted above, this is because informal-sector workers are not included in the RAIS. 

\subsection{Additional Outcomes of Bunching Firms} \label{app:outcomes}

In this Appendix, I study additional outcomes of firms that only hired workers at round numbers in my sample (``bunching firms''). As in the main text, to analyze worker separation likelihood (Appendix Tables \ref{reg_firm_performance2} and \ref{reg_firm_performance3}), I estimate the regressions at the worker-by-firm-by-year level. To analyze the firm growth (Appendix Tables \ref{reg_firm_performance4}), I estimate the regressions at the firm-by-year level (and exclude the worker controls). I cluster the standard errors at the firm level.

\subsubsection{Worker Separation and Resignation Rates Over Time.}

Appendix Table \ref{reg_firm_performance2} shows the separation and resignation rates of new workers hired by bunching firms over time. Each column shows the separation or resignation rate within $t$ years of a new hire joining a bunching firm. Columns 1 and 4 reproduce the one-year separation and resignation rates shown in the main text. In columns 2 and 5, the outcome equals one if the new hire separated/resigned within two years of joining the firm. Finally, in columns 3 and 6, the outcome equals one if the new hire separated/resigned within three years of joining the firm. 

The higher separation and resignation rates of new workers hired by bunching firms persist over time. Within one year of joining the firm, new hires in bunching firms are 4.1 percentage points and 1.2 percentage points more likely to separate and resign than new hires in non-bunching firms ($p<0.01$). These figures represent increases of 11.6\% and 10.4\%, respectively, relative to the sample mean. The corresponding figures within two and three years of joining the firm are of a similar magnitude. For instance, new hires in bunching firms are 3.0 percentage points more likely to separate within two years and 1.9 percentage points more likely to separate within three years of joining the firm ($p<0.01$). These results are robust to excluding small firms (Panel B).

\begin{table}[H]{\footnotesize
		\begin{center}
			\caption{Over-time worker separation rate of firms that tend to hire workers at round numbers} \label{reg_firm_performance2}
			\newcommand\w{1.63}
			\begin{tabular}{l@{}lR{\w cm}@{}L{0.45cm}R{\w cm}@{}L{0.45cm}R{\w cm}@{}L{0.45cm}R{\w cm}@{}L{0.45cm}R{\w cm}@{}L{0.45cm}R{\w cm}@{}L{0.45cm}}
				\midrule
				&& \multicolumn{12}{c}{Dependent variable: $=1$ if worker separated/resigned within $t$ years of joining the firm} \\ \cmidrule{3-14}
				
				&& \multicolumn{6}{c}{Separation rate} & \multicolumn{6}{c}{Resignation rate} \\ \cmidrule{3-7} \cmidrule{9-13}
				
				&& 1-year && 2-years && 3-years && 1-year && 2-years && 3-years \\
				
				&& (1) && (2) && (3) && (4) && (5) && (6)  \\
				\midrule
				\multicolumn{10}{l}{\hspace{-1em} \textbf{Panel A. All firms}}  \\
				Bunching firm&&0.041&\sym{***}&0.030&\sym{***}&0.019&\sym{***}&0.012&\sym{***}&0.011&\sym{***}&0.009&\sym{***}\\
&&(0.003&)&(0.003&)&(0.003&)&(0.001&)&(0.001&)&(0.002&)\\
Mean Dep. Var.&&0.353&&0.478&&0.554&&0.115&&0.137&&0.148&\\
N&&20,281,312&&20,281,312&&20,281,312&&20,281,312&&20,281,312&&20,281,312&\\
  \midrule
				
				\multicolumn{6}{l}{\hspace{-1em} \textbf{Panel B. Firms with more than five workers}} \\
				Bunching firm&&0.057&\sym{***}&0.045&\sym{***}&0.025&\sym{*}&0.020&\sym{***}&0.017&\sym{***}&0.014&\sym{**}\\
&&(0.012&)&(0.015&)&(0.015&)&(0.006&)&(0.006&)&(0.006&)\\
Mean Dep. Var.&&0.344&&0.466&&0.541&&0.116&&0.138&&0.150&\\
N&&17,074,154&&17,074,154&&17,074,154&&17,074,154&&17,074,154&&17,074,154&\\
  \midrule
			\end{tabular}
		\end{center}
		\begin{singlespace}  \vspace{-.5cm}
			\noindent \justify \textit{Notes:} This table displays estimates of $\beta$ from equation \eqref{reg:firm-outcomes}. Each column shows the result of a regression using the dependent variable listed in the column header. 
			
			In column 1, the outcome equals one if a new hire separated from the firm within one year of being hired, i.e., whether she separated during the year she was hired (year $t$) or the following year (year $t+1$), and zero otherwise. In column 2, the outcome equals one if a new hire separated from the firm within two years of being hired. In column 3, the outcome equals one if a new hire separated from the firm within three years of being hired. Columns 4--6 are defined analogously but using worker resignation likelihood instead of separation likelihood. 
			
			I use the firm random sample to estimate all regressions. See notes to Table \ref{reg_firm_performance} for the list of controls and variable definitions. Heteroskedasticity-robust standard errors clustered at the firm level in parentheses. $^{***}$, $^{**}$ and $^*$ denote significance at the 1\%, 5\% and 10\% levels.
		\end{singlespace} 	
	}
\end{table}

\subsubsection{The Separation and Resignation Rates of High-Skilled Workers.}

Appendix Table \ref{reg_firm_performance3} shows the separation and resignation rates of new workers hired by bunching firms as a function of their educational attainment. I divide workers into those with a completed high-school degree (53.2\% of workers, see Table \ref{tab_rais_summ}), and those without a high-school degree (46.8\% of workers). For conciseness, I refer to these workers as ``high-skilled'' and ``low-skilled,'' respectively. Columns 1 and 3 show the one-year separation/resignation rates of high-skilled workers, while columns 2 and 4 show the corresponding rates for low-skilled workers. The coefficient on the high- and low-skilled workers adds up to the coefficient estimated on the regression for all workers.

The higher separation and resignation rates of new workers hired by bunching firms tend to be driven by high-skilled new hires. Columns 1 and 3 show that high-skilled new hires in bunching firms are, on average, 2.9 percentage points (a 14.6\% increase relative to the sample mean) and 0.7 percentage points (or 7.6\%) more likely to separate and resign, respectively, than new high-skilled hires in non-bunching firms ($p<0.01$). The corresponding figures for low-skilled new hires are 1.2 percentage points and 0.4 percentage points (or 10.1\% and 8.6\%). These results are robust to excluding small firms (Panel B).

\begin{table}[H]{\footnotesize
		\begin{center}
			\caption{Separation rates of new workers hired by bunching firms by worker skill level} \label{reg_firm_performance3}
			\newcommand\w{2}
			\begin{tabular}{l@{}lR{\w cm}@{}L{0.45cm}R{\w cm}@{}L{0.45cm}R{\w cm}@{}L{0.45cm}R{\w cm}@{}L{0.45cm}}
				\midrule
				&& \multicolumn{8}{c}{Dependent variable: =1 if the new hire separated within one year} \\ \cmidrule{3-10}
				&& \multicolumn{4}{c}{Separation rate} & \multicolumn{4}{c}{Resignation rate} \\ \cmidrule{3-5} \cmidrule{7-9}				
				&& High-skilled  && Low-skilled    && High-skilled  && Low-skilled  \\
				&& workers && workers    && workers  && workers \\
				&& (1) && (2) && (3) && (4)  \\
				\midrule
				\multicolumn{10}{l}{\hspace{-1em} \textbf{Panel A. All firms}}  \\
				Bunching firm&&0.029&\sym{***}&0.012&\sym{***}&0.007&\sym{***}&0.004&\sym{***}\\
&&(0.002&)&(0.002&)&(0.001&)&(0.001&)\\
Mean Dep. Var.&&0.198&&0.156&&0.069&&0.046&\\
N&&20,281,312&&20,281,312&&20,281,312&&20,281,312&\\
  \midrule
				
				\multicolumn{6}{l}{\hspace{-1em} \textbf{Panel B. Firms with more than five workers}} \\
				Bunching firm&&0.040&\sym{***}&0.017&\sym{**}&0.012&\sym{***}&0.008&\sym{***}\\
             &&(0.008&)&(0.008&)&(0.004&)&(0.003&)\\
Mean Dep. Var.&&0.189&&0.156&&0.069&&0.046&\\
N&&17,074,154&&17,074,154&&17,074,154&&17,074,154&\\
  \midrule
			\end{tabular}
		\end{center}
		\begin{singlespace}  \vspace{-.5cm}
			\noindent \justify \textit{Notes:} This table displays estimates of $\beta$ from equation \eqref{reg:firm-outcomes}. Each column shows the result of a regression using the dependent variable listed in the column header. 
			
			In column 1, the outcome equals one if a new hire has a high-school degree and separated from the firm within one year of being hired, and zero otherwise. In column 2, the outcome equals one if a new hire does not have a high-school degree and separated from the firm within one year of being hired, and zero otherwise. Columns 3 and 4 are defined analogously but using worker resignation likelihood instead of separation likelihood. 
			
			I use the firm random sample to estimate all regressions. See notes to Table \ref{reg_firm_performance} for the list of controls and variable definitions. Heteroskedasticity-robust standard errors clustered at the firm level in parentheses. $^{***}$, $^{**}$ and $^*$ denote significance at the 1\%, 5\% and 10\% levels.
		\end{singlespace} 	
	}
\end{table}

\subsubsection{The Job Growth Rate of High-Skilled and Highly-Paid Workers.}

Appendix Table \ref{reg_firm_performance4} shows the job growth rate of workers across the educational attainment distribution. As before, I divide workers into those with and without a high school degree. In addition, I divide workers into ``high-paid'' workers---defined as those whose average monthly salary is at or above the median monthly salary of all workers in the sample---and ``low-paid'' workers, analogously defined. Columns 1 and 2 show the growth rate in the firm's number of high- and low-skilled employees. Columns 3 and 4 show the growth rate in the firm's number of high- and low-paid employees. 

Bunching firms have lower job growth rates for both high-skilled and low-skilled workers. Column 1 shows that bunching firms have a 4.0 percentage points lower high-skilled workers growth rate, on average, than non-bunching firms ($p<0.01$). The corresponding figure for low-skilled workers is 3.4 percentage points (column 2). Similarly, bunching firms have a 3.6 and 3.8 percentage points lower job growth rate of high- and low-paid workers, on average, than non-bunching firms (columns 3 and 4). The results are quantitatively smaller and not statistically different from zero for large firms (Panel B), indicating that these lower job growth rates are mainly driven by smaller firms.

\begin{table}[H]{\footnotesize
		\begin{center}
			\caption{Job growth rate of firms that tend to hire workers at round numbers} \label{reg_firm_performance4}
			\newcommand\w{2}
			\begin{tabular}{l@{}lR{\w cm}@{}L{0.45cm}R{\w cm}@{}L{0.45cm}R{\w cm}@{}L{0.45cm}R{\w cm}@{}L{0.45cm}}
				\midrule
				&& \multicolumn{8}{c}{Dependent variable: Firm job growth rate} \\ \cmidrule{3-10}
				&& High-skilled  && Low-skilled    && High-paid    && Low-paid  \\
				&& workers && workers    && workers  && workers \\
				&& (1) && (2) && (3) && (4)  \\
				\midrule
				\multicolumn{10}{l}{\hspace{-1em} \textbf{Panel A. All firms}}  \\
				Bunching firm&&$-$0.040&\sym{***}&$-$0.034&\sym{***}&$-$0.036&\sym{***}&$-$0.038&\sym{***}\\
&&(0.006&)&(0.003&)&(0.004&)&(0.005&)\\
Mean Dep. Var.&&0.032&&0.018&&0.062&&$-$0.044&\\
N&&804,727&&1,130,123&&1,108,057&&857,992&\\
  \midrule
				
				\multicolumn{6}{l}{\hspace{-1em} \textbf{Panel B. Firms with more than five workers}} \\
				Bunching firm&&$-$0.012&&$-$0.021&&$-$0.002&&$-$0.034&\\
&&(0.038&)&(0.025&)&(0.036&)&(0.025&)\\
Mean Dep. Var.&&0.184&&0.166&&0.223&&0.084&\\
N&&402,306&&420,177&&441,043&&390,534&\\
  \midrule
			\end{tabular}
		\end{center}
		\begin{singlespace}  \vspace{-.5cm}
			\noindent \justify \textit{Notes:} This table displays estimates of $\beta$ from equation \eqref{reg:firm-outcomes}. Each column shows the result of a regression using the dependent variable listed in the column header. 
			
			The dependent variable is the percent change in the number of workers employed between consecutive years. Each column shows the growth rate of workers with different observable characteristics. In column 1, I compute the growth rate of workers with a high-school degree; in column 2, without a high-school degree; in column 3, with a monthly salary at or above the median in the sample; and in column 4, with a monthly salary below the median in the sample.
			
			I use the firm random sample to estimate all regressions. See notes to Table \ref{reg_firm_performance} for the list of controls and variable definitions. Heteroskedasticity-robust standard errors clustered at the firm level in parentheses. $^{***}$, $^{**}$ and $^*$ denote significance at the 1\%, 5\% and 10\% levels.
		\end{singlespace} 	
	}
\end{table}
 	
	\clearpage 
\section{Theoretical Appendix} \label{app:theory}

\setcounter{table}{0}
\setcounter{figure}{0}
\setcounter{equation}{0}	
\renewcommand{\thetable}{C\arabic{table}}
\renewcommand{\thefigure}{C\arabic{figure}}
\renewcommand{\theequation}{C\arabic{equation}}

\subsection{Canonical Wage-setting Models in Labor Economics}  \label{app:labor-models}

There are two broad classes of wage-determination models. The first class of models is wage-posting models. In these models, firms choose what wage to post to maximize profit, in which the optimal wage depends on the worker's productivity and the firm's market power, as measured by the elasticity of labor supply. If both worker productivity and firm market power have smooth distributions, then wages should display no bunching. The textbook model of competitive labor markets---in which firms hire workers up to the point that the marginal product of labor equals the market-determined wage---is a special case of wage-posting models. In perfectly competitive models, firms cannot pay a wage below the equilibrium one since no worker would join the firm. Likewise, firms have no incentive to pay a wage above the equilibrium wage. Therefore, in this framework, there is a unique wage determined in equilibrium. Differences in wages across firms and industries might exist due to compensating differentials that arise from job amenity differences. However, as long as these differentials are smoothly distributed across firms, the resulting wage distribution should also be smooth.

The second class of models is wage-bargaining or search-match models. Central to these models are search frictions. The canonical search model is the McCall model \citep{mccall1970economics}. In this model, job offers are characterized by a wage, which is the realization of a random variable distributed according to some exogenous distribution. Since firms offer every possible value in the support of the (exogenous) wage distribution, the resulting distribution of wages is smooth. More generally, in wage-bargaining models, firms match with workers and each match generates a surplus that is divided between the firm and the worker. The amount of surplus workers capture in the form of wages depends on their bargaining power. As long as bargaining power is smoothly distributed across workers, there should not be bunching in the wage distribution.

The following section presents a model that can account for the bunching of wages at round numbers observed in the data.

\subsection{Setup of the Model} \label{app:model}

Consider an economy populated by firms using a linear production technology. Firms face an upward-sloping labor supply curve, $l(w)$. The positive slope of the labor supply means that firms have to increase the wage they offer to increase the probability that a worker will accept the offer. Let $p$ be worker productivity and for now assume that the firm observes $p$. Each time the firm wants to hire a worker, the firm's problem is to choose the wage offer $w$ that maximizes profit
\begin{align} \label{eq_profit}
	\pi = l(w)(p-w).
\end{align}
\subsubsection{Market equilibrium in the frictionless model.} 

Before introducing optimization frictions, consider first the solution of the standard frictionless model. Suppose workers are randomly matched to firms. In an interior solution, the profit-maximizing wage is
\begin{align} \label{eq_opt_w}
	w^* = p \frac{\eta}{1+\eta},
\end{align}	
where $\eta \equiv l'(w^*) \frac{w^*}{l(w^*)}$ is the elasticity of labor supply. Equation \eqref{eq_opt_w} is the standard solution of the frictionless wage-posting model. This equation tells us that the firm pays workers a fraction $\frac{\eta}{1+\eta}$ of their productivity and earns a profit equal to $\pi(w^*) = \frac{p}{1+\eta} l(w^*)$. As $\eta$ increases, workers get compensated for a higher fraction of their productivity. In the limit, as  $\eta \to \infty$, we get the standard solution of competitive markets: firms pay workers their productivity ($w^* = p$) and earn zero profits. For simplicity, I will refer to $w^*$ as the ``fully-optimal wage,'' although it is optimal only insofar there are no optimization costs. 

The shape of the wage distribution in the frictionless model depends on the distribution of market-power-adjusted productivity, $\tilde{p} \equiv  p \frac{\eta}{1+\eta}$, across firms. Let $F_w$ be the cumulative distribution function (CDF) of observed wages and $F_{\tilde{p}}$ the CDF of $\tilde{p}$. Then,
\begin{align}\label{eq_joint_peta}
	F_w(w) = \Pr(w^* \leq w) &= \Pr\Big(p \frac{\eta}{1+\eta} \leq w \Big) = F_{\tilde{p}}(w).
\end{align} 

Equation \eqref{eq_joint_peta} indicates that, if $F_{\tilde{p}}$ is a smooth distribution, then the distribution of observed earnings, $F_w(w)$, is also smooth.

\subsubsection{Introducing optimization frictions.} 

I depart from the standard formulation by modeling coarse wage-setting as a consequence of optimization frictions. I assume that firms' initial estimate of the fully-optimal wage is a coarse round-numbered wage, $w_r$. For example, $w_r$ might be the fully-optimal wage rounded to the nearest 1,000. I also assume that firms can pay an optimization cost to learn the fully-optimal wage $w^*$. While these assumptions should not be viewed as a perfect description of firm behavior---but rather as useful approximations---they are consistent with evidence from numerical cognition research reviewed in Section \ref{sub_psych}.

Departing from the fully-optimal wage is costly. When the firm offers a coarse wage above the fully-optimal wage ($w_r > w^*$), the probability that a worker will accept the job offer is higher than the one under the fully-optimal wage, i.e., $l(w_r) > l(w^*)$. This leads to the firm hiring workers faster than what would take them if they offered the fully-optimal wage and paying them a wage higher than is optimal. Symmetrically, when a firm offers a coarse wage below the fully-optimal one ($w_r < w^*$), the firm will be slow to hire workers and the workers will receive a lower wage than is optimal.

Firms will compute $w^*$ when they believe it is profitable to do so, namely, whenever the profit gain from computing the fully-optimal wage exceeds the optimization cost. The expected profit difference between paying the fully-optimal and the coarse wage is
\begin{align} \label{eq_diff_profit}
	G(\cdot) &\equiv \E[\pi(w^*)] - \E[\pi(w_r)] \notag \\
	&= (p-w^*) l(w^*) - (p-w_r) l(w_r).
\end{align}
where the expectation is taken over the possible realizations of worker productivity. A first-order Taylor approximation of $l(w_r)$ around $w^*$ yields
\begin{align} \label{eq_taylor} 
	l(w_r) \simeq l(w^*) + l'(w^*)(w_r - w^*). 
\end{align}

Plugging \eqref{eq_taylor} back into \eqref{eq_diff_profit} and using the FOC, we can write the gain function as follows 
\begin{align} \label{eq_G_sol}
	G(\cdot) &\simeq  (p-w^*) l(w^*) - (p-w_r)\Big(l(w^*) + \frac{l(w^*)}{p-w^*}(w_r - w^*)\Big) \notag \\
	&= p l(w^*) \frac{\eta^2}{1+\eta} \Big(\frac{w_r - w^*}{w^*}\Big)^2 \notag \\
	&= \pi(w^*) \eta^2 \tilde{w}^2,
\end{align}
where $\tilde{w} \equiv \frac{w_r - w^*}{w^*}$ is the percentage deviation of $w_r$ about $w^*$ or the wedge between the optimal and the round-numbered wage. 

The firm will optimize whenever the profit gain (given by equation \eqref{eq_G_sol}) is greater than the optimization cost. I assume that firms have to forego a fraction $\tau$ of their profits to optimize.\footnote{There are two main approaches to modeling the optimization cost. First, as a fixed cost $c$. In the context of attention to final prices when some taxes are not salient, this is the approach taken by \cite{chetty_salience_2009}. Under a fixed cost of optimizing, firms compute the optimal wage whenever the profit gain (equation \eqref{eq_G_sol}) exceeds $c$. Second, as a fraction $\tau$ of profits. In a context analogous to mine, this is the approach taken by \cite{dube_monopsony_2020}.} Hence, firms fully optimize whenever $\eta^2 \tilde{w}^2 \geq \tau$.

\subsubsection{Heterogeneity in the optimization cost.} 

Suppose that the optimization cost $\tau$ is heterogeneously distributed across firms according to the CDF $F_\tau$. The probability that a firm will offer a coarse wage is
\begin{align} \label{eq_theta}
	\theta &= \Pr\Big(\tau > \eta^2 \tilde{w}^2  \Big) = 1- F_\tau\Big(\eta^2 \tilde{w}^2 \Big).
\end{align}

Using equation \eqref{eq_theta}, one can characterize the distribution of observed wages in the model with frictions. A fraction $\theta$ of workers are hired at a coarse round-numbered wage. The remaining workers are hired by firms that optimize according to the distribution of the fully-optimal wage, $F_{\tilde{p}}$. The CDF of observed wages, $F_w$, is a convex combination of the distribution of the fully-optimal wage, $F_{\tilde{p}}$, and the distribution of the coarse round-numbered wage, $F_{w_r}$, with mixture weight $\theta$:
\begin{align} \label{eq_dist_wage}
	F_w = \theta F_{w_r} + (1-\theta) F_{\tilde{p}}.
\end{align}

Consistent with the data, the distribution of observed wages in the model with frictions exhibits bunching at $w_r$. The size of the bunching is given by the fraction of workers hired through coarse wage-setting, $\theta$. The standard wage-posting model is a special case of the model with optimization frictions, in which $\tau = 0$ (which implies $\theta = 0$).

\subsection{Optimization with Varying Degrees of Precision} \label{app:varying-precision}

The baseline model with frictions assumes that the decision of the firm is binary: the firm either offers a wage equal to $w_r$ or pays an optimization cost and offers the fully-optimal wage, $w^*$. In this subsection, I extend the model to incorporate different degrees of precision in refining the initial estimate of the fully-optimal wage. In the generalized model, the wage distribution exhibits bunching at multiple round numbers. The size of the bunching at each round number reflects the relative marginal benefit and cost of making a better approximation to the fully-optimal salary. 

Without loss of generality, assume that wages can have at most four digits.\footnote{In the new-hires sample, less than one percent of all salaries are equal or greater than R\$10,000 (i.e., have more than four digits).} Suppose, furthermore, that the firm's initial estimate of the fully-optimal wage is such a wage rounded to the coarsest round number. In this case, the fully-optimal wage rounded to the nearest 1,000, $w_{1000}$. By paying $\tau_{100}$, they can learn the second digit of the optimal wage and offer the optimal wage rounded to the nearest 100, $w_{100}$. After learning the second digit, the firm can pay $\tau_{10}$ to learn $w_{10}$, the optimal wage to the nearest ten, and finally, pay $\tau_1$ to learn exactly the fully-optimal wage.\footnote{The optimal wage is a continuous variable, so the firm can continue learning the decimals of the fully-optimal wage following the same logic just described. Salaries with cents are rare in the data, which probably reflects the fact that the gain from learning the decimal digits is small.} 

\begin{figure}[H]
	\caption{Firm's profit as a function the worker's optimal wage} \label{fig_profit_graph}
	\centering
	\includegraphics[width=.7\linewidth]{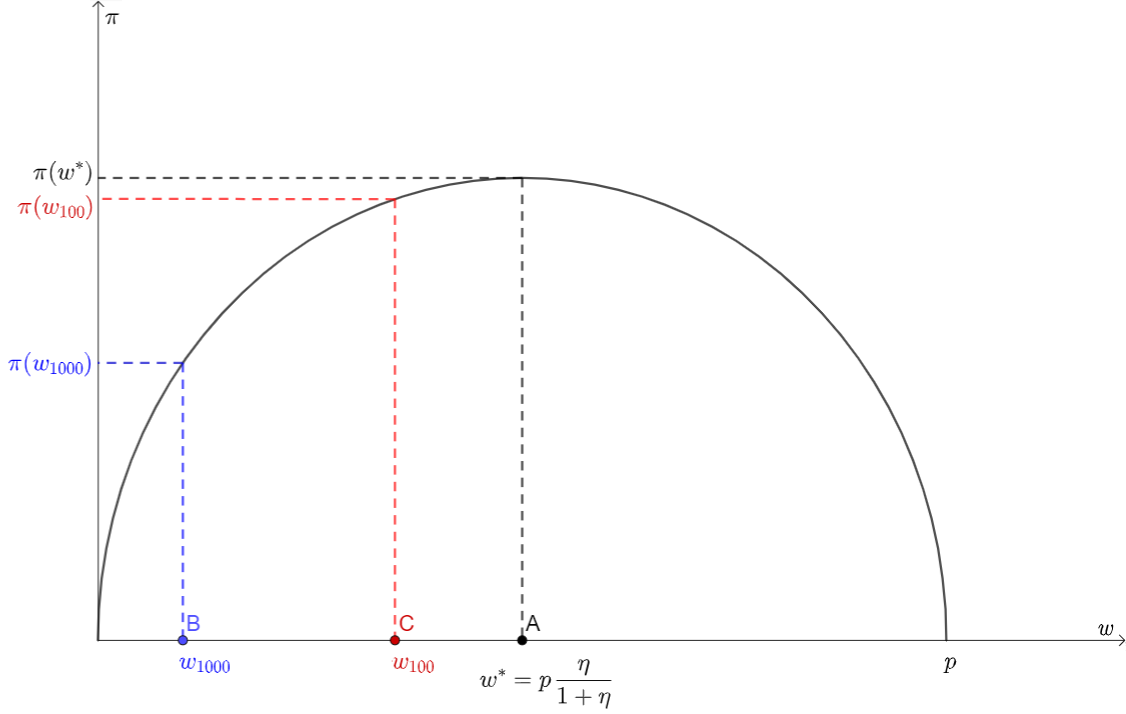}
	\footnotesize \singlespacing \justify \textit{Notes:} This figure illustrates the problem of a firm deciding how many digits of a worker's fully-optimal wage to learn. The figure plots the profit of the firm as a function of the wage posted. The optimal wage of the frictionless model, $w^*$, is ex-ante unknown to the firm and shown in  \circled{A}. For illustration purposes, the figure displays the case in which $w_{1000} < w^*$ is the firm's initial estimate of the fully-optimal wage (point \textcolor{blue}{\circled{B}}). The firm can forego a fraction $\tau_{100}$ of its profits to compute the second digit of the optimal wage (i.e., the optimal wage up to the nearest hundred) and learn $w_{100}$, shown in point \textcolor{red}{\circled{C}}. The firm will do so as long as $\pi(w_{100})(1 - \tau_{100}) \geq \pi(w_{1000})$.
	
\end{figure}

To illustrate the trade-offs faced by the firm, Appendix Figure \ref{fig_profit_graph} plots a firm's profit as a function of the wage posted. The fully-optimal wage (ex-ante unknown to the firm) is at point A. Without loss of generality, suppose that $w_{1000} < w^*$ is the firm's initial estimate of the fully-optimal wage, shown at point B (i.e., the fully-optimal wage rounded to the nearest 1,000). The firm could forfeit a fraction $\tau_{100}$ of its profits to compute the second digit of the optimal wage and learn $w_{100}$ (i.e., the optimal wage up to the nearest 100), shown at point C. The firm will do so as long as $\frac{\pi(w_{100})}{\pi(w_{1000})} \geq \frac{1}{1 - \tau_{100}}$.

The firm will continue refining its estimate of the fully-optimal salary as long as the marginal benefit of learning an additional digit is greater than the marginal optimization cost. Observe that learning further digits of the fully-optimal wage shrinks the mispricing wedge at a decreasing rate. If the initial estimate is equal to the fully-optimal wage up to the nearest 1,000, the error from not learning the second digit is at most 500, the error from not learning the following digit is at most 50, and the error from not learning the final digit is at most 5. 

Let $\theta_{1000}$, $\theta_{100}$, and $\theta_{10}$ be the fraction of workers hired at coarse wages divisible by 1,000, 100, and 10, respectively. The distribution of observed wages in this model has the following mixing distribution:
\begin{align} \label{eq_dist_wage_precision}
	F_w = \sum_{\mathclap{j \in \{10, 10^2, 10^3\}}} \theta_{j} F_{w_{j}} +  (1 - \sum_{\mathclap{j \in \{10, 10^2, 10^3\} }} \theta_{j}) F_{\tilde{p}}.
\end{align}

Equation \eqref{eq_dist_wage_precision} is the generalization of equation \eqref{eq_theta} for the case in which firms learn with different degrees of precision. In this case, we observe bunching at several round numbers. The size of the bunching at each round number reflects the fact that different firms learn a different number of digits, depending on how costly it is to do so and how much they stand to gain.

	\clearpage 
\section{Data Appendix} \label{app:data} 

\setcounter{table}{0}
\setcounter{figure}{0}
\renewcommand{\thetable}{D\arabic{table}}
\renewcommand{\thefigure}{D\arabic{figure}}

\subsection{Worker Record Booklet and RAIS Orientation Handbook} \label{app:handbook} 

\begin{figure}[h!]
	\caption{Example of a worker record booklet or CTPS}\label{fig_ctps}
	\centering
	\includegraphics[width=\linewidth]{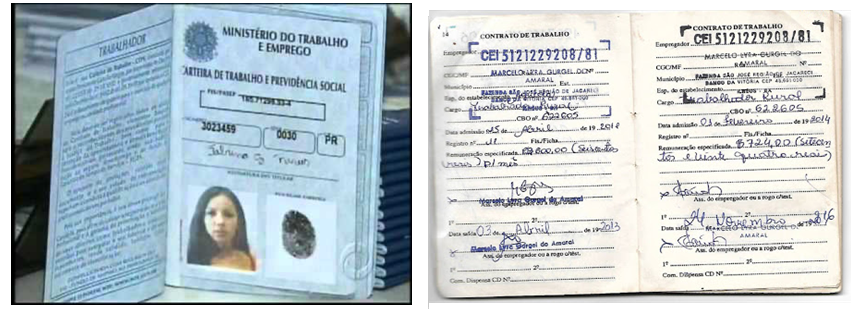}
\end{figure}

The main variable in the analysis is the contracted salary of each new hire. The contracted salary is the salary contained in the worker record booklet (CTPS). The CTPS lists the employment record of all workers employed in the formal sector and includes information on the worker's admission date, initial salary, and salary increases. Appendix Figure \ref{fig_ctps} shows an example of a worker record booklet and the information contained in it.

There are good reasons to believe that workers' contracted salary is accurately measured in the RAIS. First, firms have available an orientation handbook that details how to complete the information required by the RAIS. The following box shows an English translation of the section that explains how to complete the information regarding the contracted salary, taken from the 2019 orientation handbook (p.p. 29-30).

\begin{centering}
	\fbox{\begin{minipage}{\textwidth}
			\textbf{B.4) Contracted salary.---} Indicate the basic salary specified in the employment contract or recorded in the Work Card, resulting from the last salary change, which can correspond to the last month worked in the base year. In the case of a public servant, indicate the basic wage, as set by law.\\
			
			\textbf{B.4.1) Amount} - Should be provided in Brazilian reais (with cents).
			
			Notes:				
			\begin{enumerate}
				\item For employees whose salary is paid by commission or by various tasks with different remunerations, the monthly average of salaries paid in the base year should be indicated;
				\item For directors without an employment contract, who opt for the FGTS, indicate the last income in effect in the base year;
				\item For employees whose CTPS includes the salary plus commission, provide the base salary plus the monthly average of commissions paid in the base year;
				\item For hourly employees, indicate the hourly rate as defined in the employment contract.
			\end{enumerate}

	\end{minipage}	}\\~\\
\end{centering}

\begin{figure}[h!]
	\caption{Video explaining how to complete RAIS contracted salary information}\label{fig_youtube}
	\centering
	\includegraphics[width=.5\linewidth]{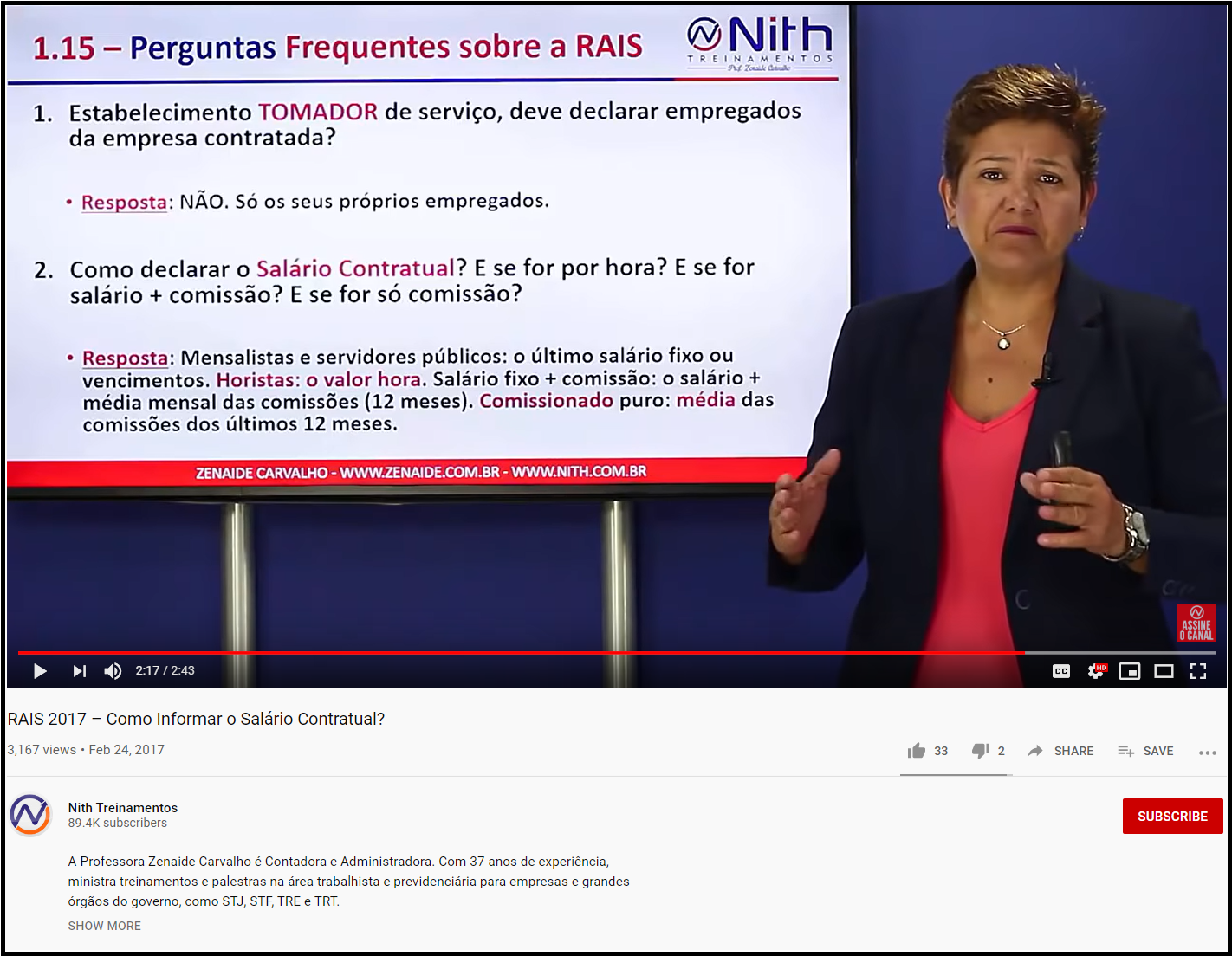}
	\footnotesize \singlespacing \justify \textit{Notes:} Source is \href{https://www.youtube.com/watch?v=JyHenUbFMUE}{RAIS 2017 – Como Informar o Salário Contratual?} 
\end{figure}

In addition to the handbook, there are several online resources that provide further assistance. Appendix Figure \ref{fig_youtube} exhibits an example of a publicly-available video that explains how to complete the contracted salary section of the RAIS.

\clearpage
\subsection{Variable Definitions} \label{app:var-def}

This section describes the variables that I use in the regressions presented in Sections \ref{sec:firm-behavior} and \ref{sec:implic}.

\begin{itemize}
	
	\item \textbf{Educational attainment of the firm manager.} This variable measures the schooling level of the highest-ranking person in each firm. I first assess if a firm has a chief executive officer (CEO). To identify a firm's CEO, I use the Brazilian occupational code classification (\textit{Classificação Brasileira de Ocupações}, or CBO for short). The CBO identifies CEOs with the code \texttt{121010}. If a firm does not employ any worker with this code, I use the educational attainment of the managers of the firm (identified by a first CBO-digit equal to one) and supervisors (identified by the third CBO-digit equal to zero). In case the firm has no managers or supervisors, I define the highest-ranking person in each firm as the worker with the highest wage.
	
	\item \textbf{Firm age.} This variable measures the number of years since the firm was created. I do not directly observe the firm creation date in the data. I proxy the foundation year as the minimum between (i) the first year in which the firm appears in the RAIS (using data since 1995) and (ii) the oldest admission year among all workers employed by the firm. I calculate the firm age as the difference between the current year and the firm creation year.
	
	\item \textbf{Firm size growth.} This variable measures the growth rate in the firm's number of employees. To compute this measure, I calculate the percent change in the number of workers employed by each firm between consecutive years.
	
	\item \textbf{Firm survival rate.} This variable indicates whether the firm exited the market. I identify a firm as exiting the market if it does not have any active workers at the end of the year.
	
	\item \textbf{Has a human resources department.} This variable indicates whether a firm has a human resources (HR) department. I identify firms as having an HR department if one of its employees is an HR manager (CBO codes \texttt{123205}, \texttt{123210}, \texttt{142210},\texttt{142205}) or an HR support staff (CBO codes \texttt{252105}, \texttt{252405}, \texttt{411030}).
	
	\item \textbf{Mean earning of firm employees.} This variable measures the average earnings of a firm's workers in a given year. I use workers' average monthly salary throughout a year as the relevant earnings measure and compute the average of this measure across all workers. I use the yearly consumer price index (CPI) to express earnings in real terms.
	
	\item \textbf{New hire separated.} This variable measures whether a new hire separated from the firm during the year the worker was hired or the following year. This variable is equal to one if a new hire is not employed at the end of the hiring year or at the end of the following year and is equal to zero if the new hire remains employed at the end of both years.
	
	\item \textbf{New hire resigned.} This variable is computed analogously to the one that measures new hires' separation, but using resignations (i.e., worker-initiated separations) instead of overall separations.
	
	\item \textbf{Number of hires.} This variable measures the number of workers hired by the firm during 2003--2017. To compute this variable, I only consider hires with a monthly contract and hired at a salary above the federal minimum wage. This sample restriction makes the analyses of the firm random sample comparable to the analyses of the new-hires sample.
	
	\item \textbf{Ratio between percentiles of the new hires' salary distribution.} This variable measures the ratio between salaries in different percentiles of the contracted salary distribution among the new hires of a given firm during 2003--2017. Before computing the ratio, I adjust all salaries using the yearly CPI. I winsorize the ratios at the 99th percentile.
	
	\item \textbf{Salary increase in percent is an integer.} This variable indicates whether the percent salary increase of a worker is an integer number. To compute this measure, I calculate the percent change in workers' contracted salary between the year the firm hired the worker and the following year. The indicator variable is equal to one if the percent change is an integer and zero otherwise. 
	
	\item \textbf{Salary increase in Brazilian Reals is a round number.} This variable indicates whether the salary increase of a worker, measured in Brazilian Reals, is divisible by ten. To compute this measure, I calculate the difference in a worker's contracted salary between the year the worker was hired and the following year. The indicator variable is equal to one if this difference is a round number and zero otherwise. 
	
	\item \textbf{Share of employees with completed high school.} This variable measures the fraction of a firm's employees that completed at least high school. To compute this variable, I first calculate the number of workers in each firm with educational data available over the 2003--2017 period. Next, I compute the number of workers who finished high school over the same period. Finally, I compute the ratio between these two variables.  
	
	\item \textbf{Share of employees with completed college.} This variable is computed analogously as the share of employees with completed high school.
	
	\item \textbf{Worker contracted salary.} The contracted salary represents a worker's salary as per the worker's contract at the end of each year. For a new hire, the contracted salary is the same as the initial salary. For other workers, the contracted salary is equal to the current salary, which might differ from the initial salary due to promotions or other wage adjustments.
	
\end{itemize}

\subsection{Measurement Error in the Contracted Salaries of 2016 and 2017}

In the 2016 and 2017 RAIS, the contracted salary variable contains substantial measurement error. The RAIS reports two measures of a worker's contracted salary that are equivalent before 2016:

\begin{itemize}
	\item The first measure is the contracted salary in Brazilian Reals. This is the variable that I use throughout the paper.
	
	\item The second measure is the contracted salary measured in multiples of the federal monthly minimum wage.
\end{itemize}

In 2016 and 2017, these two measures are not equivalent. Half of the workers earn monthly salaries \textit{below} the minimum wage according to the contracted salary in Reals but earn salaries \textit{above} the minimum wage according to the second measure. Upon further exploration, it appears that many firms reported their employees'  earnings in units of hundreds of Brazilian Reals. In other words, for many workers, the contracted salary reported in multiples of the minimum wage is equal to the contracted salary reported in Reals divided by the minimum wage and multiplied by 100. I adjusted the reported earnings for these workers to correct this discrepancy. Excluding 2016 and 2017 from the analysis does not change the main results of the paper.

\subsection{Sample Restrictions} \label{app:samp-rest}

In this section, I describe the sample restrictions that I impose on the new-hires sample. Appendix Table \ref{tab_metadata} shows the number of observations (contracts) at the beginning and at the end of each step of the data cleaning process. The analysis begins in 2003 since this is the first year in which the characteristics of workers' contracts are available in the RAIS.

\begin{enumerate}[leftmargin=*]
	
	\item I include only new hires in each year. I exclude the contracts of workers hired during previous years to avoid double-counting the same worker.
		
	\item I only consider workers with a valid identifier. Workers in the private sector are uniquely identified by their ID in the Social Integration Program (PIS, for its name in Portuguese, \textit{Programa de Integração Social}). The eleven-digit PIS ID of a worker is constant throughout the worker's career. I only keep workers with an eleven-digit ID.
	
	\item I exclude workers employed by public-sector firms. 
	
	\item I only consider workers hired at a monthly contract. In the sample, about 91\% of contracts are signed at the monthly level. The second most common type of contract is at the hourly level (about 7.5\% of all contracts). 
	
	\item Some firms report hiring workers at a salary below the federal montly minimum salary. This is likely due to measurement error. To deal with this, I drop all the contracts that are made for earnings below the federal monthly minimum salary of each year. 
	
	\item Some firms report hiring the same worker multiple times in a given year. I only keep one observation per worker-firm-year. 
	
	\item I exclude new hires for whom their contracted wage is missing.
		
\end{enumerate}

At the end of this process, I remain with data on over 210 million contracts. I group workers in R\$1 bins (roughly 30 cents of a dollar) and winsorize the right tail of the distribution at R\$10,100 (this affects about 0.3\% of the workers). 

\begin{landscape}
	\begin{table}[H]{\scriptsize
			\centering
			\caption{Sample size after each restriction} \label{tab_metadata}
			\begin{tabular}{cccccccccccccccc}
				\toprule
				& \multicolumn{3}{c}{Raw data} &   & \multicolumn{7}{c}{Fraction of observations remaining after each restriction} &   & \multicolumn{3}{c}{New-hires sample} \\
				\cmidrule{2-4}\cmidrule{6-12}\cmidrule{14-16}   
		          	 &           & Unique  & Unique && New   & Valid     & Private & Monthly  & Salary   & No multiple & Contracted     &&           & Unique  & Unique  \\
				Year & Contracts & workers & firms  && hires & worker ID & firms   & contract & above MW & positions   & wage available && Contracts & workers & firms   \\
				(1)  & (2)       & (3)     & (4)    && (5)   & (6)       & (7)     & (8)      & (9)      & (10)        & (11)           && (12)      & (13)    & (14)    \\
				\midrule

				2003 &41,969,162 &35,925,326 &2,504,099 & &0.314 &0.314 &0.254 &0.209 &0.205 &0.198 &0.198 & &8,310,668 &7,614,477 &1,193,589 \\
2004 &44,683,910 &37,856,735 &2,602,198 & &0.330 &0.330 &0.272 &0.224 &0.220 &0.212 &0.212 & &9,492,613 &8,605,493 &1,272,182 \\
2005 &47,657,099 &40,179,150 &2,700,198 & &0.339 &0.339 &0.271 &0.229 &0.223 &0.216 &0.216 & &10,310,220 &9,324,565 &1,354,526 \\
2006 &50,701,027 &42,486,868 &2,805,601 & &0.331 &0.331 &0.271 &0.229 &0.223 &0.216 &0.216 & &10,931,030 &9,835,373 &1,402,385 \\
2007 &54,649,133 &45,227,446 &2,904,935 & &0.344 &0.344 &0.283 &0.241 &0.233 &0.226 &0.226 & &12,367,591 &10,992,829 &1,485,222 \\
2008 &59,706,419 &48,573,811 &3,048,597 & &0.357 &0.357 &0.299 &0.258 &0.251 &0.244 &0.244 & &14,541,694 &12,701,540 &1,626,228 \\
2009 &61,126,896 &50,219,948 &3,185,547 & &0.345 &0.345 &0.281 &0.246 &0.240 &0.232 &0.232 & &14,206,565 &12,505,729 &1,677,559 \\
2010 &66,747,302 &53,771,613 &3,359,136 & &0.366 &0.366 &0.310 &0.271 &0.265 &0.257 &0.257 & &17,181,178 &14,785,688 &1,850,714 \\
2011 &75,671,386 &56,641,163 &3,541,200 & &0.351 &0.351 &0.298 &0.263 &0.254 &0.243 &0.243 & &18,382,282 &15,660,444 &1,969,594 \\
2012 &73,326,485 &58,738,850 &3,645,405 & &0.356 &0.356 &0.304 &0.270 &0.265 &0.258 &0.258 & &18,916,807 &16,022,063 &2,016,896 \\
2013 &75,400,510 &60,450,823 &3,785,842 & &0.361 &0.361 &0.301 &0.269 &0.263 &0.257 &0.257 & &19,365,828 &16,393,304 &2,091,097 \\
2014 &76,107,279 &61,492,767 &3,895,042 & &0.342 &0.342 &0.289 &0.261 &0.255 &0.249 &0.249 & &18,944,127 &16,107,136 &2,115,478 \\
2015 &73,582,564 &59,363,823 &3,885,156 & &0.297 &0.297 &0.247 &0.225 &0.216 &0.207 &0.207 & &15,222,765 &13,319,415 &1,951,694 \\
2016 &67,144,598 &56,906,493 &3,872,884 & &0.267 &0.267 &0.218 &0.199 &0.187 &0.182 &0.182 & &12,193,564 &10,977,265 &1,791,311 \\
2017 &65,655,882 &55,561,692 &3,845,034 & &0.281 &0.281 &0.221 &0.200 &0.180 &0.175 &0.175 & &11,487,730 &10,378,453 &1,684,555 \\
 \midrule
				&934,129,652 &763,396,508 &49,580,874 & &0.333 &0.332 &0.276 &0.242 &0.234 &0.227 &0.227 & &211,854,662 &185,223,774 &25,483,030 \\

		\end{tabular}}
		
		{\footnotesize
			\singlespacing \justify
			
			\textit{Notes:} This table shows the number of contracts, unique workers, and unique firms in each year before and after imposing the sample restrictions. See text for a description of each sample restriction.
			
		}  
	\end{table}
	
\end{landscape}

	\clearpage
\section{Estimating a Counterfactual Earnings Distribution} \label{app:ctfl-dist}

\setcounter{table}{0}
\setcounter{figure}{0}
\setcounter{equation}{0}	
\renewcommand{\thetable}{E\arabic{table}}
\renewcommand{\thefigure}{E\arabic{figure}}
\renewcommand{\theequation}{E\arabic{equation}}

In this Appendix, I explain how I construct a counterfactual earnings distribution that does not feature bunching at round-numbered wages.

The standard approach to construct a counterfactual distribution in the bunching literature involves estimating a high-degree polynomial on the observed earnings distribution \textit{excluding} the salaries that exhibit bunching and using the estimated polynomial coefficients to predict the counterfactual number of workers at the salaries where workers bunch. 

The first step consists of regressing the number of workers in bin $b$, $C_b$, on a function $f(\cdot)$ that depends on the earnings of bin $b$, $w_b$,
\begin{align} \label{eq_bunch}
	C_b = \alpha + f(w_b) + \varepsilon_b.
\end{align}	

Previous work has traditionally set $f(\cdot)$ as a high-degree parametric function of earnings, including dummy variables at the salaries of the distribution that exhibit bunching. A straightforward implementation of this approach would be to set
\begin{align*}
	f(w_b) = \sum_{p=1}^P \beta_p (w_b)^p + \sum_{r \in R} \gamma_r \mathbbm{1}{\{w_{b} = r\}},
\end{align*}
where $\sum_{r \in R} \gamma_r$ is a set of dummies, one for each round number, and $P$ is the polynomial degree. The counterfactual distribution without bunching is estimated using the predicted values from \eqref{eq_bunch}, omitting the contribution of the dummies	
\begin{align} \label{eq_counterfactual}
	\hat{C_b} = \hat{\alpha} + \sum_{p=1}^P \hat{\beta}_r (w_b)^p.
\end{align}	

This parametric approach is well-suited to estimate counterfactual distributions \textit{locally}, that is, around one particular kink or notch. However, I need to estimate a counterfactual density around \textit{each} round number. As I show below, the parametric approach tends to perform poorly in estimating \textit{global} counterfactuals. 

\begin{figure}[H]
	\caption{Comparison of parametric and non-parametric counterfactual distributions} \label{fig_para_nonpara}
	\centering
	\includegraphics[width=.75\linewidth]{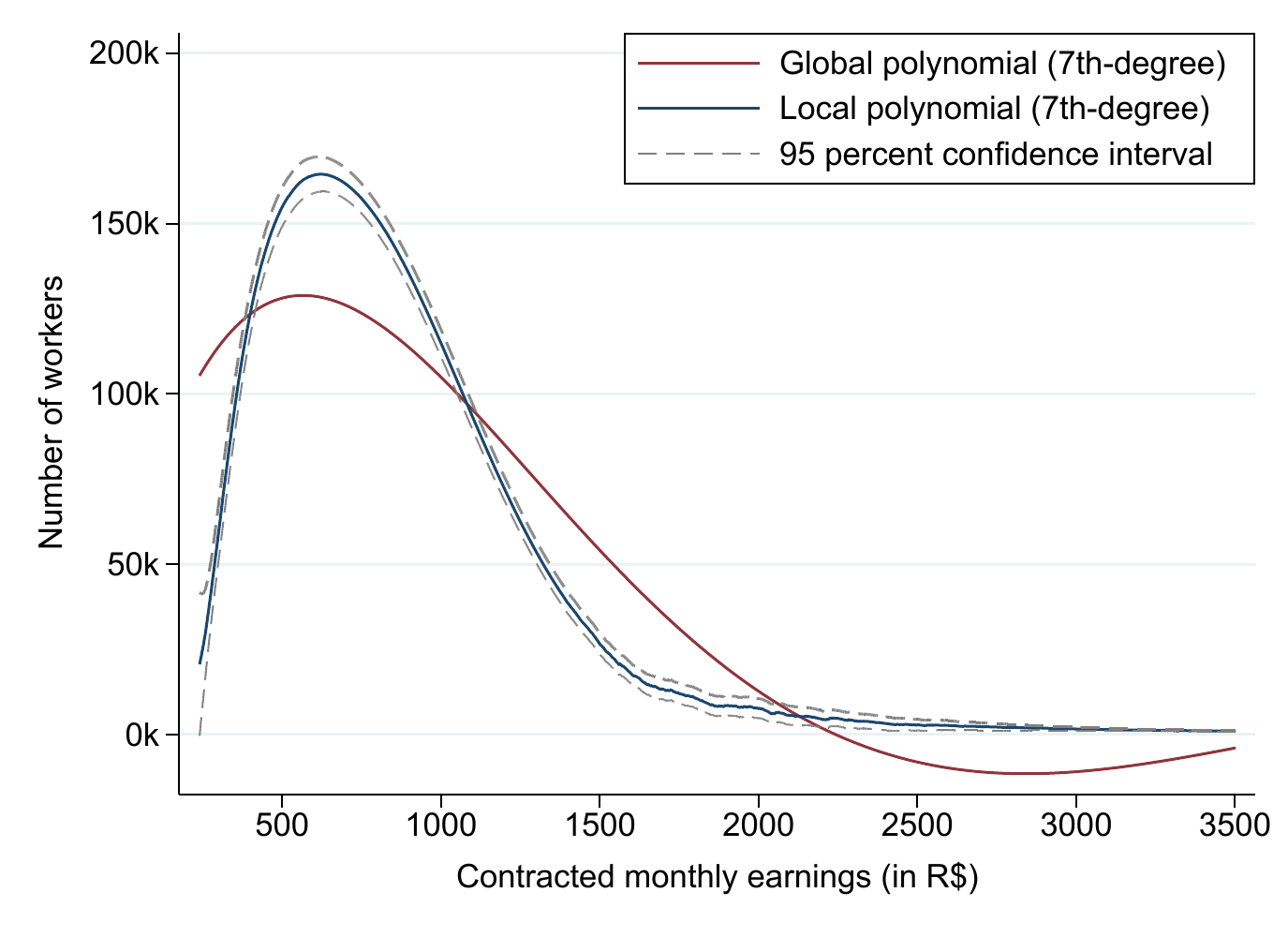}
	\hfill
	\footnotesize \singlespacing \justify
	
	\textit{Notes:} This figure compares the counterfactual earnings distribution using two different approaches. The red line denotes the counterfactual earnings distribution using a global 7th-degree polynomial. The blue line denotes the counterfactual distribution using a local 7th-degree polynomial. The gray dashed line around the local polynomial denotes the 95\% confidence interval.
\end{figure}

\begin{figure}[H]
	\caption{Robustness of the counterfactual distribution to alternative specifications} \label{fig_robust}
	\centering
	
	\begin{subfigure}[t]{.75\textwidth}
		\caption*{Panel A. Robustness to polynomial degree} \label{fig_robust_degree}
		\centering
		\includegraphics[width=\textwidth]{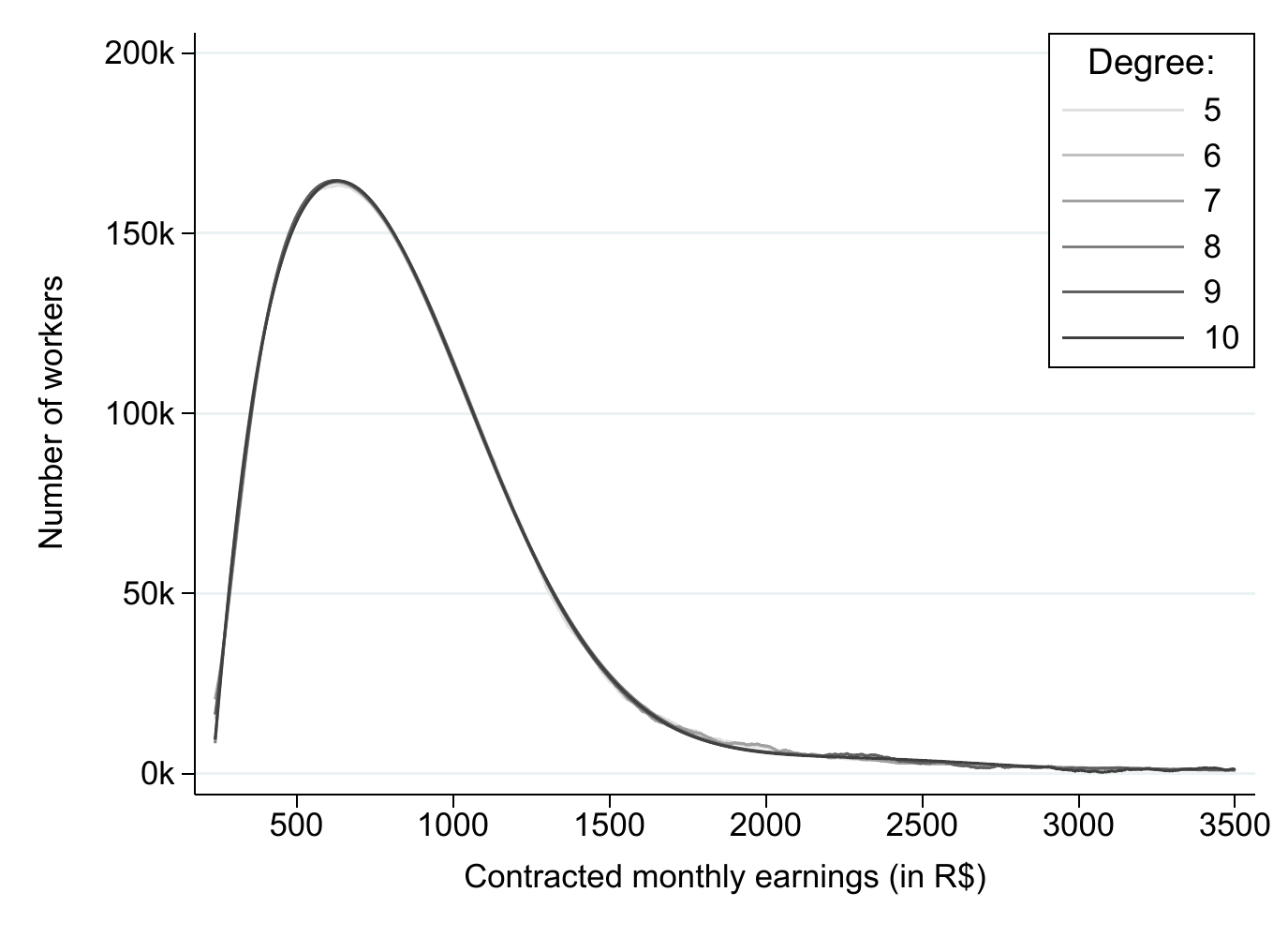}
	\end{subfigure}

	\begin{subfigure}[t]{0.75\textwidth}
		\caption*{Panel B. Robustness to kernel choice} \label{fig_robust_kernel}
		\centering
		\includegraphics[width=\textwidth]{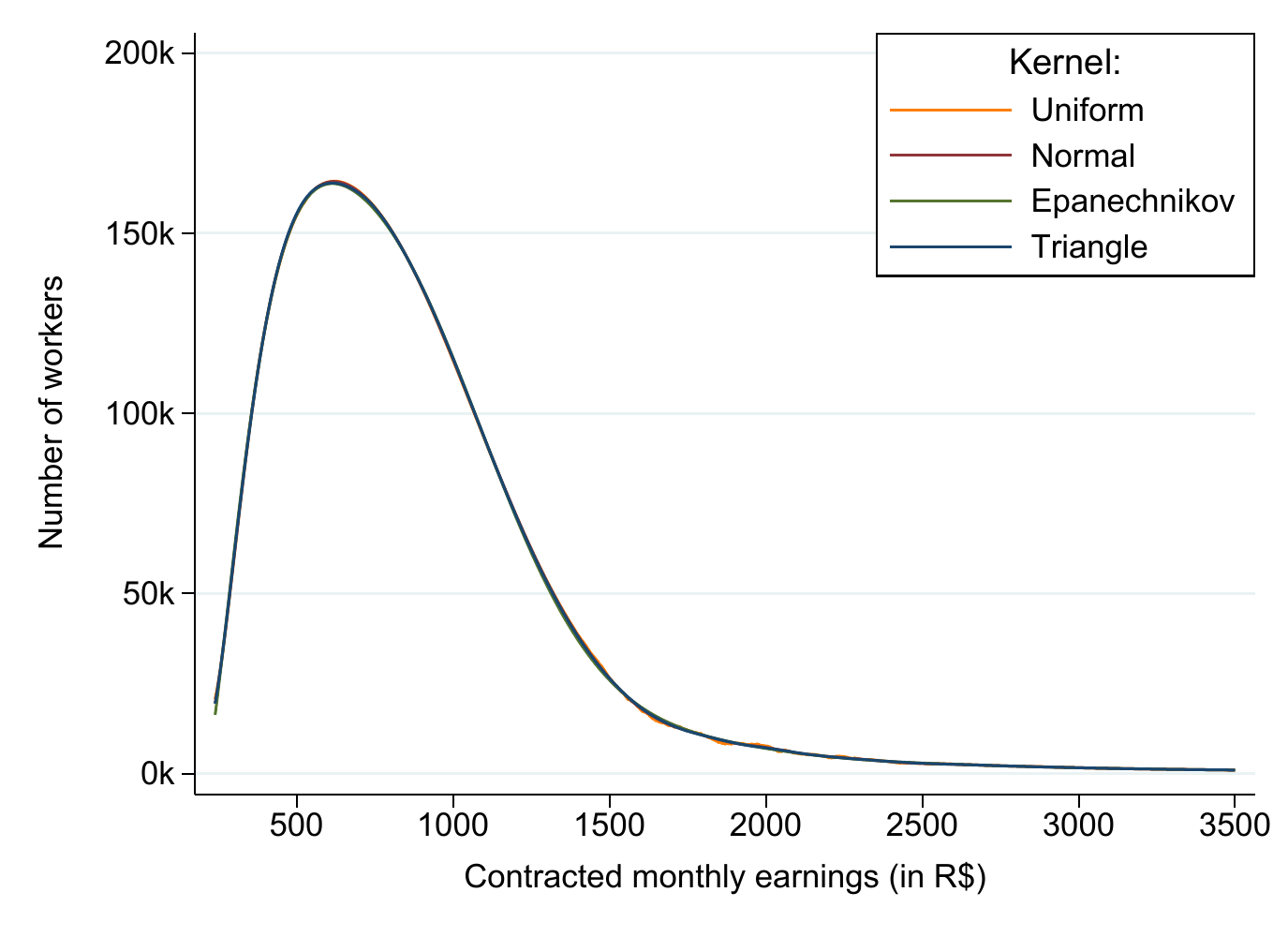}
	\end{subfigure}		
	\footnotesize \singlespacing \justify
	
	\textit{Notes:} This figure shows how the counterfactual earnings distributions estimated using a local polynomial approach changes when varying the polynomial degree (Panel A) and the type of kernel (Panel B). See Appendix \ref{app:ctfl-dist} for details on how I estimate the counterfactual distribution.
	
\end{figure}

An appealing alternative is to use a non-parametric approach. I estimate kernel-weighted local polynomial regressions using a uniform kernel on non-round-numbered earnings and use the estimates to predict the density at round-numbered wages. Intuitively, to estimate the density at each salary, I use data points ``close'' to the salary, where close is defined by the bandwidth of the kernel. For a sufficiently large bandwidth (i.e., a bandwidth that covers the entire support of the earnings distribution), the local polynomial regression yields the exact same counterfactual as the parametric one. However, for a small bandwidth, the non-parametric approach yields better-behaved estimates. To see this, Appendix Figure \ref{fig_para_nonpara} compares the counterfactual distribution of earnings using the parametric and non-parametric approaches, in both cases using a seventh-degree polynomial. Unlike the non-parametric counterfactual distribution, the parametric one yields a \textit{negative} estimated number of workers in some segments of the distribution.\footnote{The shape of the counterfactual is robust to the polynomial degree (Appendix Figure \ref{fig_robust}, Panel A) and the type of kernel (Appendix Figure \ref{fig_robust}, Panel B). All specifications include minimum wage dummies to improve the fit of the counterfactual density at the minimum wage.}

Since the counterfactual number of observations does not include the contribution of the dummies, the aggregate number of observations in the data, $N$, is necessarily higher than the predicted total number of observations, i.e., $N = \sum_b C_b > \sum_b \hat{C}_b = \hat{N}$. To account for this, I re-weight all observations by $\frac{\sum_b C_b}{\sum_b \hat{C}_b}$. This approach rules out extensive margin responses. This means that the use of coarse wages moves workers around the earnings distribution, but it does not make any worker leave or enter the labor market altogether. This implies that the excess mass at round-numbered salaries corresponds to missing mass at non-round-numbered salaries. 

\begin{figure}[H]
	\caption{Estimation of the counterfactual distribution, excess mass, and missing mass} \label{fig_approach1}
	\centering
	\begin{subfigure}[t]{.48\textwidth}
		\caption*{Panel A. Excess mass} \label{fig_eg_excess}
		\centering
		\includegraphics[width=\textwidth]{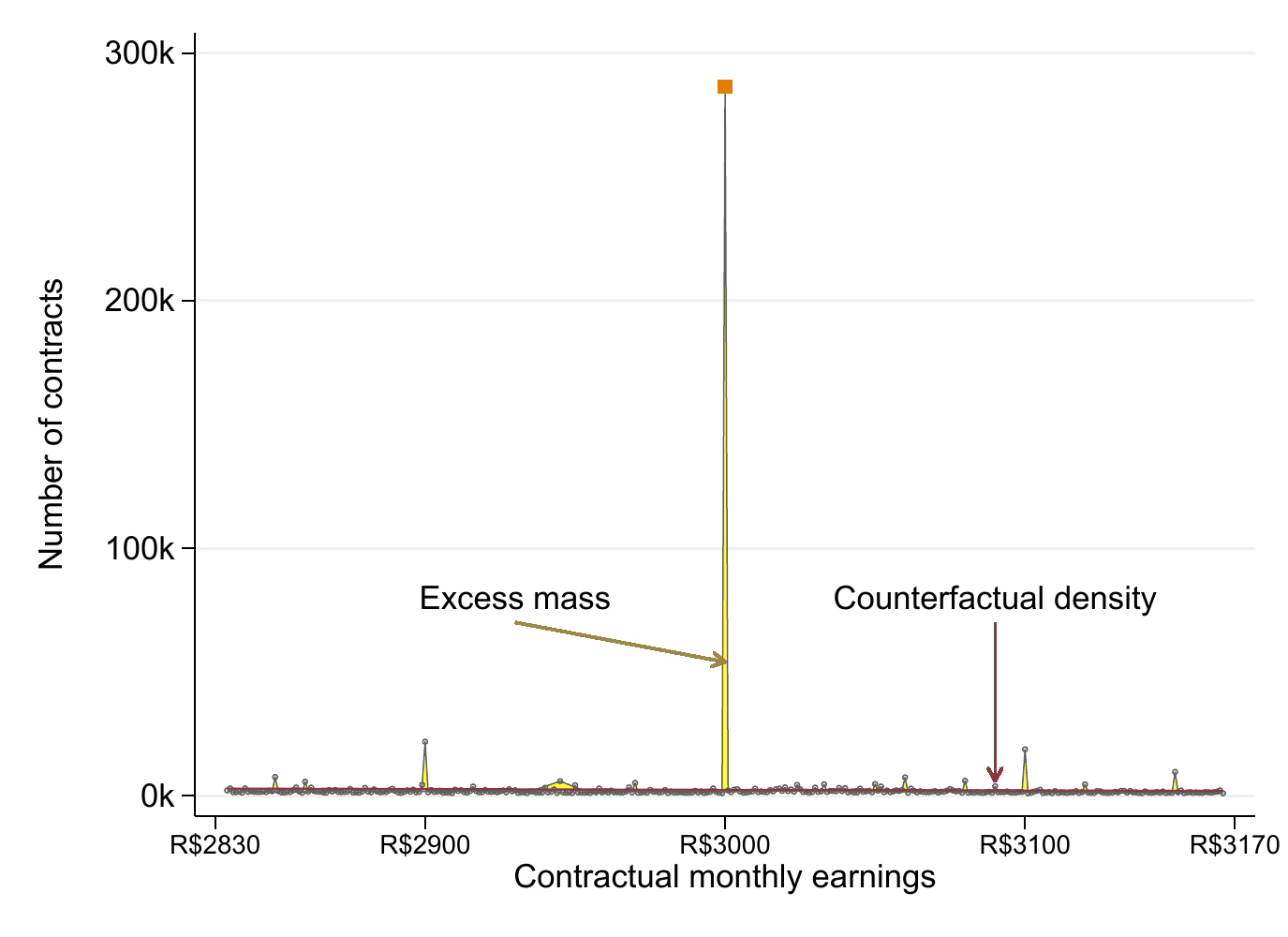}
	\end{subfigure}
	\hfill        
	\begin{subfigure}[t]{0.48\textwidth}
		\caption*{Panel B. Missing mass} \label{fig_eg_missing}
		\centering
		\includegraphics[width=\textwidth]{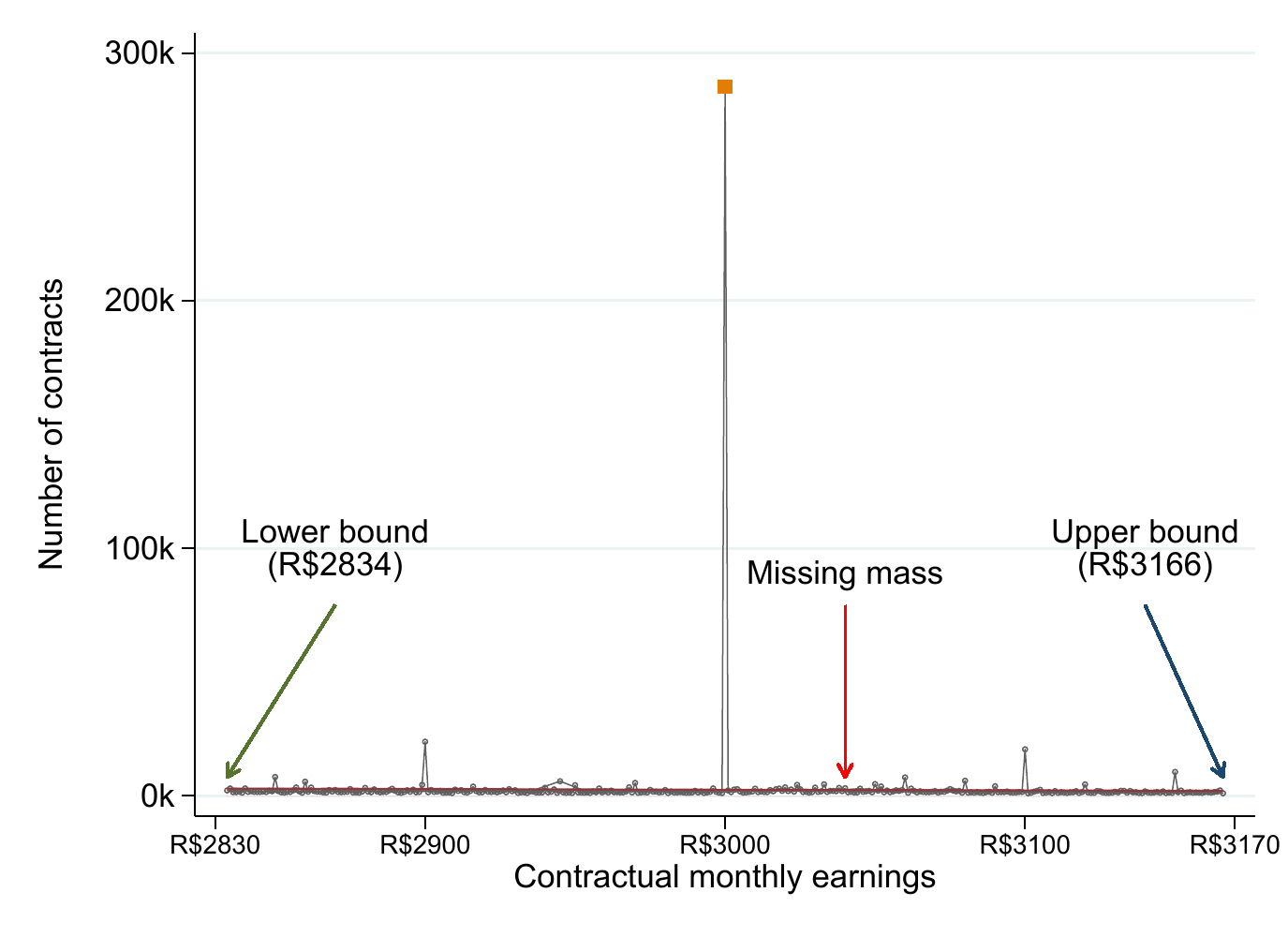}
	\end{subfigure}	
	\footnotesize \singlespacing \justify
	
	\textit{Notes:} This figure illustrates how I calculate the excess mass at R\$3,000. The figure shows the distribution of earnings between R\$2,834 and R\$3,166 in the new-hires sample. Gray dots denote the observed number of workers, while the red line denotes the counterfactual distribution estimated with a local polynomial. The yellow area in Panel A denotes the excess mass, which is equal in magnitude to the missing mass denoted by the red area in Panel B. 
	
\end{figure}

To quantify the missing mass, I follow \cite{kleven2013using} and select the narrowest manipulation region consistent with the data. To illustrate how the approach works, Appendix Figure \ref{fig_approach1} shows how the counterfactual distribution, excess mass (Panel A), and missing mass (Panel B) around R\$3000 are estimated.

\clearpage

\clearpage

\clearpage

	\clearpage 
\section{Alternative Explanations} \label{app:alt-exp} 

\setcounter{table}{0}
\setcounter{figure}{0}
\setcounter{equation}{0}	
\renewcommand{\thetable}{F\arabic{table}}
\renewcommand{\thefigure}{F\arabic{figure}}
\renewcommand{\theequation}{F\arabic{equation}}

In this Appendix, I assess five alternative explanations for the bunching observed in the data. The explanations I discuss are: worker left-digit bias, focal points in wage bargaining, fairness concerns, round wages as a signal of job quality, and changes in marginal tax rates.

\subsection{Worker left-digit bias} \label{app:ldb}

One possible explanation for the clustering of wages at round numbers is that firms use round salaries as an optimal response to a worker bias. A plausible bias that has been documented in other environments is the left-digit bias, that is, the propensity of individuals to pay more attention to the first digit of a number relative to the other digits \citep{korvorst_differential_2008, lacetera_heuristic_2012, strulov2019more}. 

I view the results in Section \ref{sec:firm-behavior} as the main evidence against firms paying round-numbered wages as an optimal response to worker left-digit bias. Specifically, I find that firms that are smaller, younger, have less hiring experience, and do not have an HR department are the ones more likely to pay round-numbered salaries to new hires. It is unlikely that these firms are paying round-numbered salaries to exploit a worker bias. Having awareness of a worker bias requires a considerable amount of sophistication, and these firms are less sophisticated in observable characteristics.

For completeness, I conduct two additional tests for worker left-digit bias. As a first test, I analyze whether workers earning just below round salaries have systematically higher separation rates than workers earning exactly a round salary or a salary just above it. This test is analogous to one conducted by \cite{dube_monopsony_2020} using observational data. Intuitively, in the presence of a left-digit bias, workers with salaries close to but below a round number would be more likely to leave a firm to pursue a better wage than workers earning a round salary or a salary just above it. A problem with separation rates is that the separations might be driven by firms exiting the market, as opposed to workers leaving because they found a better match. In the data, I observe whether the employer or the employee initiated the separation. Thus, I estimate worker \textit{resignation rates} (i.e., worker-initiated separations) in the vicinity of round salaries. 

As a second test for worker left-digit bias, I analyze whether there is an asymmetric mass of workers just below and just above round salaries. According to some models of left-digit biased workers, most of the excess mass observed at round salaries should come from salaries just \textit{below} the round number. There are alternative ways of modeling worker left-digit bias, some of which predict that the missing mass also comes from \textit{above} each round number \citep[e.g.,][]{strulov2019more}. Thus, while this test is informative of the possible existence of left-digit bias, it is by no means conclusive.

\subsubsection{Worker resignation rate \\~\\} 

\begin{figure}[H]
	\caption{Resignation rates below, at, and above salaries divisible by 100} \label{fig_res_rates}
	\centering
	\begin{subfigure}[t]{0.48\textwidth}
		\caption*{Panel A. Average resignation rate in the vicinity of round salaries}\label{fig_avgres_bw10}
		\centering
		\includegraphics[width=\linewidth]{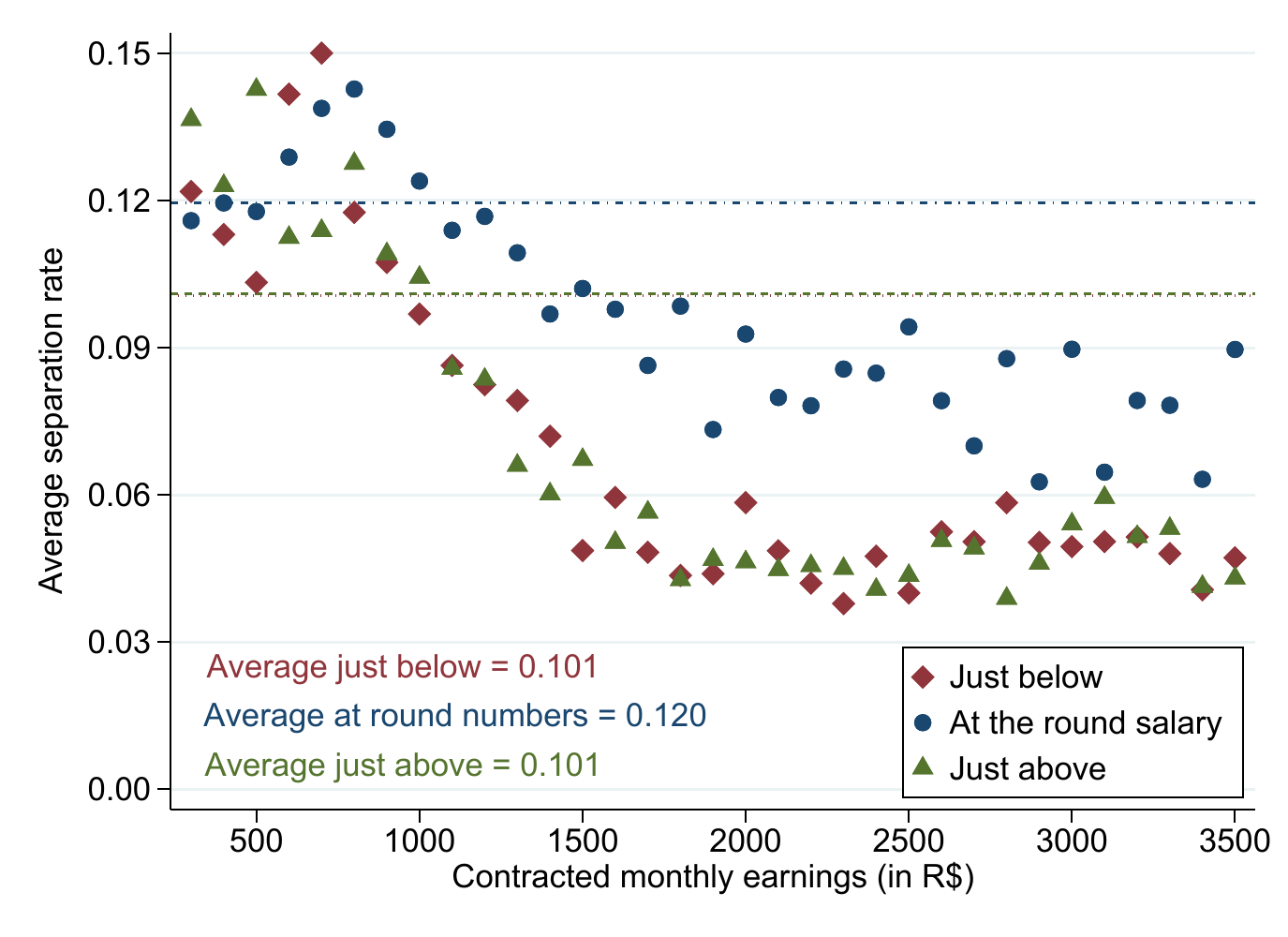}
	\end{subfigure}
	\hfill		
	\begin{subfigure}[t]{0.48\textwidth}
		\caption*{Panel B. Regression discontinuity estimates $\hat{\beta}_r$'s and $\hat{\gamma}_r$'s from equation \eqref{eq_rd_sep}} \label{fig_rdres_bw10}
		\includegraphics[width=\linewidth]{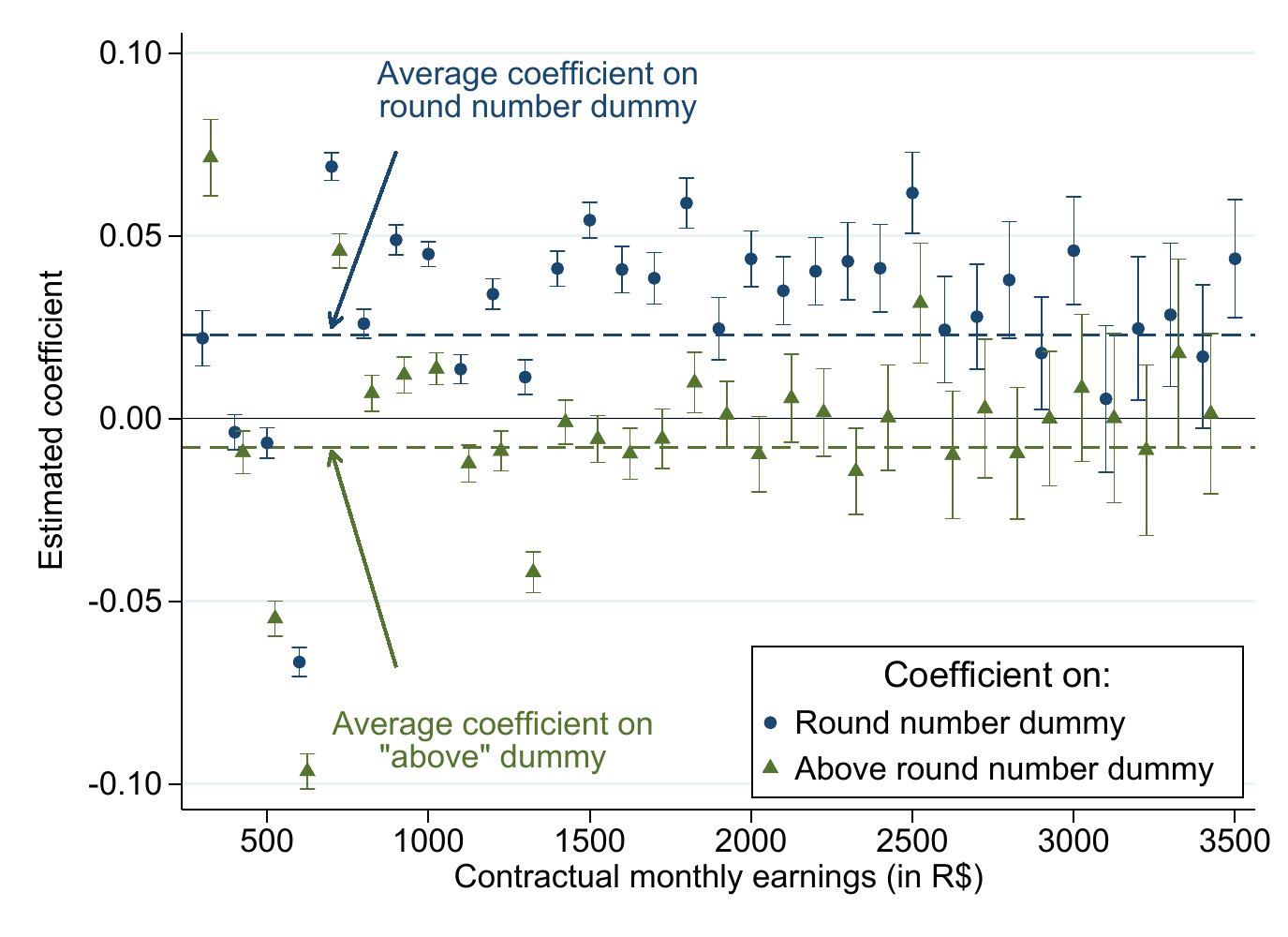}
	\end{subfigure}	
	
	{\footnotesize
		\singlespacing \justify
		
		\textit{Notes:} This figure shows whether there are systematic differences in the resignation likelihood of workers earning a salary just below and just above round numbers. To construct the figures in both panels, I use the firm random sample. The figures only display workers with earnings above the minimum wage and below R\$3,500 (which roughly corresponds to the 99th percentile of the earnings distribution above the minimum wage).
		
		Panel A shows the average resignation rate of workers earning a salary just below, equal to, or just above each salary divisible by 100, using a bandwidth $h = 10$. For example, the figure shows that the resignation rate of workers earning [R\$490, R\$500) is 4.8\%, the resignation rate of workers earning R\$500 is 5.3\%, and the resignation rate of workers earning (R\$500, R\$510] is 4.4\%. The horizontal dashed lines denote the weighted average resignation rate of each group of workers across all salaries divisible by 100, using the number of workers used to estimate each separation rate as the weight. 
		
		Panel B presents the RD estimates of regression \eqref{eq_rd_sep}, using as the outcome a dummy that equals one if the worker resigned and zero otherwise, and using a bandwidth $h = 10$. Each point in the figure comes from a separate regression using data in the vicinity of a salary divisible by 100. For example, the point estimate at $R\$500$ uses data from workers whose earnings are within a distance $10$ of R\$500 (including workers who earn exactly R\$500). The vertical lines denote 95\% confidence intervals using heteroskedasticity-robust standard errors. Standard errors are clustered at the worker level. The horizontal dashed line denotes the weighted average RD coefficients across all regressions, where the weights are the number of workers used to estimate each regression.
		
	}	
\end{figure}

\textbf{Visual evidence}. Appendix Figure \ref{fig_res_rates}, Panel A shows the resignation rate of workers hired at each salary divisible by 100, a salary just below it, and just above it. To construct this figure, I compute the resignation rate for three sets of workers: those who earn a round salary $w_r$, those whose earnings fall in the $[w_r - h, w_r)$ range where $h$ is the bandwidth (these are the workers ``just below'' $w_r$), and those who earn a salary in the $(w_r, w_r + h]$ range (these are the workers ``just above'' $w_r$). I calculate the resignation rates in the vicinity of each salary divisible by 100 and for $h = 10$. 

The average resignation rate of workers earning just above round salaries is equal to the one for workers hired at a salary just below a round number (in both cases, equal to 0.048). In turn, these workers are, on average, slightly \textit{less} likely to resign relative to workers that earn exactly a round salary. On average across round numbers, the average resignation rate of workers that earn a salary divisible by 100 is 0.051. Moreover, workers earning a round salary have higher resignation rates not just on average, but also for almost every salary divisible by 100. These results are robust to alternative bandwidths.

\textbf{Regression discontinuity analysis}. Next, I use a regression discontinuity (RD) design to assess whether the differences in resignation rates shown above are statistically significant. I estimate regressions of the form:
%
	\begin{align} \label{eq_rd_sep}
		\text{Res}_{i} &= \alpha + \nu w_{i} + \beta_r  \mathbbm{1}{\{w_{i} = w_r\}} + \gamma_r  \mathbbm{1}{\{w_{i} > w_r\}} \notag \\  &+ \delta_r w_{i} \mathbbm{1}{\{w_{i} > w_r\}} + \varepsilon_{i} \hspace{.1cm}  \text{ if } |w_{i} - w_r| \leq h,
	\end{align}
%
where $\text{Res}_{i}$ equals one if worker $i$ resigned and zero otherwise, $w_{i}$ is the contracted salary of worker $i$, $w_r$ is a round salary within distance $h$ of $w_i$, and $h$ is the bandwidth. The two coefficients of interest are $\beta_r$ and $\gamma_r$. They measure whether workers earning exactly $w_r$ and workers earning just above $w_r$, respectively, have differential average resignation likelihoods, relative to workers earning just below $w_r$.

Appendix Figure \ref{fig_res_rates}, Panel B plots the estimated $\hat{\beta}_r$'s and $\hat{\gamma}_r$'s for $h = 10$. Each coefficient comes from estimating equation \eqref{eq_rd_sep} around a different round number divisible by 100. Consistent with the visual evidence, workers earning a round salary are \textit{more} likely to resign relative to workers with earnings just below or just above one. This is true for most round numbers, although, in some cases, the standard errors are quite large. In contrast, workers earning just above each round salary do not have systematically different likelihoods of resigning than workers earning just below round salaries.

In sum, these results indicate that the workers who earn a round-numbered salary are \textit{more} likely to resign than workers who earn a salary just below or just above the round number. This provides further evidence against the hypothesis that firms pay round-numbered salaries to exploit worker left-digit bias.

\subsubsection{Mass of contracts below and above round salaries  \\~\\} \label{app:ldb_mass_work} 

\begin{figure}[H]
	\caption{Difference in the number of contracts around salaries divisible by 100} \label{fig_mass_workers}
	\centering
	\begin{subfigure}[t]{.48\textwidth}
		\caption*{Panel A. Share of contracts just below and just above each side of the round salary}\label{fig_mass_mult_100_bw10}
		\centering
		\includegraphics[width=\linewidth]{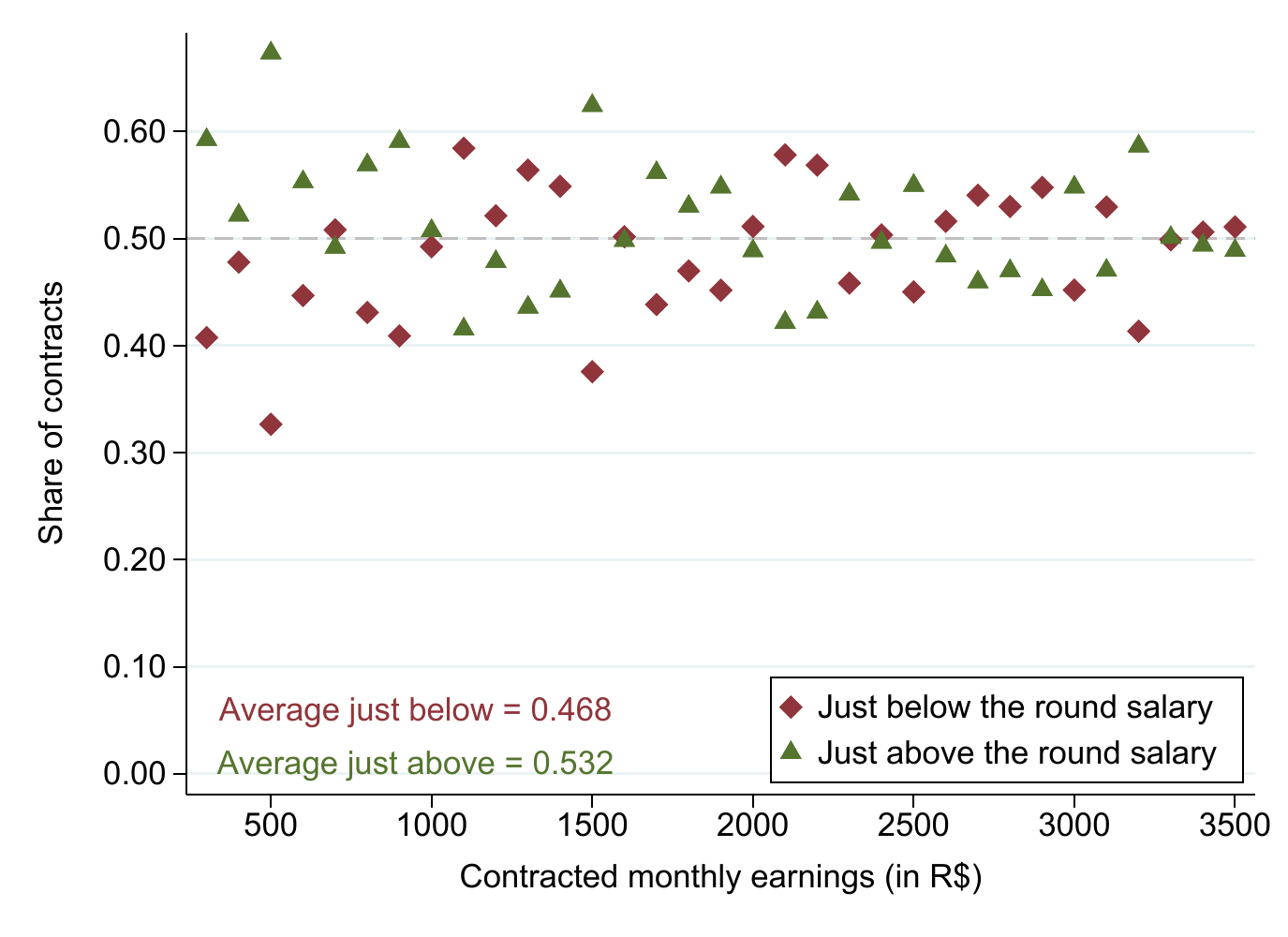}
	\end{subfigure}
	\hfill		
	\begin{subfigure}[t]{0.48\textwidth}
		\caption*{Panel B. Regression discontinuity estimates $\hat{\beta}_r$'s from equation \eqref{eq_mccrary}}\label{fig_mccrary_bw10}
		\centering
		\includegraphics[width=\linewidth]{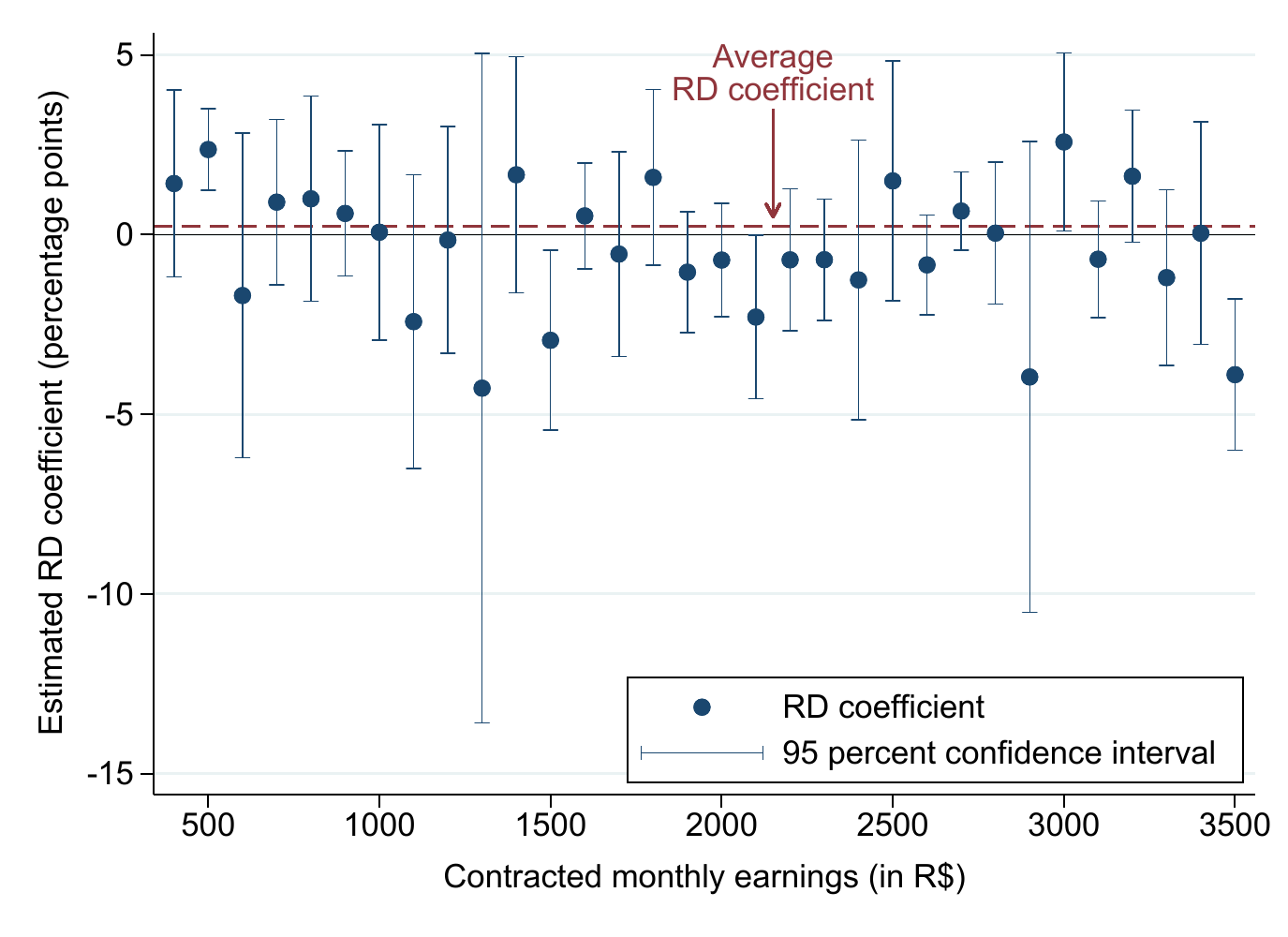}
	\end{subfigure}
	{\footnotesize
		\singlespacing \justify
		
		\textit{Notes:} Panel A shows the fraction of contracts accrued by workers earning a salary just below and just above each salary divisible by 100, using a bandwidth $h = 10$. For example, the figure shows that approximately 48\% of all workers earning [R\$490, R\$510] - \{R\$500\} are contracts just below R\$500, that is, workers earning [R\$490, R\$500), while the other 52\% come from above R\$500, i.e., workers earning (R\$500, R\$510]. If workers' earnings were uniformly distributed, the share of each side would be 50\%.
		
		Panel B presents the RD estimates of regression \eqref{eq_mccrary}, using as outcome variable the fraction of workers in each salary bin and a bandwidth $h = 10$. Each point in the figure comes from a separate regression using data in the vicinity of a salary divisible by 100. For example, the point estimate at $R\$500$  uses data from workers whose earnings are within a distance $10$ of R\$500 (excluding workers who earn exactly R\$500). The vertical lines denote 95\% confidence intervals using heteroskedasticity-robust standard errors. The horizontal red dashed line denotes the weighted average RD coefficient across all regressions, where the weights are the number of workers used to estimate each coefficient. 
		
		To construct the figures in both panels, I use the new-hires sample. The figures only display workers with earnings above the minimum wage and below R\$3,500 (which roughly corresponds to the 99th percentile of the earnings distribution above the minimum wage).
		
	}
\end{figure}

\textbf{Visual evidence}. Appendix Figure \ref{fig_mass_workers}, Panel A shows the fraction of workers whose earnings are just below and just above salaries divisible by 100. To construct this figure, I compute the number of workers whose earnings are within a bandwidth $h$ of a round salary $w_r$. Specifically, I compute the number of workers whose earnings fall in the range $[w_r - h, w_r)$---these are the workers ``just below'' $w_r$---and in the range $(w_r, w_r + h]$---these are the workers ``just above'' $w_r$. Next, I add up the number of workers just below and just above. Finally, I calculate the fraction of workers that come from each side of the round number. I do this calculation for each salary divisible by 100 and a bandwidth $h= 10$.

I find no systematic differences in the number of workers. For some round salaries (e.g., R\$500), there are more contracts above the round number, while for other salaries (e.g., R\$1,300), the opposite is true.\\ 

\textbf{Regression discontinuity analysis}. Next, I use a RD design to formally test whether the number of workers exhibits a statistically significant jump at round salaries. I follow the approach of papers that look for discontinuities in the number of observations around a target value \cite[e.g.][]{camacho2011manipulation}. Specifically, I estimate the following regression for each $w_r$ divisible by 100:
\begin{align}
	\frac{C_{b}}{N_b} = \tilde{\alpha}_r + \tilde{\beta}_r  \mathbbm{1}{\{w_{b} > w_r\}}  + \tilde{\nu}_r w_{b} + \tilde{\delta}_r w_{b} \mathbbm{1}{\{w_{b} > w_r\}} + \tilde{\varepsilon}_{b} \text{ if } |w_{b}| \leq h \text{ and } w_b \neq w_r, \label{eq_mccrary}
\end{align}
where $C_b$ is the count of contracts in bin $b$, $w_b$ is the salary of the bin, $w_r$ is a round salary, $N_b$ is the total number of contracts within distance $h$ of $w_b$, and $h$ is the bandwidth. The dependent variable is the fraction of contracts in each bin. The coefficient of interest is $\tilde{\beta}_r$. It measures whether there is a discontinuity in the fraction of observations in each bin after crossing a round salary $w_r$. Some left-digit bias models predict $\tilde{\beta}_r > 0$. 

Appendix Figure \ref{fig_mass_workers}, Panel B plots the estimated discontinuity $\tilde{\beta_r}$ at each salary divisible by 100. Each coefficient comes from estimating equation \eqref{eq_mccrary} for a different round salary. Across round numbers, the point estimates are small, in many cases negative, and always statistically indistinguishable from zero. The results are similar using alternative bandwidths. Taken together, the results of this section show that the difference between the number of workers just above and just below round salaries does not exhibit any systematic pattern, tends to be quantitatively small, and is statistically insignificant.

\subsection{Other Alternative Explanations} \label{app:oth-exp}

\subsubsection{Focal points in wage bargaining.} 
\begin{figure}[H]
	\caption{Fraction of workers hired at a coarse salary across industries and occupations} \label{fig_theta_industry_occ}
	\centering
	
	\begin{subfigure}[t]{0.7\textwidth}
		\caption*{Panel A. Industry level} \label{fig_theta_industry}
		\centering
		\includegraphics[width=\textwidth]{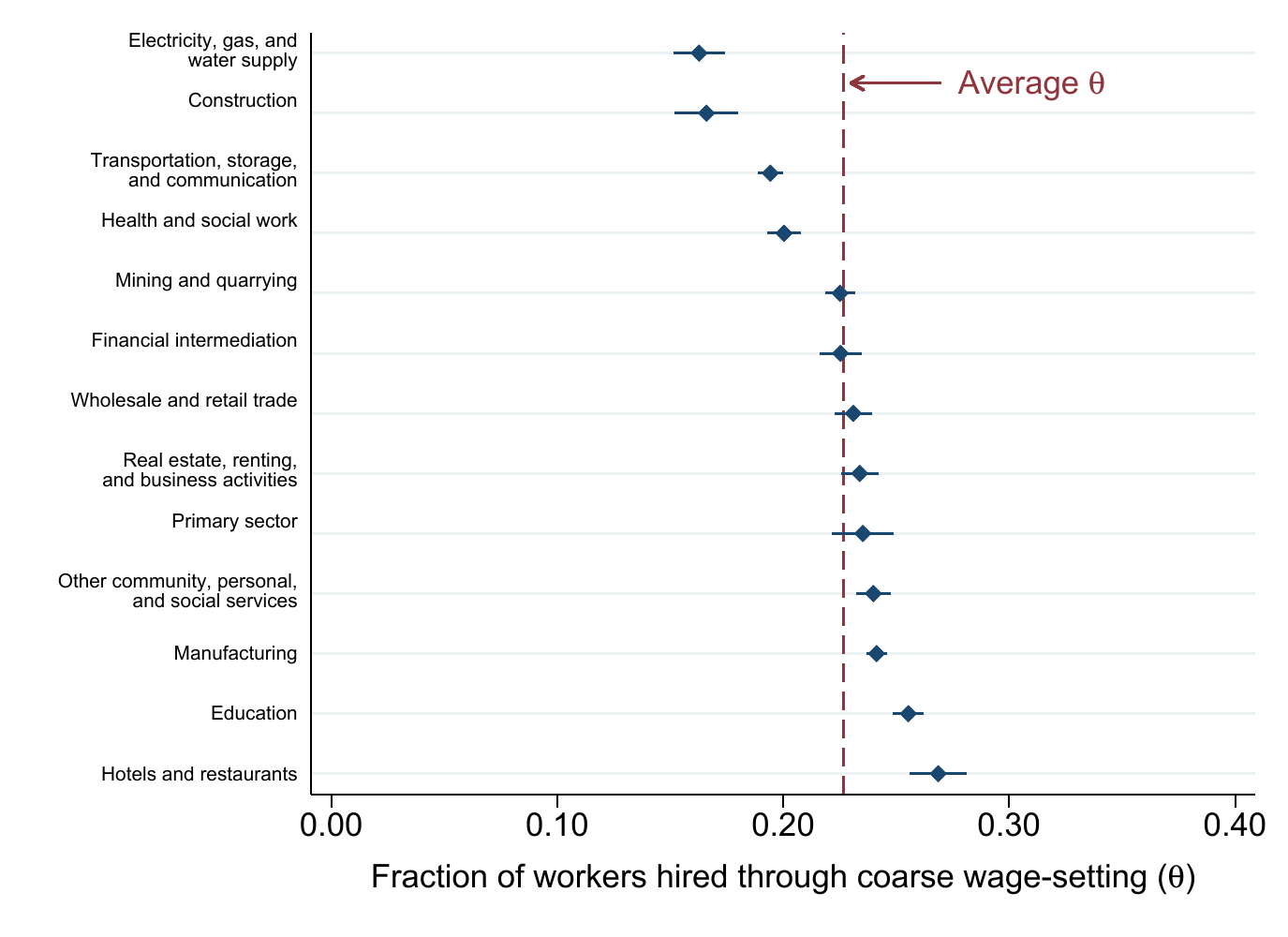}
	\end{subfigure}

	\begin{subfigure}[t]{0.7\textwidth}
		\caption*{Panel B. Occupation level} \label{fig_theta_occ}
		\centering
		\includegraphics[width=\textwidth]{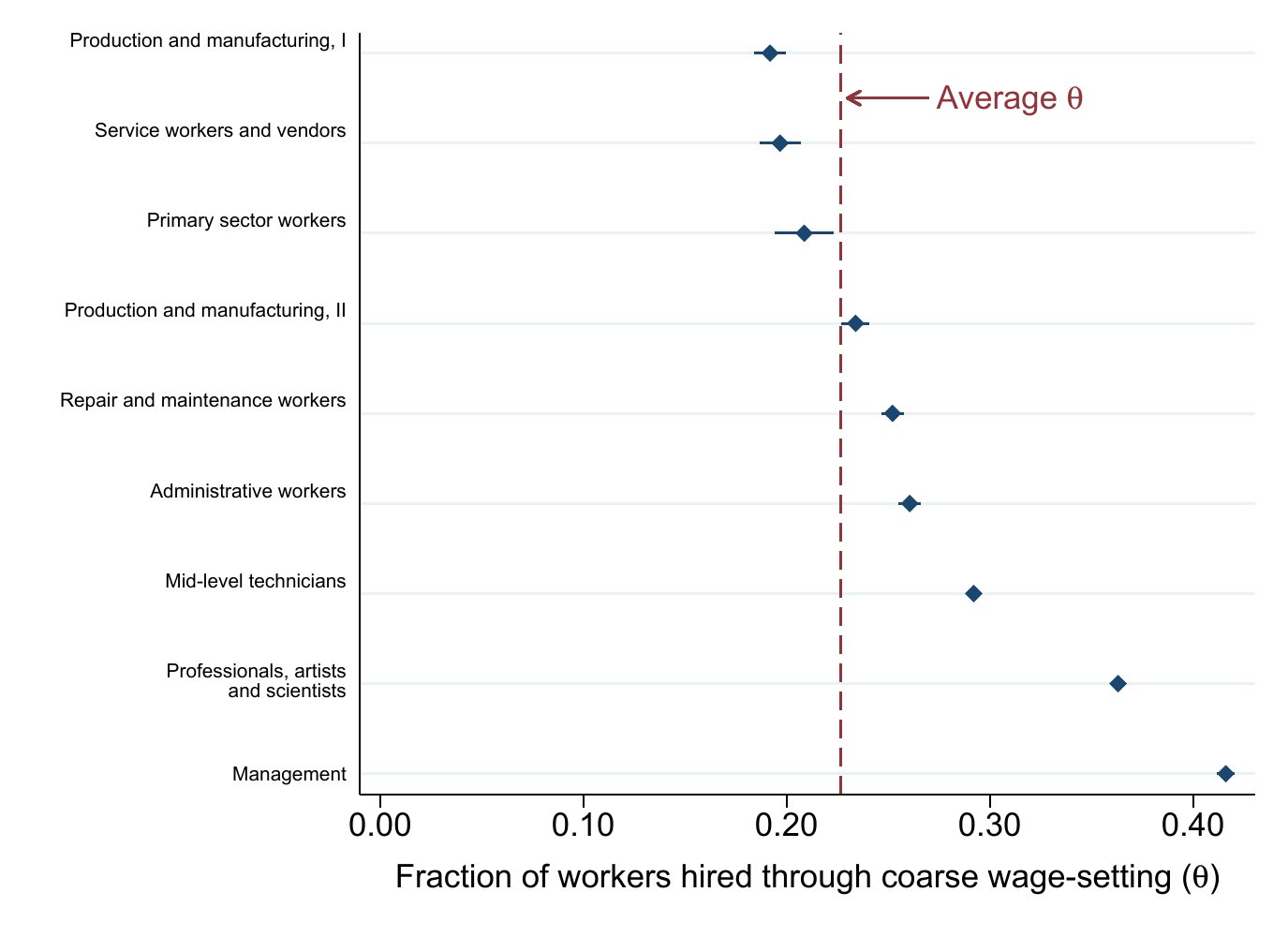}
	\end{subfigure}
	
	\footnotesize \singlespacing \justify \textit{Notes:} This figure shows the estimated fraction of workers hired at a coarse salary across two-digit industries (Panel A) and occupations (Panel B). To construct this figure, I estimate $\hat{\theta}$ conditioning on the firm industry (Panel A) or the occupation of the new hire (Panel B), following the methodology described in Section \ref{sub:bunching}. Horizontal lines represent 95\% confidence intervals. The vertical dashed red line displays the unconditional fraction of workers hired at a coarse salary.

\end{figure}

If workers and firms bargain over the initial salary and round numbers are focal points in these negotiations, then we might expect to observe bunching at round salaries. \cite{hall2012evidence} show that wage bargaining is more prevalent across high-wage knowledge workers, whereas wage posting is more frequent in low-wage blue-collar occupations. Therefore, if the bunching were driven entirely by focal points in wage bargaining, we should not expect to observe any bunching in low-wage occupations, where take-it-or-leave-it offers are more prevalent. To test this hypothesis, I estimate the fraction of workers hired at coarse wages across industries and occupations. Appendix Figure \ref{fig_theta_industry_occ} shows the results. 

Overall, coarse wages are prevalent both across industries where we should expect more wage-posting (such as manufacturing) and more wage-bargaining (such as financial intermediation). Similarly, coarse wages are pervasive across both blue-collar occupations (like administrative workers) and white-collar occupations (like professionals, artists, and scientists). Therefore, focal points in negotiations are unlikely to explain the bunching observed in the data.

\subsubsection{Focal points in collective bargaining agreements.} 
\begin{figure}[H]
	\caption{Collective bargaining agreements and fraction of workers hired through coarse wage-setting ($\hat{\theta}$)} \label{fig_theta_cba}
	\centering
	\includegraphics[width=.75\linewidth]{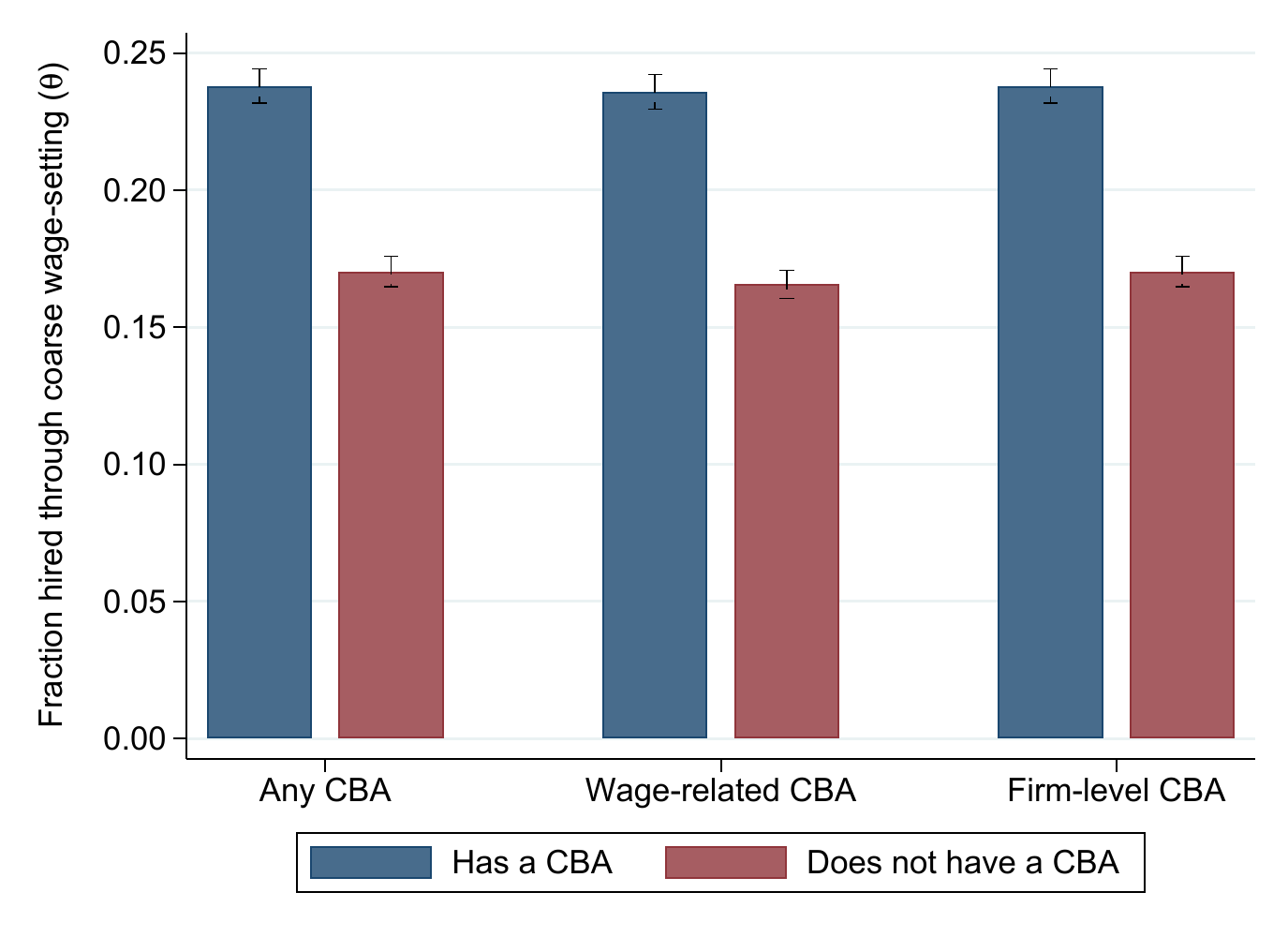}
	\footnotesize \singlespacing \justify \textit{Notes:} This figure shows the estimated fraction of workers through coarse wage-setting for firms that did/did not sign a collective bargaining agreement (CBA) during 2008--2017. To construct this figure, I estimate $\hat{\theta}$ conditioning on a firm signing a CBA following the methodology described in Section \ref{sub:bunching}. Vertical lines represent 95\% confidence intervals.
\end{figure}

The bunching of salaries at round-numbered wages could be explained by round numbers acting as focal points in collective bargaining agreements (CBAs), which are legal contracts between a firm and a union representing the workers. To test this explanation, I use data on the universe of CBAs signed during 2008--2017 \citep{lagos2023labor}. For context, 11.7\% of the workers in the new-hires sample were hired by firms that signed a CBA, and 9.1\% of the workers hired at a round-numbered wage were hired by firms that signed a CBA. I use this data to estimate the fraction of workers hired at coarse wages for firms that signed a CBA and firms that did not sign a CBA. Appendix Figure \ref{fig_theta_cba} shows the results. 

Coarse wage-setting is prevalent in both firms that signed a CBA (blue bars) and firms without a CBA (red bars). This is true for firms that signed any type of CBA, and also for  firms that signed a CBA that includes a wage clause (usually related to wage floors and salary adjustments). Therefore, focal points in CBAs are unlikely to the bunching observed in the data.

\subsubsection{Fairness concerns.} 
\begin{figure}[H]
	\caption{Firm size and fraction of workers hired through coarse wage-setting ($\hat{\theta}$)} \label{fig_theta_firm_size}
	\centering
	\includegraphics[width=.75\linewidth]{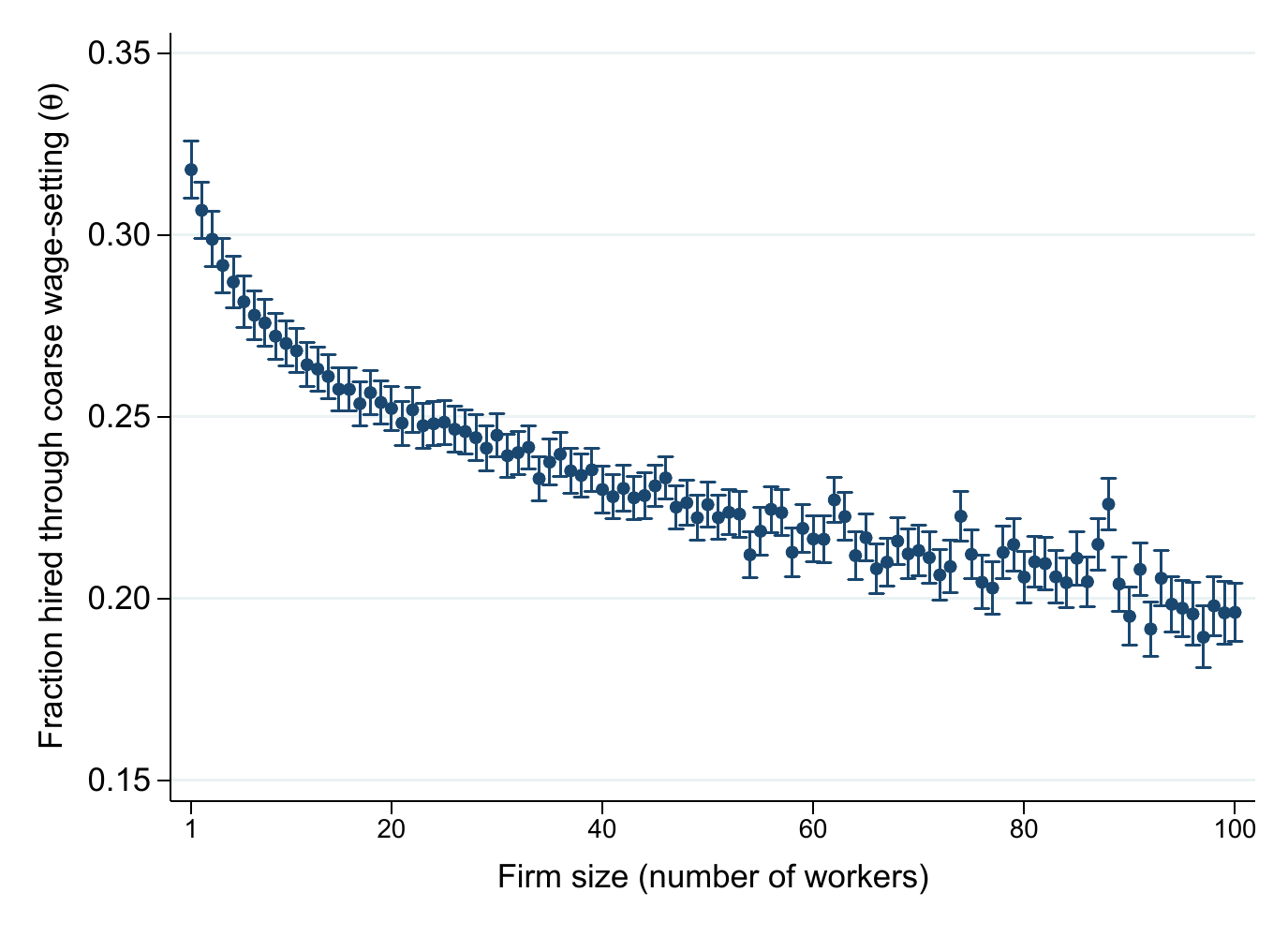}
	\footnotesize \singlespacing \justify \textit{Notes:} This figure shows the estimated fraction of workers through coarse wage-setting across firms of different sizes. To construct this figure, I estimate $\hat{\theta}$ conditioning on firm size following the methodology described in Section \ref{sub:bunching}. Vertical lines represent 95\% confidence intervals.
\end{figure}

Inequity aversion and fairness concerns might induce firms to pay the same salary to coworkers performing the same tasks, even if their productivity is different. By definition, fairness concerns should only matter in firms that employ multiple employees. However, firms with just one employee are the ones most likely to pay coarse wages (Appendix Figure \ref{fig_theta_firm_size}).

\subsubsection{Round wages as a signal of job quality.} In the consumer market, some high-quality firms price their products at round numbers to signal their quality. Some evidence suggests high-end retailers are more likely to round their prices relative to low-end retailers \citep{stiving_price-endings_2000}. In the labor market, firms might also use the roundness of the salary to signal the job's quality. Crucial to this information-based explanation is that consumers or job-seekers, correspondingly, lack information about the quality of relative products or jobs. Otherwise, there would not be a need to use prices to signal quality. If workers become better at assessing the quality of a job as they gain more experience, we should expect firms hiring more experienced workers to be less likely to bunch. However, this is the opposite of what I find. As worker experience increases, firms are more likely to pay a coarse wage.

\subsubsection{Changes in marginal tax rates.} 
\begin{figure}[H]
	\caption{Distribution of contracted salaries and kinks in the income tax schedule during 2015} \label{fig_mtr_2015}
	\centering
	\includegraphics[width=.75\linewidth]{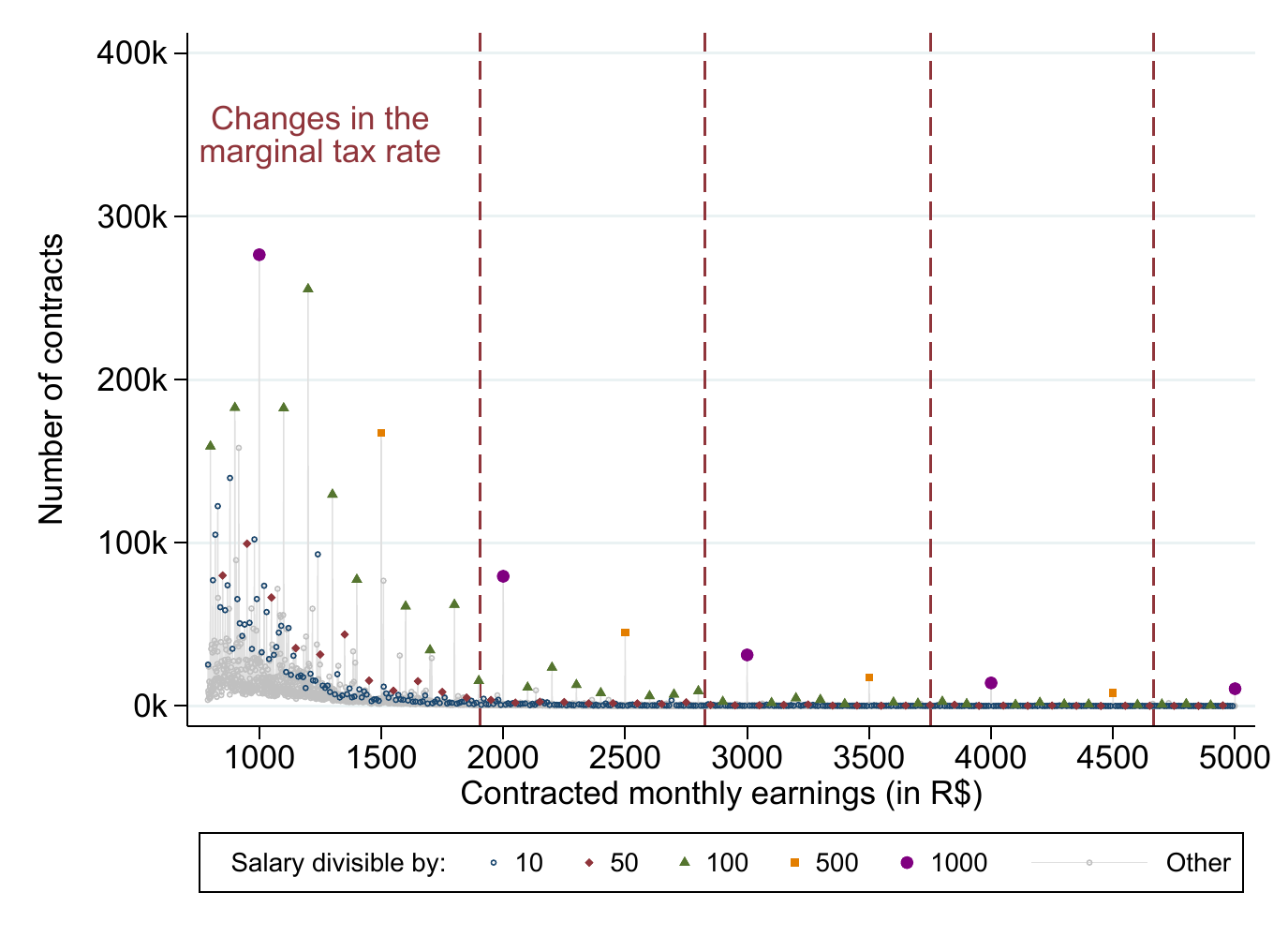} %
	\footnotesize \singlespacing \justify \textit{Notes:} This figure shows the distribution of contracted salaries in the new-hires sample during 2015. Red dashed lines indicate kinks in the personal income tax rate during 2015. To construct this figure, I first group workers in R\$1 bins and then count the number of workers in each bin. Workers whose contracted salary is a round number are denoted with colored markers. The figure only displays workers with earnings above the minimum wage and below R\$3,500 (which roughly corresponds to the 99th percentile of the distribution of earnings above the minimum wage). See Appendix \ref{app:data} for the sample restrictions.  
	
\end{figure}

Beginning with \cite{saez_taxpayers_2010}, several papers have shown that changes in marginal incentives---particularly, changes in marginal tax rates---can generate bunching. Thus, one possible concern is that the estimate of $\theta$ might be confounded by changes in the marginal tax rate. To assess this, I collected data on all the changes in the personal income tax rate in Brazil from 2007--2015. I find that none of the kink points in this period were at round numbers. Furthermore, there is no detectable bunching at any of the kink points. For example, Appendix Figure \ref{fig_mtr_2015} shows the distribution of earnings and kink points using data from 2015. For monthly earnings below R\$1,903.98, the marginal tax rate is zero. The marginal tax rate jumps to 7.5\% for earnings between R\$1,903.99 and R\$2,826.65 and keeps increasing by 7.5 percentage points at each of the following income thresholds: R\$2,826.66, R\$3,751.06, and R\$4,664.68. There is no bunching at any of these thresholds. The lack of bunching at the kink points is consistent with the findings of \cite{saez_taxpayers_2010} and \cite{chetty_adjustment_2011}, who show that the bunching observed in tax data is driven by the self-employed---who have more scope to manipulate their earnings---rather than wage employees.

\clearpage

\clearpage

\clearpage

\clearpage

\clearpage

	\clearpage 
\section{Changes in the Minimum Wage and Coarse Wage-Setting} \label{app:min-wage} 

\setcounter{table}{0}
\setcounter{figure}{0}
\setcounter{equation}{0}	
\renewcommand{\thetable}{G\arabic{table}}
\renewcommand{\thefigure}{G\arabic{figure}}
\renewcommand{\theequation}{G\arabic{equation}}

In this Appendix, I study how coarse wage-setting interacts with changes in the minimum wage (MW). \cite{dube_monopsony_2020} note that whenever a minimum wage is equal to a round number, two types of firms hire at the minimum wage: those that are constrained by the wage floor and those that are misoptimizing with respect to wages and pay the minimum wage because it is a round number. An increase in the minimum wage affects both types of firms and possibly causes the second type of firm to fully optimize wages. A similar logic follows for firms that pay a round-numbered wage below the new minimum wage.

I observe hiring decisions under fifteen different federal minimum wages, seven of which are round numbers (see Appendix Table \ref{tab_min_wages}). I also observe the year $t+1$ salary of workers hired in year $t$. Thus, to shed light on this potential spillover effect, I analyze the fraction of workers who earn a non-round salary in year $t+1$ as a function of the salary at which they were hired.

Table \ref{tab_mw_spillover} summarizes all possible wage transitions. Panel A shows the transitions for workers that were hired at the minimum wage, $w_t =$ MW$_t$; Panel B for workers hired at a wage above the minimum wage, but below the minimum wage of the following year, $w_{t} \in ($MW$_t,$ MW$_{t+1})$; and Panel C for workers hired at a wage above the year $t+1$ minimum wage, $w_{t} \geq $MW$_{t+1}$. By construction, only workers in Panels A and B are directly affected by the change in the minimum wage between year $t$ and year $t+1$. Hence, the transitions in Panel C are useful as a comparison group to assess how different types of wages tend to change irrespective of the direct effect due to a change in the minimum wage.

For conciseness, I focus on how a change in the minimum wage affects the round salaries that it crosses. Panel B shows that 47.7\% of the workers hired at a round salary between MW$_t$ and MW$_{t+1}$ in year $t$ earn a non-round salary in year $t+1$ (excluding the new minimum wage). One way to benchmark this magnitude is to compare it to the fraction of workers hired at a round salary \textit{above} MW$_{t+1}$ who earn a non-round salary the following year (excluding the new minimum wage). This figure equals 42.3\%. This benchmark can be thought of as the counterfactual fraction of workers that would earn a non-coarse wage in year $t+1$ had the minimum wage not changed. Comparing these two transitions following a ``differences-in-differences'' approach, suggests that a change in the minimum wage decreases the share of coarse wages by 5.4 percentage points (or 11.3\%). 

An alternative comparison group is the fraction of workers hired at a non-round salary above MW$_{t+1}$ who also earn a non-round salary in year $t+1$. This figure is akin to the likelihood that a firm that optimized salaries in year $t$ also optimizes in year $t+1$. Since this benchmark uses firms that fully optimized wages in the first period, it can be thought of as an upper bound for firms that initially paid coarse wages. The second row of Panel C show that this figure is 88.3\% (column 6). The increase in the minimum wage achieves 54.0\% (= 47.7\%/88.3\%) of this benchmark.

These findings suggest that changes in the minimum wage can have sizable spillover effects on firm wage-setting behavior.

\begin{table}[H]{\footnotesize
		\begin{center}
			\caption{Federal minimum wages in Brazil: 2003--2017} \label{tab_min_wages}
			\newcommand\w{2}
			\begin{tabular}{rcc}
				\midrule
				& \multicolumn{2}{c}{Federal minimum wage } \\
				\cmidrule{2-3}      & In nominal terms & In real terms \\
				& (current R\$) &  (2003 R\$) \\
				\midrule
				
				2003 & \textbf{240} & 240.00 \\
2004 & \textbf{260} & 243.91 \\
2005 & \textbf{300} & 263.34 \\
2006 & \textbf{350} & 294.90 \\
2007 & \textbf{380} & 308.92 \\
2008 & 415 & 319.25 \\
2009 & 465 & 341.04 \\
2010 & \textbf{510} & 356.10 \\
2011 & 545 & 356.86 \\
2012 & 622 & 386.40 \\
2013 & 678 & 396.58 \\
2014 & 724 & 398.28 \\
2015 & 788 & 397.59 \\
2016 & \textbf{880} & 408.32 \\
2017 & 937 & 420.29 \\
 \midrule
			\end{tabular}
		\end{center}
		\begin{singlespace} \vspace{-.5cm}
			\noindent \justify \textit{Notes:} This table indicates the federal minimum monthly salary in R\$ at the end of each calendar year. \textbf{Bolded} figures indicate minimum wages that are round numbers.
		\end{singlespace}
	}
\end{table}

\clearpage 
\begin{landscape}
	\begin{table}[H]{\footnotesize
			\begin{center}
				\caption{Fraction of workers earning a round salary in year $t+1$ as a function of their initial salary} \label{tab_mw_spillover}
				\begin{tabular}{lcccccc}
					\midrule
					&   & \multicolumn{5}{c}{Fraction of workers in year $t+1$ earning:} \\	\cmidrule{3-7}  
					& Fraction of & The new min. & A round  & A non-round &  A round salary   &  A non-round salary   \\
					&  workers in $t$  &  wage (MW$_{t+1}$) &  salary  &  salary  & excluding MW$_{t+1}$ & excluding MW$_{t+1}$ \\
					& (1) & (2) & (3) & (4) & (5) & (6) \\
					\midrule
					\multicolumn{7}{l}{\textbf{\hspace{-1em}Panel A. Workers hired at $w_{t}$ = MW$_t$}} \\  
					Initial salary is a round number &0.020 &0.798 &0.287 &0.713 &0.074 &0.128 \\
Initial salary is not a round number &0.032 &0.865 &0.126 &0.874 &0.047 &0.089 \\
 \midrule					  
					\multicolumn{7}{l}{\textbf{\hspace{-1em}Panel B. Workers hired at $w_{t} \in ($MW$_t,$ MW$_{t+1})$}}         \\
					Initial salary is a round number &0.125 &0.882 &0.057 &0.943 &0.047 &0.071 \\
Initial salary is not a round number &0.396 &0.905 &0.028 &0.972 &0.023 &0.072 \\
 \midrule					  
					\multicolumn{4}{l}{\textbf{\hspace{-1em}Panel C. Workers hired at $w_{t} \geq$  MW$_{t+1}$}} &   &   &  \\
					Initial salary is a round number &0.126 &0.009 &0.554 &0.446 &0.546 &0.445 \\
Initial salary is not a round number &0.302 &0.005 &0.097 &0.903 &0.096 &0.899 \\
 \midrule					  
				\end{tabular}
			\end{center}
			\begin{singlespace} \vspace{-.5cm}
				\noindent \justify \textit{Notes:} This table shows worker transitions between different types of salaries. The rows in each panel indicate the salary at which the firm hired the worker. Panel A includes workers hired at the federal minimum wage. Panel B includes workers hired at a salary above the federal minimum wage of the hiring year (year $t$) but below the federal minimum wage of the following year ($t+1$). Panel C includes workers hired at a salary above the year $t+1$ federal minimum wage. Workers that appear to be hired at a salary below the minimum wage are excluded. I present the transitions separately for workers hired at a round salary (first row of each panel) and a non-round salary (second row of each panel). In Panel A, this is equivalent to splitting the sample based on whether the federal minimum wage is a round number.
				
				Column 1 shows the fraction of workers hired at each type of salary. The sum of the rows in column 1 equals one. The subsequent columns indicate the salary earned by the worker in year $t+1$. Column 2 shows the fraction of workers that earn the $t+1$ federal minimum wage. Columns 3 and 5 show the fraction of workers that earn a round salary in $t+1$. In column 5, this fraction is calculated using salaries different from the new minimum wage (only relevant for years in which the new minimum wage is a round salary, see Appendix Table \ref{tab_min_wages}). Columns 4 and 6 show the fraction of workers that do not earn a round salary in $t+1$. In column 6, this figure is calculated using salaries different from the new minimum wage (only relevant for years in which the new minimum wage is not a round salary). Columns 2, 5, and 6 add up to one. Similarly, columns 3 and 4 also add up to one. 
			\end{singlespace}
		}
	\end{table}
	
\end{landscape}

\end{document}